\documentclass[referee]{aa} 
\usepackage{natbib}
\usepackage{verbatim}
\usepackage{subfig}
\usepackage{amsmath}
\usepackage{multirow}
\usepackage{multicol}
\usepackage{longtable}
\usepackage[varg]{txfonts}
\usepackage{graphicx}
\usepackage{lscape}
\usepackage{hyperref}

\newcommand{\Msun}{\mbox{$M_{\odot}$}}
\newcommand{\Mearth}{\mbox{$M_{\oplus}$}}

\newcommand{\Lsun}{\mbox{$L_{\odot}$}}
\newcommand{\Mjup}{\mbox{$M_{Jup}$}}

\def\farcs{\hbox{$.\!\!^{\prime\prime}$}}

\def\ref{\hangindent=15pt     
        \hangafter=1}

\def\lesssim{\mathrel{\hbox{\rlap{\hbox{%
  \lower4pt\hbox{$\sim$}}}\hbox{$<$}}}}
\def\gtrsim{\mathrel{\hbox{\rlap{\hbox{%
  \lower4pt\hbox{$\sim$}}}\hbox{$>$}}}}

\def\farcs{\hbox{$.\!\!^{\prime\prime}$}}

\begin{document}

\title{A \textit{Herschel} PACS survey of the dust and gas in Upper Scorpius disks \thanks{{\it Herschel} is an ESA space observatory with science instruments provided by European-led Principal Investigator consortia and with important participation from NASA.} }


\author{Geoffrey S. Mathews\inst{\ref{inst1}}, Christophe Pinte\inst{\ref{inst2}}, Gaspard Duch\^ene\inst{\ref{inst2},\ref{inst3}}, Jonathan P. Williams\inst{\ref{inst4}}, Fran\c cois M\'enard\inst{\ref{inst2},\ref{inst5}}} 

\institute{Leiden Observatory, Leiden University, PO Box 9513, 2300 RA, Leiden, The Netherlands \email{gmathews@strw.leidenuniv.nl} \label{inst1}
\and
CRNS-INSU / UJF-Grenoble 1, Institut de Plan\'etologie et d'Astrophysique de Grenoble (IPAG) UMR 5274, Grenoble, F-38041, France \label{inst2}
\and
Astronomy Department, University of California, Berkeley CA 94720-3411 USA \label{inst3}
\and
Institute for Astronomy, University of Hawaii, 2680 Woodlawn Dr., Honolulu HI 96826 \label{inst4}
\and
UMI-FCA (UMI 3386), CNRS/INSU France and Universidad de Chile, Santiago, Chile \label{inst5}
}

\date{}

\abstract{We present results of far-infrared photometric observations with \textit{Herschel} PACS of a sample of Upper Scorpius stars, with a detection rate of previously known disk-bearing K and M stars at 70, 100, and 160 $\mu$m of 71\%, 56\%, and 50\%, respectively.  We fit power-law disk models to the spectral energy distributions of K \& M stars with infrared excesses, and have found that while many disks extend in to the sublimation radius, the dust has settled to lower scale heights than in disks of the less evolved Taurus-Auriga population, and have much reduced dust masses.    We also conducted \textit{Herschel} PACS observations for far-infrared line emission and JCMT observations for millimeter CO lines.  Among B and A stars, 0 of 5 debris disk hosts exhibit gas line emission, and among K and M stars, only 2 of 14 dusty disk hosts are detected.  The OI 63 $\mu$m and CII 157 $\mu$m lines are detected toward $[$PZ99$]$ J160421.7-213028 and $[$PBB2002$]$ J161420.3-190648, which were found in millimeter photometry to host two of the most massive dust disks remaining in the region.  Comparison of the OI line emission and 63 $\mu$m continuum to that of Taurus sources suggests the emission in the former source is dominated by the disk, while in the other there is a significant contribution from a jet.  The low dust masses found by disk modeling and low number of gas line detections suggest that few stars in Upper Scorpius retain sufficient quantities of material for giant planet formation.  By the age of Upper Scorpius, giant planet formation is essentially complete.  } 


\titlerunning{PACS survey of Upper Scorpius}
\authorrunning{G. Mathews et al.}

\maketitle

\section{Introduction}
\label{sec:intro}
Circumstellar disks of gas and dust are the sites of planet formation.  They are most readily detected from  continuum emission of cool dust at wavelengths from the (near-)infrared to millimeter wavelengths. At least initially, the dust only accounts for 1\% of the disk mass, and a more complete understanding of disk structure, chemistry, evolution, and -- not least -- the formation of giant planets, requires that we measure and accurately interpret the line emission from disks.  The interpretation of line data requires a complete understanding of the dust, which establishes the temperature baseline for gas models.  

The detection and characterization of disk spectral line emission is far more challenging than observations of the continuum.  Due to atmospheric opacity, ground based observations have been restricted to (sub-)millimeter molecular rotational lines in the cold outer disk, where depletion of molecules (i.e. freeze-out onto dust grains) is important and in the near-infrared (NIR) from the hot, central few AU, regions of disks \citep{van-Dishoeck:2006a}.

Several surveys have been conducted using the ESA \textit{Herschel} Space Observatory \citep{2010A&A...518L...1P} to examine line emission from circumstellar disks.  The Water In Star-forming regions with \textit{Herschel} (WISH) guaranteed time key program focused on high resolution spectroscopy to trace water-related chemistry in objects ranging from collapsing clouds to protoplanetary disks \citep{van-Dishoeck:2011zr}.  The Dust, Ice, and Gas In Time (DIGIT) open time key project has focused on broad spectral surveys of a smaller number of objects ranging from embedded to the Class II disk stage \citep[e.g.][]{van-Kempen:2010rt}

Two other large surveys have focused on photometry to characterize the dust of circumstellar disks.  DUst around NEarby Stars (DUNES), has carried out 100 and 160 $\mu$m, and in a few cases, longer wavelength photometry to search for debris disks and proto-Kuiper belts \citep{Eiroa:2013pd}.  Disk Emission via a Bias-free Reconnaissance in the Infrared/Submillimeter (DEBRIS) has focused on a photometric survey for debris disks \citep{Matthews:2010fr}.

The Gas in Protoplanetary Systems \citep[GASPS,][]{Dent:2013qy} \textit{Herschel} open time key project, on the other hand, has focused on a mix of photometry and spectroscopy, with one key goal being to trace the evolution of disk mass.  It is the largest circumstellar disk survey carried out with \textit{Herschel}.
Using the Photoconductor Array Camera and Spectrometer \citep[PACS,][]{2010A&A...518L...2P} integral field unit spectrometer and bolometer array, we are surveying nearly 200 young stars in three emission lines, [OI] 63$\mu$m, [OI] 145$\mu$m, and [CII] 158$\mu$m.  The GASPS team is modeling the data using a 3 dimensional radiative transfer code MCFOST \citep{Pinte:2006, Pinte:2009} and the gas thermal balance and chemistry code ProDiMo \citep{Woitke:2009,2010A&A...510A..18K}.  We have also built a grid of 300,000 disk models \citep[DENT,][]{2010MNRAS.405L..26W,Pinte:2010,Kamp:2011a} exploring a variety of disk parameters which has proven a useful tool for analysis of observations.  

This survey covers young stars in 6 nearby clusters spanning a range of ages 1--30\,Myr.  At a distance of 145 pc \citep{de-Zeeuw:1999}, \object{Upper Scorpius} (Upper Sco) is one of the closest sites of recent star formation.  At an age of 5 Myr \citep{Preibisch:1999}, it serves as an intermediate age sample for GASPS.  However, recent work questions the 5 Myr age of the group.  \cite{Pecaut:2012} find evidence from examination of B, A, F, and G stars that the age for Upper Scorpius may be $\sim$10 Myr.  In our discussion below, we examine the implications either age presents for planet formation.  

Whatever the exact age, recent studies have indicated that the disks in Upper Scorpius are greatly evolved.  Few stars are still accreting \citep[e.g.][]{Walter:1994,2009AJ....137.4024D}, suggesting that most disks have depleted their inner disk gas content.  \textit{\textit{Spitzer}} photometry at 4.5 to 70 $\mu$m \citep{2006ApJ...651L..49C,Carpenter:2009b}, as well as spectroscopy \citep{2009AJ....137.4024D}, revealed that the overall population of Upper Sco disks is more evolved than that in younger star forming regions, with only $\sim20$\% of stars exhibiting an infrared excess at any wavelength.  Moreover, many of these excesses are consistent with debris disks \citep[$L_{disk} / L_{star} < 10^{-2}$,][]{Wyatt:2008} rather than primordial disks.  Under the assumption of the primordial 100-to-1 gas-to-dust mass ratio, millimeter photometry indicates that less than 2\% of Upper Scorpius disks have sufficient mass to support giant planet formation, and disk masses have become so low as to end the correlation between accretion and NIR excess seen in younger regions \citep{Mathews:2011}.

We have carried out the largest survey to date for far-infrared photometry and gas-line emission from a group of stars in the late stages of disk evolution, attaining a 3$\sigma$ gas-mass sensitivity of $\sim$1 \Mjup ~and determining the large scale changes in disk geometry.  In Sect. \ref{sec:sample}, we describe our target sample.  In Sect. \ref{sec:observations}, we describe our observations and data reduction, and in Sect. \ref{sec:results}, we describe our observational results.  In Sect. \ref{sec:models}, we model the disks and discuss the improved estimates of their dust characteristics enabled by our observations.  We conclude with a discussion of the implications for planet formation and disk evolution in Sect. \ref{sec:discussion}, and summarize our findings in Sect. \ref{sec:summary}.  

\section{Sample}
\label{sec:sample}

The GASPS Upper Scorpius sample emphasizes sources having a NIR excess at 8 or 16 $\mu m$ in the NIR continuum survey of 218 Upper Sco members of \cite{2006ApJ...651L..49C}.  We include a small number of nonexcess sources to both place limits on the number of sources with excesses starting at longer wavelengths and to search for disks with gas line emission but lacking observable dust emission.  We observed 8 of the 9 B and A stars with an 8 or 16 $\mu m$ excess (excluding two known Be stars), all 7 NIR-excess K stars, and all 17 M stars with a NIR-excess.  For our control sample, we included 3 B and A, 4 G and K, and 6 M stars with no excess at 8 or 16 $\mu$m.  In Table \ref{tab:stellar} we present an overview of the previously determined properties of our sample, including the spectral type, extinction, luminosity, surface temperature, H$\alpha$ equivalent width, and disk classification as Class II, III, or debris disk \citep[using the NIR slope criteria of ][]{Greene:1994}, and classification as Classical or Weak-line T Tauri star \citep[using the spectral type dependent criteria of ][]{White:2003}.  We also give each object a number by which it is referred in the tables and text.  In the remainder of the text, we refer to $[$PBB2002$]$ J161420.3-190648 (object 36) and $[$PZ99$]$ J160421.7-213028 (object 41) as J1614-1906 and J1604-2130, respectively.

\begin{table*}
  \caption{ \label{tab:stellar}Target properties} 
\begin{tabular}{r l c c c c c c c c}
\hline\hline

Object    &   Name\tablefootmark{a}  &    Sp.Type\tablefootmark{b} &    A$_V$          &  log (L)  &    log (T)  &    M$_{star}$  &    W$_{\lambda}$(H$\alpha$)\tablefootmark{c} &    SED &   Accretion\tablefootmark{d}    \\
       &         &                                                   &     Mags  &  L$_{sun}$            &    K    &    $M_{\rm{sun}}$  &    \AA                                                                &    class &   state   \\
  \hline

1\phantom{0}  & HIP 76310                             	   			&         A0V &      0.1 &     1.43 &    3.960 &      2.2 &              $<-0.1$  &                         debris        & No	\\
2\phantom{0}  & HIP 77815                                				&         A5V &      0.7 &     1.06 &    3.914 &      2.0 &                  ... &                 none                     & ...	\\
3\phantom{0}  & HIP 77911                                				&         B9V &      0.3 &     1.88 &    4.050 &      2.8 &              $<-0.1$  &                         debris        & No	\\
4\phantom{0}  & HIP 78099                                				&         A0V &      0.5 &     1.43 &    3.960 &      2.2 &              $<-0.1$  &                         none            & No	\\
5\phantom{0}  & HIP 78996                                				&         A9V &      0.4 &     1.10 &    3.870 &      1.8 &              $<-0.1$  &                      debris        & No	\\
6\phantom{0}  & HIP 79156                                				&         A0V &      0.5 &     1.42 &    3.980 &      2.2 &              $<-0.1$  &                         debris        & No	\\
7\phantom{0}  & HIP 79410                                				&         B9V &      0.6 &     1.63 &    4.020 &      2.5 &              $<-0.1$  &                       debris        & No	\\
8\phantom{0}  & HIP 79439                                				&         B9V &      0.6 &     1.65 &    4.020 &      2.5 &              $<-0.1$  &                        debris        & No	\\
9\phantom{0}  & HIP 79878                                				&         A0V &      0.0 &     1.51 &    3.980 &      2.3 &              $<-0.1$  &                       debris        & No	\\
10\phantom{0}  & HIP 80088                                				&         A9V &      0.4 &     0.91 &    3.880 &      1.7 &              $<-0.1$  &                        debris        & No	\\
11\phantom{0}  & HIP 80130                                				&         A9V &      0.6 &     1.20 &    3.920 &      1.9 &              $<-0.1$  &                          none             & No	\\
12\phantom{0}  &  RX J1600.7-2343                         			&              M2 &      0.5 &    -0.95 &    3.557 &      0.5 &                  ... &                          none & ...	\\
13\phantom{0} & ScoPMS 31                                				&           M0.5V &      0.9 &    -0.43 &    3.570 &      0.4 &             -21.06 &                    II &      C	\\
14\phantom{0}  &      $[$PBB2002$]$ J155624.8-222555     &              M4 &      1.7 &    -1.12 &    3.516 &      0.3 &      -5.4, -5.51 &                          II &      W	\\
15\phantom{0}  &      $[$PBB2002$]$ J155706.4-220606     &              M4 &      2.0 &    -1.31 &    3.513 &      0.2 &      -3.6, -9.92 &                         II &      W	\\
16\phantom{0}  &      $[$PBB2002$]$ J155729.9-225843     &              M4 &      1.4 &    -1.30 &    3.511 &      0.2 &      -7.0, -6.91 &                       II &      W	\\
17\phantom{0}  &      $[$PBB2002$]$ J155829.8-231007     &              M3 &      1.3 &    -1.63 &    3.532 &      0.3 &      -250, -158 &                         II &      C	\\
18\phantom{0}  &      $[$PBB2002$]$ J160210.9-200749     &              M5 &      0.8 &    -1.40 &    3.501 &      0.2 &             -3.5 &                             none &      W	\\
19\phantom{0}  &      $[$PBB2002$]$ J160245.4-193037     &              M5 &      1.1 &    -1.31 &    3.495 &      0.2 &             -1.1 &                              none &      W	\\
20\phantom{0} &      $[$PBB2002$]$ J160357.9-194210     &              M2 &      1.7 &    -0.87 &    3.546 &      0.5 &      -3.0, -2.70 &                        II &      W	\\
21\phantom{0}  &      $[$PBB2002$]$ J160525.5-203539     &              M5 &      0.8 &    -1.30 &    3.499 &      0.2 &      -6.1, -8.61 &                         III &      W	\\   
22\phantom{0}  &      $[$PBB2002$]$ J160532.1-193315     &              M5 &      0.6 &    -1.88 &    3.499 &      0.1 &      -26.0, -152 &                        III &      C	\\   
23\phantom{0}  &      $[$PBB2002$]$ J160545.4-202308     &              M2 &      2.2 &    -0.98 &    3.560 &      0.6 &      -35.0, -2.04 &                          II &      C	\\
24\phantom{0}  &      $[$PBB2002$]$ J160600.6-195711     &              M5 &      0.6 &    -0.82 &    3.505 &      0.3 &      -7.5, -4.11 &                         II &      W	\\
25\phantom{0}  &      $[$PBB2002$]$ J160622.8-201124     &              M5 &      0.2 &    -1.47 &    3.499 &      0.2 &      -6.0, -3.12 &                         II &      W	\\
26\phantom{0}  &      $[$PBB2002$]$ J160643.8-190805     &              K6 &      1.8 &    -0.31 &    3.623 &      0.8 &             -2.39 &                        III &      W	\\   
27\phantom{0}  &      $[$PBB2002$]$ J160702.1-201938     &              M5 &      1.0 &    -1.37 &    3.501 &      0.2 &       -30.0, -8.3 &                      II &      C	\\
28\phantom{0}  &      $[$PBB2002$]$ J160801.4-202741     &              K8 &      1.5 &    -0.45 &    3.601 &      0.7 &             -2.3 &                          none &      W	\\
29\phantom{0}  &      $[$PBB2002$]$ J160823.2-193001     &              K9 &      1.5 &    -0.49 &    3.586 &      0.7 &      -6.0, -2.75 &                        II &      W	\\
30\phantom{0}  &      $[$PBB2002$]$ J160827.5-194904     &              M5 &      1.1 &    -1.11 &    3.495 &      0.2 &      -12.3, -14.5 &                       III &      W	\\
31\phantom{0}  &      $[$PBB2002$]$ J160900.0-190836     &              M5 &      0.7 &    -1.45 &    3.503 &      0.2 &      -15.4, -12.8 &                     II &      W	\\
32\phantom{0}  &      $[$PBB2002$]$ J160900.7-190852     &              K9 &      0.8 &    -0.60 &    3.592 &      0.7 &      -12.7, -20.0 &                     II &      C	\\
33\phantom{0}  &      $[$PBB2002$]$ J160953.6-175446     &              M3 &      1.9 &    -1.03 &    3.539 &      0.4 &      -22.0, -22.2 &                       II &      C	\\
34\phantom{0}  &      $[$PBB2002$]$ J160959.4-180009     &              M4 &      0.7 &    -1.26 &    3.518 &      0.3 &      -4.0, -4.41 &                     II &      W	\\
35\phantom{0}  &      $[$PBB2002$]$ J161115.3-175721     &              M1 &      1.6 &    -0.42 &    3.574 &      0.6 &      -2.4, -4.47 &                   II &      W	\\
36\tablefootmark{e}   &      $[$PBB2002$]$ J161420.3-190648     &              K5 &      1.8 &    -0.59 &    3.630 &      0.8 &      -52.0, -43.7 &                             II &      C	\\

37\phantom{0}  &      $[$PZ99$]$ J153557.8-232405             &              K3 &      0.7 &    -0.12 &    3.649 &      0.9 &              0.0 &                       none &      W	\\
38\phantom{0}  &      $[$PZ99$]$ J154413.4-252258             &              M1 &      0.6 &    -0.43 &    3.564 &      0.4 &             -3.15 &                        none &      W	\\
39\phantom{0}  &      $[$PZ99$]$ J160108.0-211318             &              M0 &      0.0 &    -0.36 &    3.576 &      0.4 &                 -2.4 &                         none &      W	\\
40\phantom{0}  &      $[$PZ99$]$ J160357.6-203105             &              K5 &      0.9 &    -0.24 &    3.630 &      0.8 &             -11.57 &                     II &      C	\\
41\tablefootmark{f}  &      $[$PZ99$]$ J160421.7-213028             &              K2 &      1.0 &    -0.12 &    3.658 &      1.0 &            -0.57 &                 II &      W	\\
42\phantom{0}  &      $[$PZ99$]$ J160654.4-241610             &              M3 &      0.0 &    -0.34 &    3.538 &      1.0 &             -3.62 &                          none &      W	\\
43\phantom{0}  &      $[$PZ99$]$ J160856.7-203346             &              K5 &      1.4 &    -0.11 &    3.630 &      0.7 &            -0.47 &                        none &      W	\\
44\phantom{0}  &      $[$PZ99$]$ J161402.1-230101             &              G4 &      2.0 &     0.30 &    3.724 &      1.5 &              0.0 &                         none & No	\\
45\phantom{0}  &      $[$PZ99$]$ J161411.0-230536             &              K0 &      2.4 &     0.74 &    3.676 &      1.0 &      0.96, 0.8, 0.38 &                    II &      W	\\

  \hline
\end{tabular}
\tablefoot{
  \tablefoottext{a}{Names with [PZ99] indicate naming based on \cite{Preibisch:1999}, and [PBB2002] indicates naming based on \cite{Preibisch:2002}.}
  \tablefoottext{b}{Spectral types from \cite{Hernandez:2005,Preibisch:1998,Preibisch:2002}}
  \tablefoottext{c}{W$_{\lambda}$(H$\alpha$) from \cite{Hernandez:2005, Preibisch:1998,Preibisch:2002,Riaz:2006}}
  \tablefoottext{d}{Accretion determinations from \cite{Mathews:2011}.  For A and G stars, this indicates whether the star has detectable H$\alpha$ emission possibly tracing accretion.  For K \& M stars, this indicates whether the star exhibits strong enough H${\alpha}$ emission to be classified as accreting according to the spectral type dependent criteria of \cite{White:2003}.  Objects with no H$\alpha$ measurement in the literature are shown with no value.}
  \tablefoottext{e}{Referred to as J1614-1906 in the text.}
  \tablefoottext{f}{Referred to as J1604-2130 in the text.}
}
 \end{table*}

The sources for our spectral types, H$\alpha$ equivalent widths, stellar masses, and disk classifications are described in \cite{Mathews:2011}.  Luminosity, temperature, and extinction estimates are from \cite{Preibisch:1999}, \cite{Preibisch:2002}, and \cite{Hernandez:2005}.

\section{Observations}
\label{sec:observations}

In the following section, we describe new photometry and spectroscopy of these Upper Scorpius sources, as well as our data reduction.  Observations with the {\textit{Herschel} Space Observatory's PACS instrument \citep{2010A&A...518L...2P} were carried out as part of the \textit{\textit{Herschel}} open time key project GASPS (P.I. Dent).  Additional observations were carried out using the Receiver A and HARP instruments on the \textit{James Clerk Maxwell Telescope} (JCMT).  In Table \ref{tab:obs}, we report the observation identification numbers (obsid) and settings for \textit{\textit{Herschel}} observations, as well as the date, instrument, and integration times for JCMT observations.

\subsection{PACS photometry}

We carried out a full survey of our sample using the \textit{Herschel} PACS instrument's photometric scan map mode (PacsPhoto).  Scans were carried out at the medium scan speed of 20$''$/s, with 10 legs of 3$'$ length and separations of 4$''$.  Targets were observed in both the ``blue'' passband centered at 70 $\mu$m, and the ``green'' passband centered at 100 $\mu$m, with each simultaneously observing in the ``red'' passband at 160 $\mu$m.  All but 4 early observations were carried out in a pair of 276 second scan-map observations at 70\degr and 110\degr, so as to provide more even coverage of the field of view and reduce smearing of the point spread function in the scan direction.   

The data were reduced with HIPE 9.0 \citep{Ott:2010} using the standard reduction pipeline for photometric data.  For most objects, there are two scans each in the 70 and 100 $\mu$m bands, with a total of 4 accompanying scans in the 160 $\mu$m band.  For the subset of objects detected at 70 $\mu$m with \textit{\textit{\textit{Spitzer}}} \citep{Carpenter:2009b}, there are only two scans in the 100 $\mu$m band and 2 scans in the 160 $\mu$m band.  The scans for each object were reduced separately, with high-pass filtering conducted using a 1$\sigma$ threshold to mask sources.  We then combined separate scans and projected them at 2'', 2'', and 3'', in the 70, 100, and 160 $\mu$m bands.  These oversample the native PACS pixel scales of 3\farcs2, 3\farcs2, and 6\farcs4, respectively.

We measured the flux in the stacked images using the procedure recommended by the \textit{Herschel} Science Center\footnote{http://herschel.esac.esa.int/twiki/pub/Public/PacsCalibrationWeb /pacs\_bolo\_fluxcal\_report\_v1.pdf}.  This is aperture photometry with an aperture radius of 10.5''.  We used outer sky annuli of 20$''$--30$''$ in the 70 and 100 $\mu$m bands, with aperture corrections of 0.77 and 0.73 based on PACS observations of Vesta.  For 160 $\mu$m photometry, we used a sky annulus of 30$''$--40$''$, with an aperture correction of 0.63.  We calculated the pixel error, $\sigma_{px}$, as the standard deviation of values in the sky annulus, which we then divided by a correlated noise factor.  For our pixel sizes, this term is 0.72, 0.72, and 0.57.   
We assume the noise is background dominated, for which the source flux error was then calculated as $\sigma_{px} n_{ap}^{0.5}$, where $n_{ap}$ is the number of pixels within the aperture.  We then applied the aperture corrections to our error measurements.  There are additional flux calibration errors of 3, 3, and 5\% at 70, 100, and 160 $\mu$m, respectively, which we do not include in our reported errors.  

\subsection{PACS spectroscopy}
\label{sec:obs:PACSspec}

We used the PACS integrated field unit, which consists of a 5$\times$5 filled array of 9\farcs4$\times9$\farcs4 spaxels, to carry out spectroscopic observations of 5 B and A stars, 5 K stars, and 9 M stars with NIR excesses.   Observations were carried out in chop-nod mode in order to remove the telescope emission, with the central spaxel centered on the target in both nods.  We carried out PacsLineSpec observations of these 19 spectroscopic targets, which includes simultaneous observations from 62.94--63.44 and 188.79--190.32 $\mu$m.  

We also carried out PacsRangeSpec observations of 4 K stars and 3 M stars with either detected millimeter continuum or strong infrared excess emission.  These observations consisted of simultaneous observations from 71.82--73.31 and 143.62--146.61 $\mu$m, followed by simultaneous observations at  78.38--79.72 and 156.73--159.44 $\mu$m.  One source, J1604-2130 (object 41), was also observed at 89.30--90.71 and 178.61--181.36 $\mu$m.  

We used the standard PACS spectroscopy reduction pipeline within HIPE 9 to reduce the data, with the addition of flags in intermediate steps to preserve information regarding the random error within wavelength bins.  

For each spaxel, we extracted the A and B nods.  For each nod, we then binned the data in wavelength with nonoverlapping bins half the width of the instrumental resolution.  The mean of the two nods restores the observed flux in each spaxel, and we propagate the errors of the individual nods to determine the error-per-bin in the final spectrum.  We note that due to telescope roll, off-center spaxels do not see the exact same location on sky between nods.  At the corners of the IFU, these offsets can be as large as 2$''$.  Despite these offsets, the off-center spaxels are useful for checking for extended emission or potentially contaminating sources.  

For point source flux measurements, we use the flux in the central spaxel and apply wavelength dependent flux and aperture corrections released by the PACS development team\footnote{http://herschel.esac.esa.int/twiki/pub/Public/PacsCalibrationWeb /PacsSpectroscopyPerformanceAndCalibration\_v2\_4.pdf}, as well as their reported flux calibration uncertainty of 30\%.  These aperture corrections are considered valid for well centered sources, and the flux calibration uncertainty includes the effects of pointing uncertainty.  As a check, we compare the reported position on sky of the central spaxel with the target positions.  These offsets are generally smaller than the 1$\sigma$ pointing uncertainty of 2$''$.

To extract line fluxes and measure continuum fluxes from our reduced, nod-combined, flux corrected spectra, we carried out an inverse-error weighted fit of a gaussian and first-order polynomial to the spectra, attempting fits to the wavelengths of species listed in Table \ref{tab:species}.  The primary targets of our survey are listed in bold type, and few of these emission lines are detected.   
For each line region, we also list the instrumental full width at half maximum (FWHM) and aperture correction.  Using the polynomial fit to each spectral region, we estimated the continuum emission at the rest wavelength for each line.  To estimate the error on the continuum, we calculated the error of the mean for the residual of the polynomial fit in a region from 2 to 10 instrumental FWHM from the line rest wavelength.  For emission lines, we calculate the flux of detected lines as the integrated flux of the gaussian line fit.  We calculate 1$\sigma$ fluxes as the integral of a gaussian with height equal to the continuum RMS, and width equal to the instrumental FWHM.

\begin{table}
  \caption{ \label{tab:species} Transitions observed with PACS}
\tiny
\begin{tabular}{l l l  c  c  }
\hline\hline
  Species &      Transition     &  Wave     &  FWHM	&	Ap.  \\
              &               	 	  &      $\mu$m       &  $\mu$m			&	corr. \\
  \hline
  \multicolumn{5}{l}{Line scan} \\
  \hline
  \textbf{[OI]}	&	\textbf{3P1-3P2}	&	\textbf{63.1837}	&	0.0184	&	0.70		\\	
  o-H$_2$O	& 	8$_1$8-7$_0$7	& 	63.3236		\\
  \hline
  DCO+		&	---		&	189.57			&	0.118	&	0.41	\\	
  \hline
  \hline
  \multicolumn{5}{l}{Range scan} \\
  \hline
  o-H$_2$O	&	7$_0$7-6$_1$6	&	71.9460			&	0.0391	&	0.70		\\	
  CH+		&	5-4				&	72.1870		\\
  \textbf{CO}	&	\textbf{36-35}		&	\textbf{72.8429}	\\
  \hline
  o-H2O 	 	&	4$_2$3-3$_1$2 	&	78.7414 			&	0.0386	&	0.70		\\	
  OH 1/2-3/2\tablefootmark{a} 	&	1/2--3/2+ 1-2		& 	79.1173 	\\ 					
  OH 1/2-3/2  	&	1/2--3/2+ 0-1 		&	79.1180 	\\ 					
  OH 1/2-3/2\tablefootmark{a} 	&	1/2+ -3/2- 1-2 		&	79.18173	\\ 					
  OH 1/2-3/2 	&	1/2+ -3/2- 0-1 		&	79.1809	\\ 					
  \textbf{CO}	&	\textbf{33-32}		&	\textbf{79.3598} 	\\
  \hline
  p-H2O 		&  	4$_1$3-3$_2$2	&	144.5181 			&	0.125	&	0.56		\\	
  CO 			&	18-17 			&	144.7842	\\ 			
  \textbf{[OI]}	&	\textbf{3P0-3P1}	&	\textbf{145.5254}	\\
  \hline
  \textbf{[CII]}	& 	\textbf{2P3/2-2P1/2}	&	\textbf{157.7409} 		&	0.126	&	0.51	\\	
  p-H2O 		&	3$_3$1-4$_0$4	&	158.3090 		\\
  \hline
  \hline
  \multicolumn{5}{l}{Extended range scan (J1604-2130 only)} \\
  \hline
  p-H2O 		&	3$_2$2-2$_1$1	& 	89.9878 			&	0.0362	&	0.69		\\				
  CH+ 		&	4-3 				&	90.0731 		\\ 			
  CO 			&	29-28 			&	90.1630 	\\		
  \hline
  \textbf{o-H2O}	& 	\textbf{2$_1$2-1$_0$1}	& 	\textbf{179.5265}	&	0.122	&	0.44	\\	
  CH+ 		&	2-1 				&	179.61 		\\			
  o-H2O 		&	2$_2$1-2$_1$2	&	180.4880		\\ 			
  o-H$_2 ^{18}$O	& 	2$_1$2-1$_0$1 	&	181.051 		\\ 
  \hline
  \end{tabular}
  
\tablefoot{
  Bold text indicates the primary targets of our spectroscopic survey.
  \tablefoottext{a}{Blended with nearby line at instrumental resolution.}
} 
\end{table}

The spectral resolution at 63 $\mu$m is R $=3400$, while longer wavelength observations have a resolution R$\sim1500$.  Typical continuum RMS for our LineSpec observations at 63 and 190 $\mu$m are $\sim0.2$--0.4 Jy, and $\sim0.05$--0.1 Jy for our RangeSpec observations.  These result in line flux uncertainties of 1--3$\times$10$^{-18}$ W/m$^2$, and errors on the mean continuum value of $\sim$10 -- 50 mJy .   

\subsection{JCMT spectroscopy}

We also observed a subset of our sample for (sub)millimeter CO emission with the JCMT.  These observations were a mix of CO J=3-2 observations using HARP and observations for CO J=2-1 using Receiver A (RxA).  

HARP is a 4$\times$4 heterodyne array operating in the $\sim$350 GHz atmospheric window, which allows for simultaneous sampling of the targets and their large scale environs.  Observations were done in an on-off mode, centering the targets in the 14'' FWHM beam of a single receiver and using a 60$''$ offset for background subtraction.  HARP observations were carried out under conditions with a typical $\tau_{225 GHz} \sim0.1-0.2$.  Calibration was carried out hourly on nearby bright line sources known to have low-variability in line strength and line shape.  

RxA is a single heterodyne detector with a 21$''$ FWHM beam operating in the 230 GHz atmospheric window.  RxA observations were carried out in queue mode under conditions of $\tau_{225 GHz} \sim 0.3$, with hourly calibration observations of bright line sources.  

The data were reduced using standard routines in the Starlink Kappa and Smurf packages, the facility data reduction tools.  We binned data to a 0.25 km/s resolution.  Fluxes were converted from antenna temperature to Jy according to the formula $F_{\nu} = (2kT) / (A \eta)$, where $k$ is the Boltzmann constant, $T$ is the antenna temperature, $A$ is the telescope area, and $\eta$ is the aperture efficiency.  We used reported aperture efficiencies from JCMT documentation \footnote{http://docs.jach.hawaii.edu/JCMT/OVERVIEW/tel\_overview/} of 0.63 at 230 GHz and 0.56 at 345 GHz.  Typical continuum RMS for both the CO 2-1 and CO 3-2 regions are $\sim$1--1.5 Jy in 0.25 km/s wide channels.


\section{Results}
\label{sec:results}

\subsection{Spectroscopy}
 
Two sources, J1614-1906 and J1604-2130, are detected in [OI] and [CII] emission, though neither source is resolved at the 90--200 km/s resolution of the PACS instrument.  An additional 6 sources are detected in the 63 $\mu$m continuum.  In addition, J1614-1906 is marginally detected in both the CO J=18-17 and the OH 79.1 $\mu$m lines.  We list line fluxes in Table \ref{tab:lineresults}, and mean continuum fluxes and error on the mean in Table \ref{tab:photresults}.  We report nondetections as 3$\sigma$ upper limits, and we show our continuum subtracted line detections in Figure \ref{fig:spectra-detects}.  Our reported errors and upper limits do not include the calibration uncertainties.

Our IFU observations allow us to check for large scale extended emission over the entire $\sim45''\times45''$ field of view.  Some extended emission in the [CII] line may be present in the off-source spaxels of object 34, but not at the source location itself.  Off-source emission is also detected at the $\sim3\sigma$ level in two and three spaxels at large separations (18''--25'') toward J1604-2130 and J1614-1906, respectively.  Maps of the [CII] 157.7 $\mu$m emission toward these sources is shown in Appendix \ref{sec:CII-maps}.

For our JCMT observations,  only J1604-2130 is detected.  Assuming a FWHM of 1 km/s for nondetected sources, we find typical line flux uncertainties of 1.3$\times$10$^{-20}$ W/m$^2$.  Integrated line strength and 3$\sigma$ upper limits are included in Table \ref{tab:lineresults}.

\begin{table*}[bh]
\begin{center}

\caption{\label{tab:lineresults}Line fluxes}
\begin{tabular}{r ccccccc}

\hline\hline
  Object						& [OI] & CO & CO & [OI] & [CII] & CO & CO \\  
             						& $63.2\mu$m & 72.8$\mu$m  &  79.4$\mu$m &  145.5$\mu$m		&   157.7$\mu$m		&     $J$=3-2		&  $J$=2-1    	\\
               						& $10^{-18}W/m^2$  &   $10^{-18} W/m^2$	&   $10^{-18} W/m^2$	&   $10^{-18} W/m^2$	&   $10^{-18} W/m^2$	&   $10^{-20} W/m^2$	&  $10^{-20} W/m^2$ \\

\hline

1	&  $<$13.1	&  ...		&  ...		& ...		&  ...		&  ...		&  $<$1.5		\\
2	&  ...		&  ...		&  ...		& ...		&  ...		&  $<$2.9		&  $<$2.8		\\
3	&  $<$5.7		&  ...		&  ...		& ...		&  ...		&  $<$3.1		&  ...		\\
8	&  $<$8.6		&  ...		&  ...		& ...		&  ...		&  ...		&  ...		\\
9	&  $<$6.2		&  ...		&  ...		& ...		&  ...		&  ...		&  ...		\\
10	&  $<$6.5		&  ...		&  ...		& ...		&  ...		&  ...		&  ...		\\
11	&  ...		&  ...		&  ...		& ...		&  ...		&  ...		&  $<$3.6		\\
12	&  $<$13.4	&  ...		&  ...		& ...		&  ...		&  ...		&  ...		\\
13	&  $<$10.4	&  $<$9.0		&  $<$8.5		&  $<$2.4		&  $<$3.2		&  ...		&  $<$3.8		\\
15	&  ...		&  ...		&  ...		& ...		&  ...		&  $<$3.0		&  ...		\\
16	&  $<$5.6		& ...		&  ...		& ...		&  ...		&  ...		&  ...		\\
17	&  $<$11.5	&   ...	&  ...		& ...		&  ...		&  ...		&  $<$3.0		\\
18	&  ...		&  ...		&  ...		& ...		&  ...		&  $<$3.1		&  ...		\\	
19	&  ...  	&  ...		&  ...		& ...		&  ...		&  $<$5.9		&  ...		\\	
20	&  $<$7.9		& ...		&  ...		& ...		&  ...		&  ...		&  ...		\\	
21	&  ...  	&  ...		&  ...		& ...		&  ...		&  $<$5.3		&  ...		\\	
22	&  $<$15.2	&   ...		&  ...		& ...		&  ...		&  ...		&  ...		\\	
23	&  $<$7.1		&  $<$7.2		&  $<$12.0		&  $<$4.1		&  $<$3.0	&  ...		&  $<$3.3		\\	
24	&  $<$8.3  	&  ...		&  ...		& ...		&  ...		&  ...		&  ...		\\	
29	&  $<$9.5		&  $<$6.6		&  $<$10.2	&  $<$2.7		&  $<$3.0		&  ...		&  $<$1.8		\\	
31	&  ...  	&  ...		&  ...		& ...		&  ...		&  ...		&  $<$4.4		\\	
34	&  $<$8.5		&  $<$7.6		&  $<$10.0	&  $<$2.5		&  $<$3.0		&  ...		&  $<$4.2		\\	
36	& 46.3$\pm$4.4 &  $<$7.0	&  $<$10.0\tablefootmark{a}	&  $<$3.7\tablefootmark{b}		& 5.1$\pm$1.5	&  ...		& $<$2.1		\\	
37	&  ...  	&  ...		&  ...		& ...		&  ...		&  $<$3.0		&  $<$4.2		\\	
38	&  ...		&  ...		&  ...		& ...		&  ...		&  $<$3.0		&  ...		\\	
39	&  ...  	&  ...		&  ...		& ...		&  ...		&  $<$3.0		&  ...		\\	
40	&  $<$9.8		&  $<$8.6		&  $<$8.8		&  $<$2.6		&  $<$2.8		&  $<$4.9		&  $<$3.6		\\	
41	&  28.6$\pm$1.9  &  $<$11.2	& $<$13.5		&  $<$3.6  	& 5.9$\pm$1.7	&  6.5$\pm$1.5	&  25.0$\pm$0.9	\\	
45	&  $<$7.8  	&  ...		&  ...		& ...		&  ...		&  ...		&  $<$1.6		\\	
\hline

\end{tabular}

\tablefoot{
\tablefoottext{a}{Blended OH lines are marginally detected at 79.1 $\mu$m in this band, with a combined flux of (6.6$\pm$2.8) $\times10^{-18} W/m^2$.} 
\tablefoottext{b}{The CO J=18-17 line is marginally detected in this band, with a flux of (3.4$\pm$1.2) $\times10^{-18} W/m^2$.}
}
  
\end{center}
\end{table*}

\begin{figure*}
\vspace{2cm}
  \includegraphics[angle=0,width=0.40\textwidth]{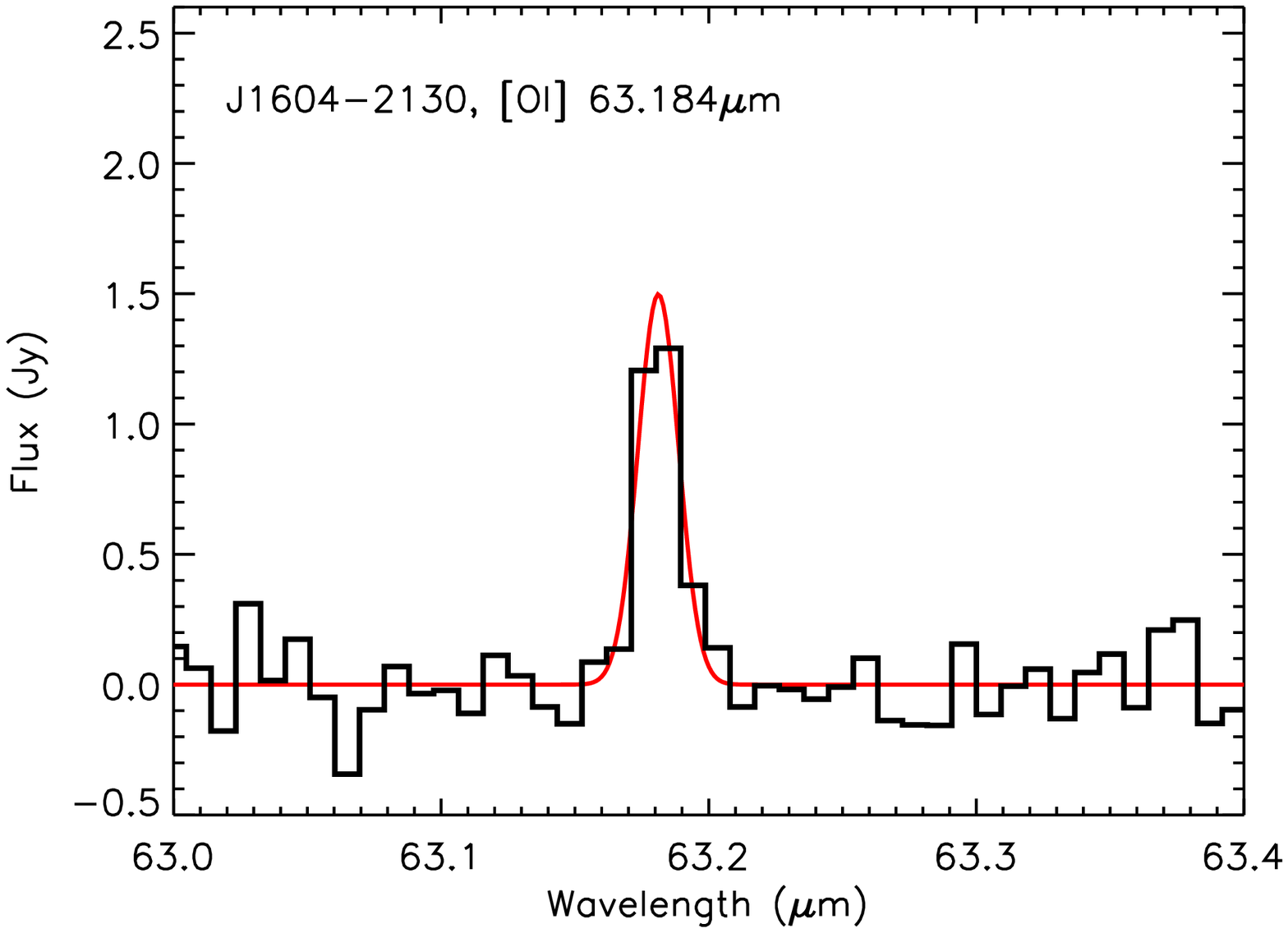}  \includegraphics[angle=0, width=0.40\textwidth]{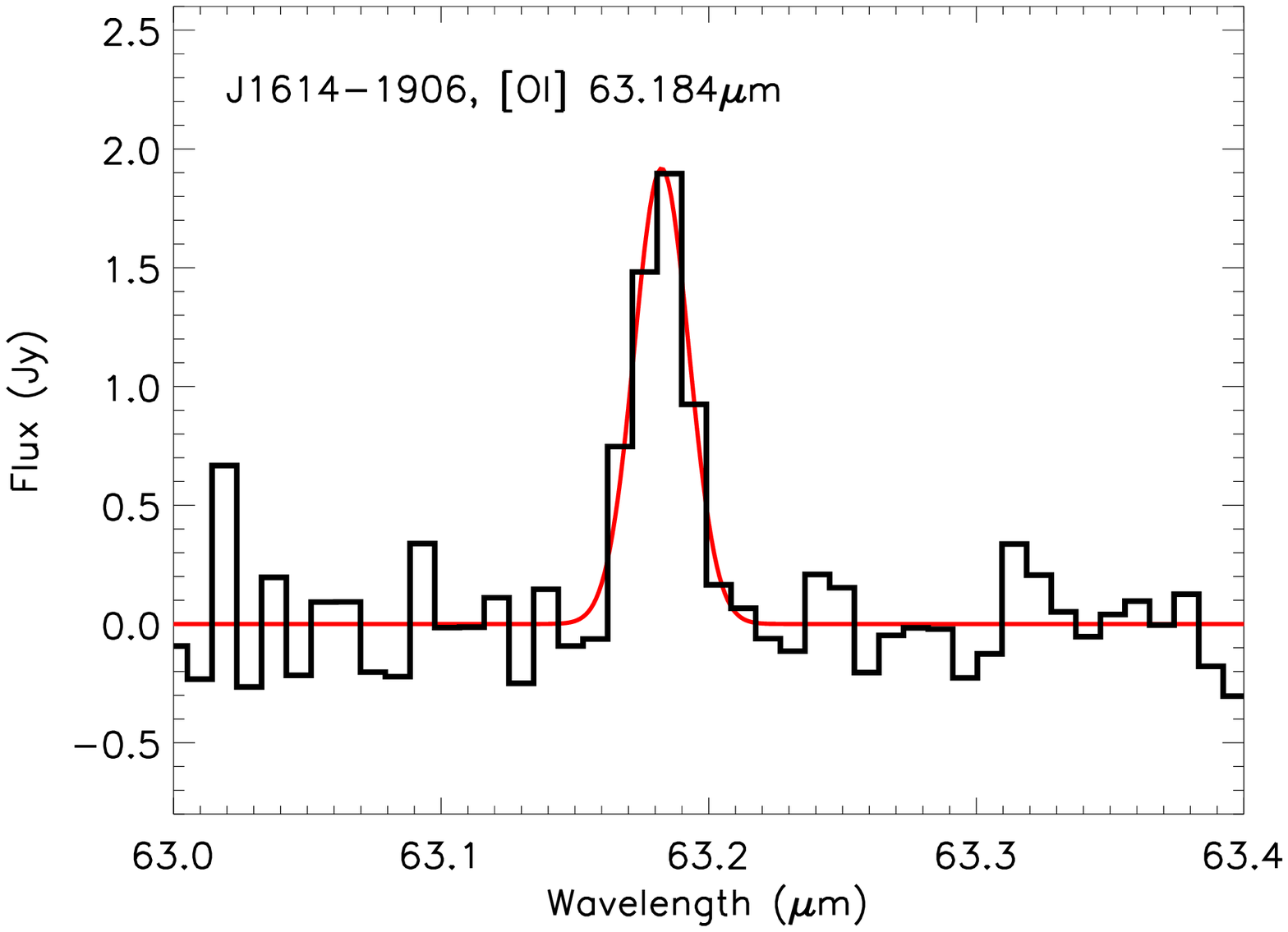} \\
  \includegraphics[angle=0, width=0.40\textwidth]{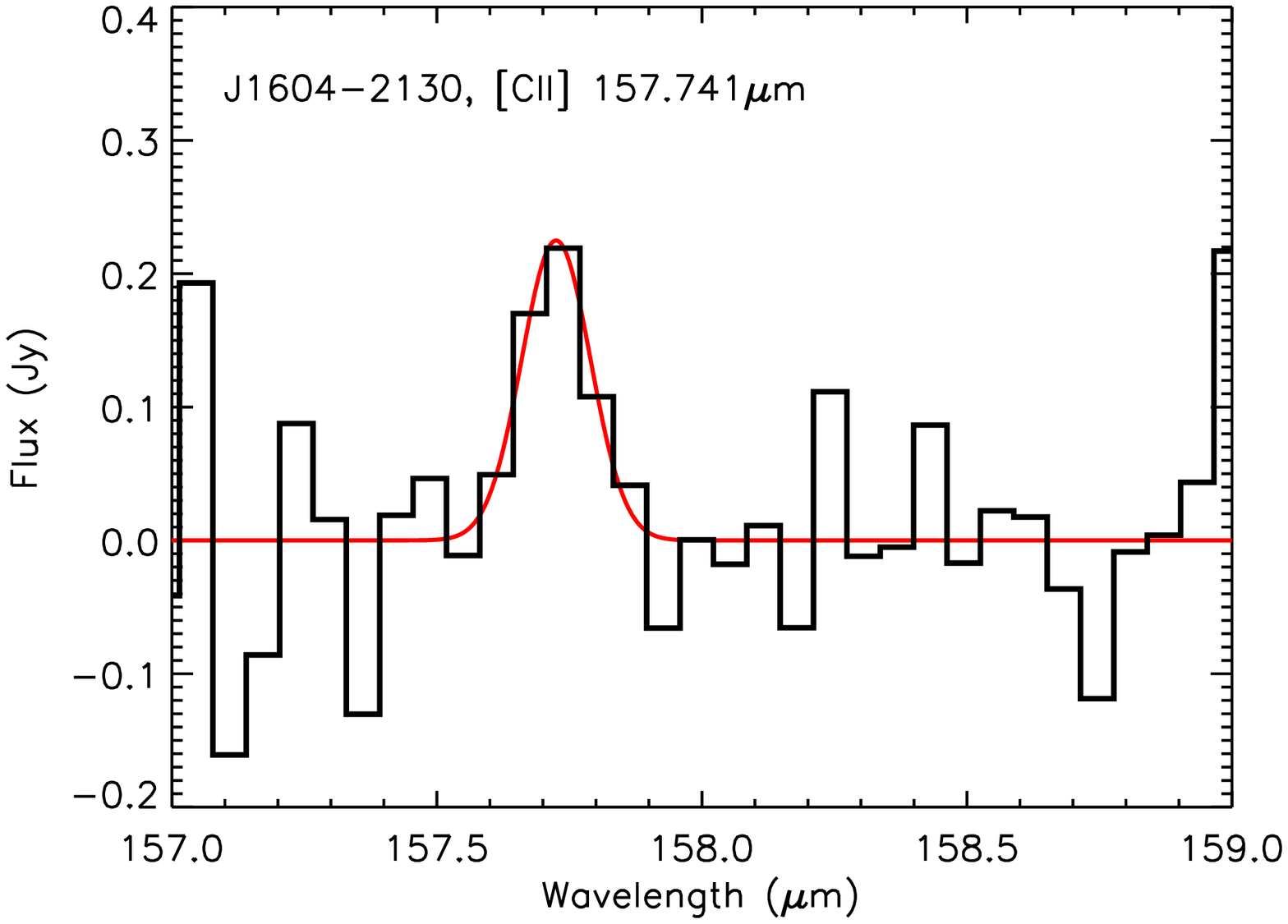}  \includegraphics[angle=0, width=0.40\textwidth]{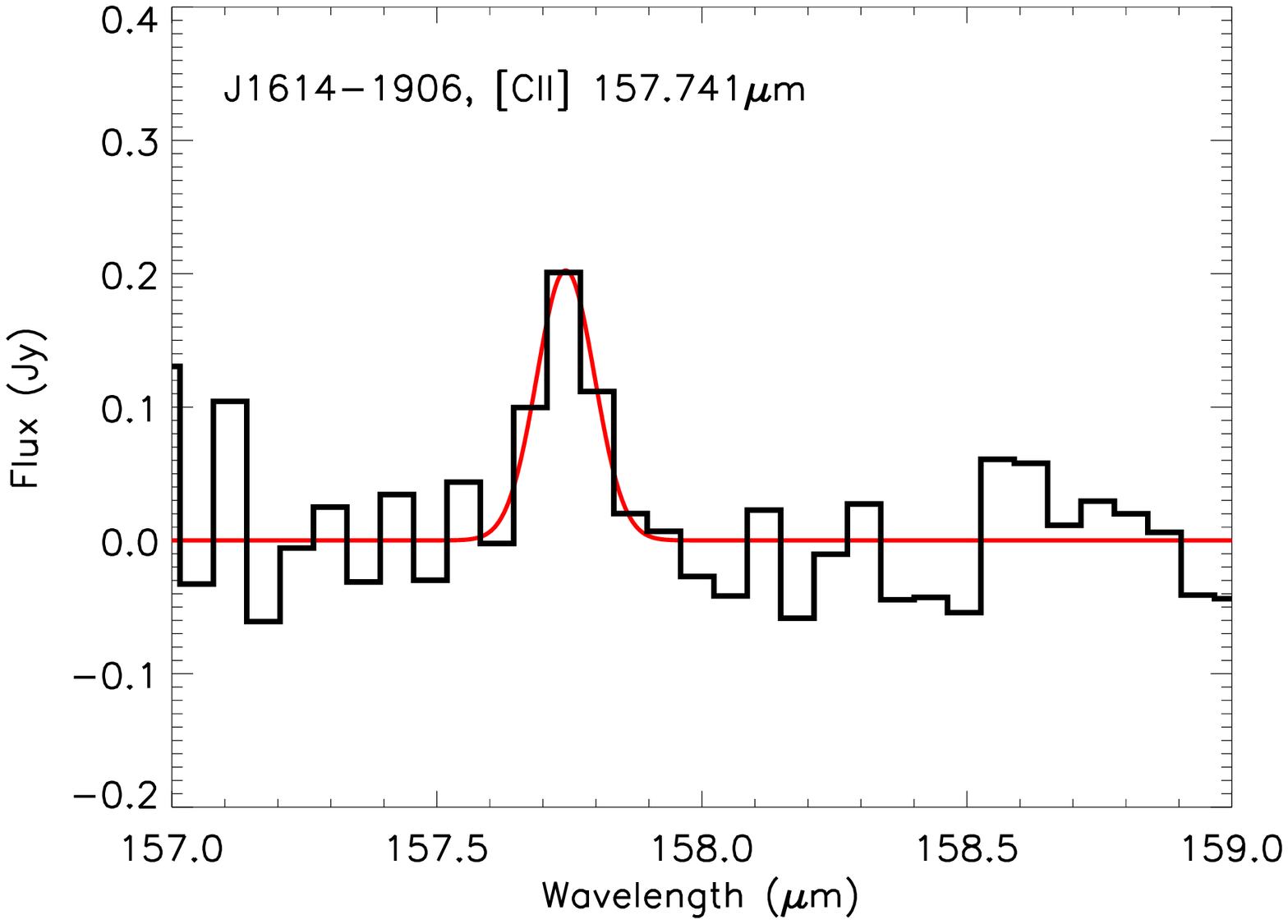} \\
  \includegraphics[angle=0, width=0.40\textwidth]{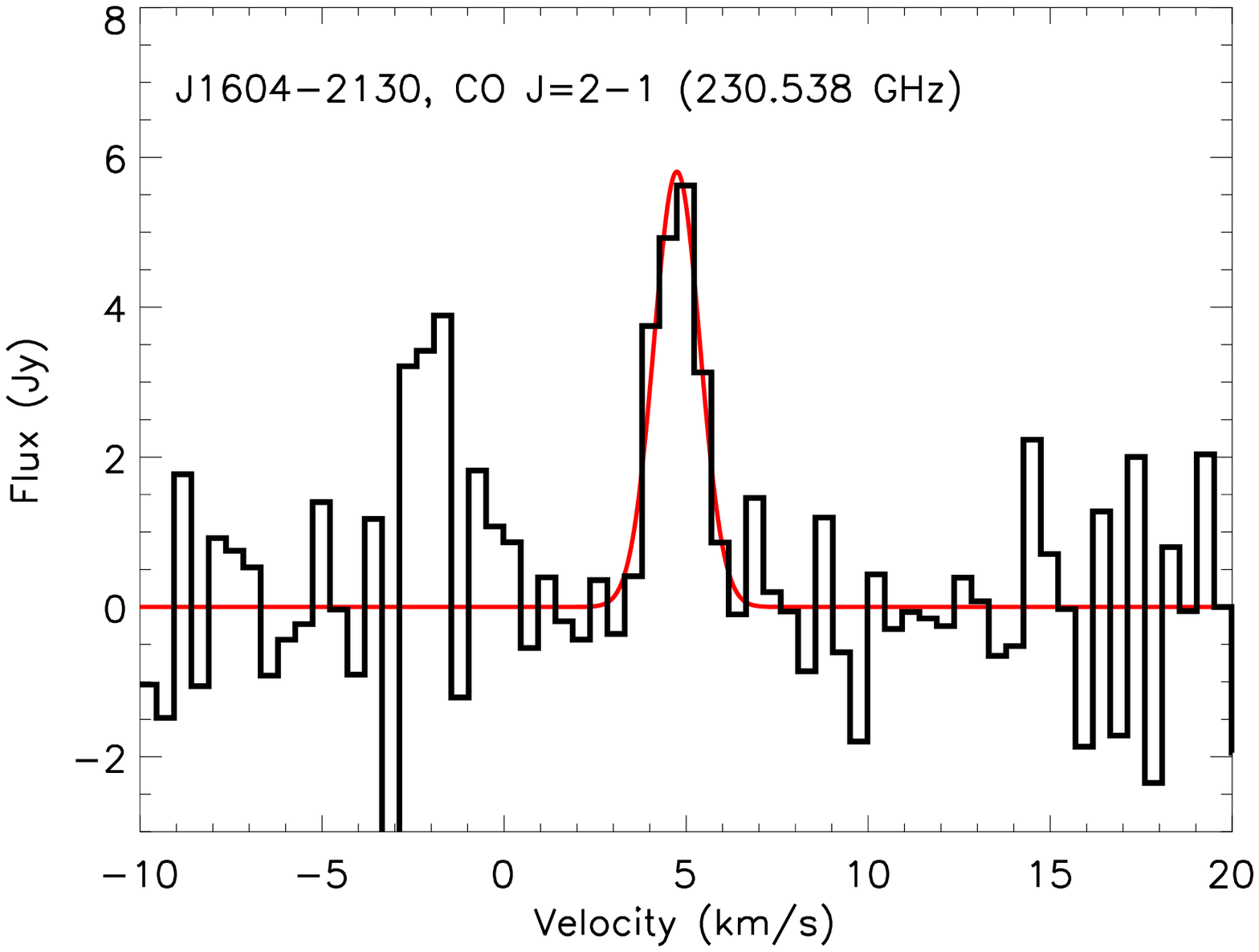}  \includegraphics[angle=0, width=0.40\textwidth]{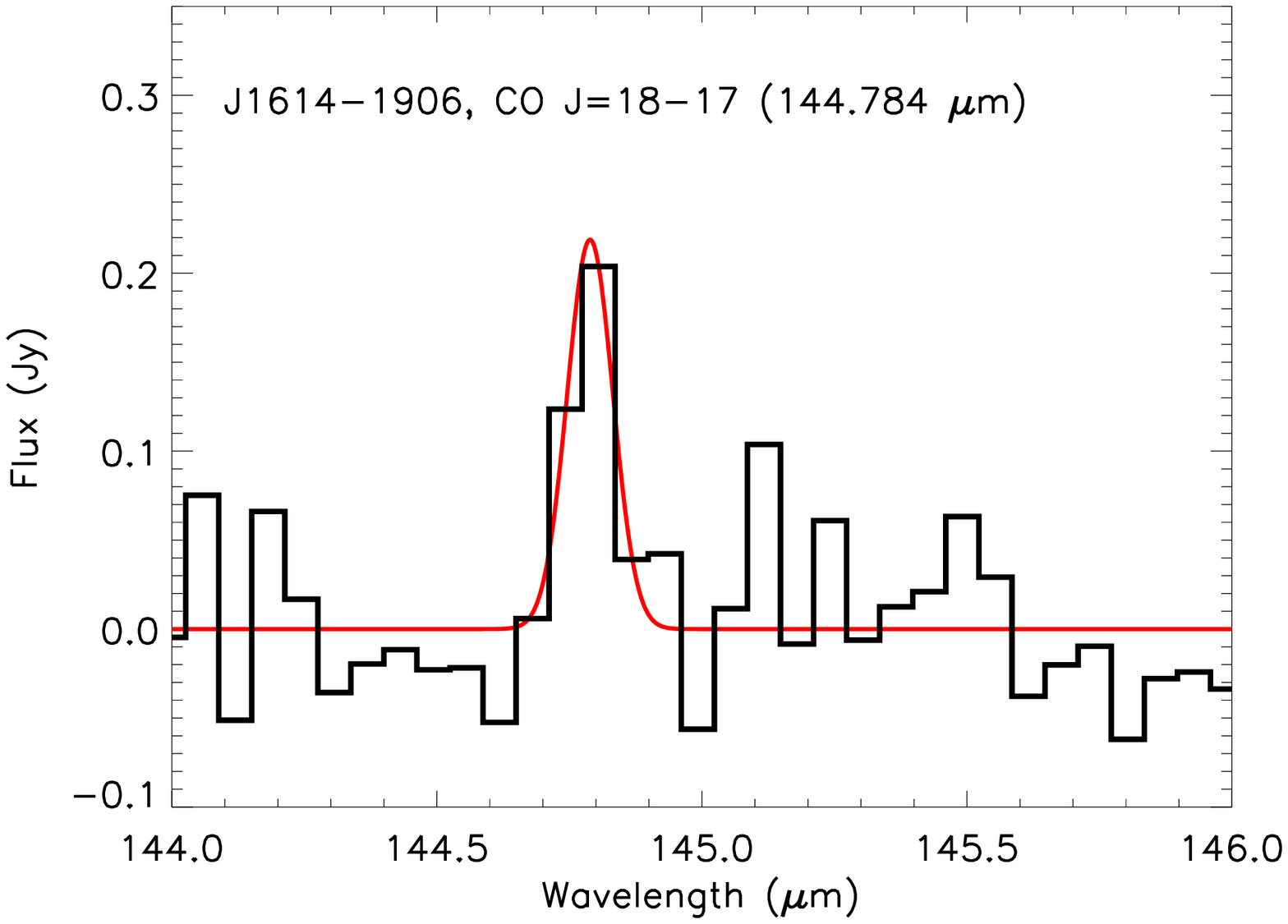} \\
  \includegraphics[angle=0, width=0.40\textwidth]{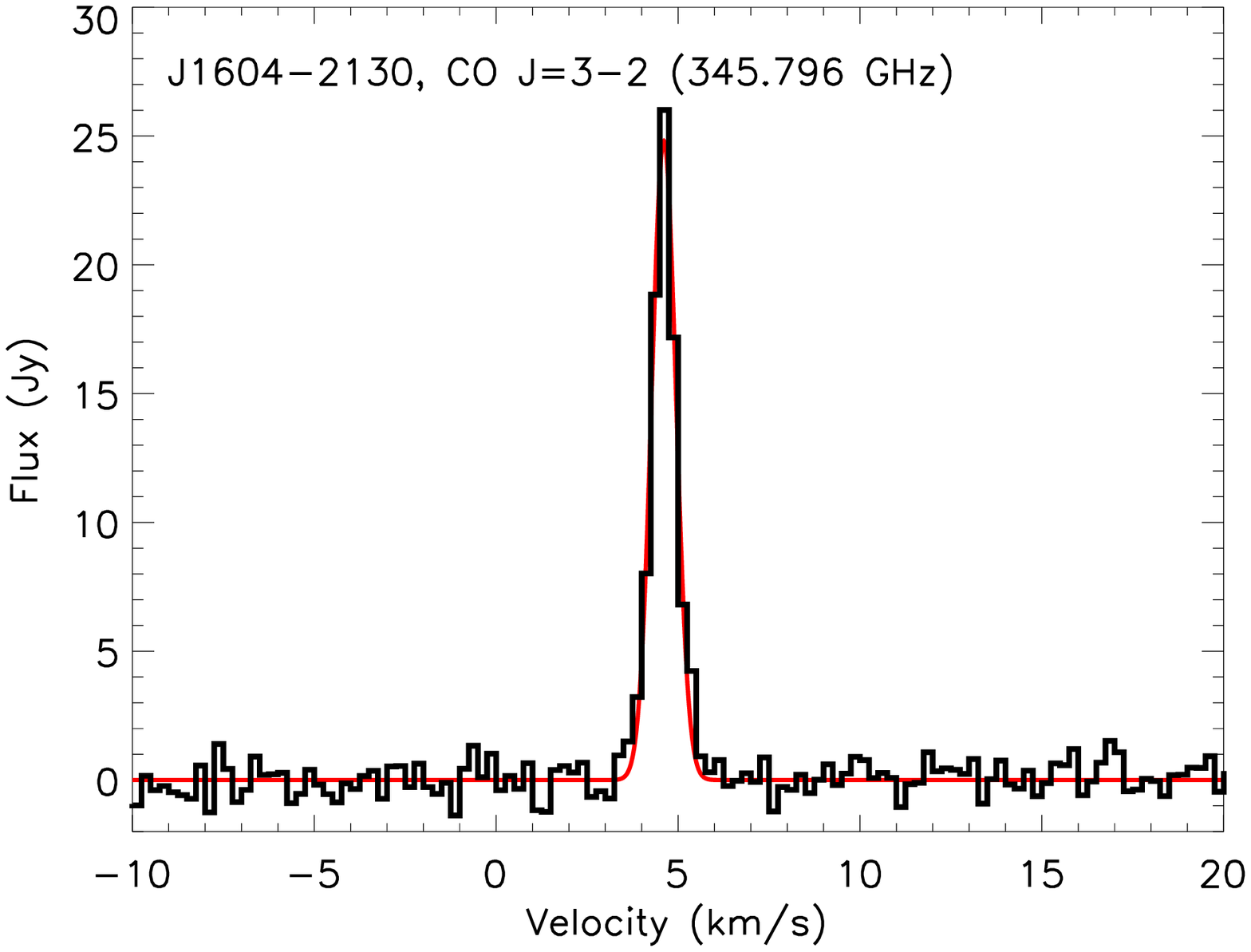}  \includegraphics[angle=0, width=0.40\textwidth]{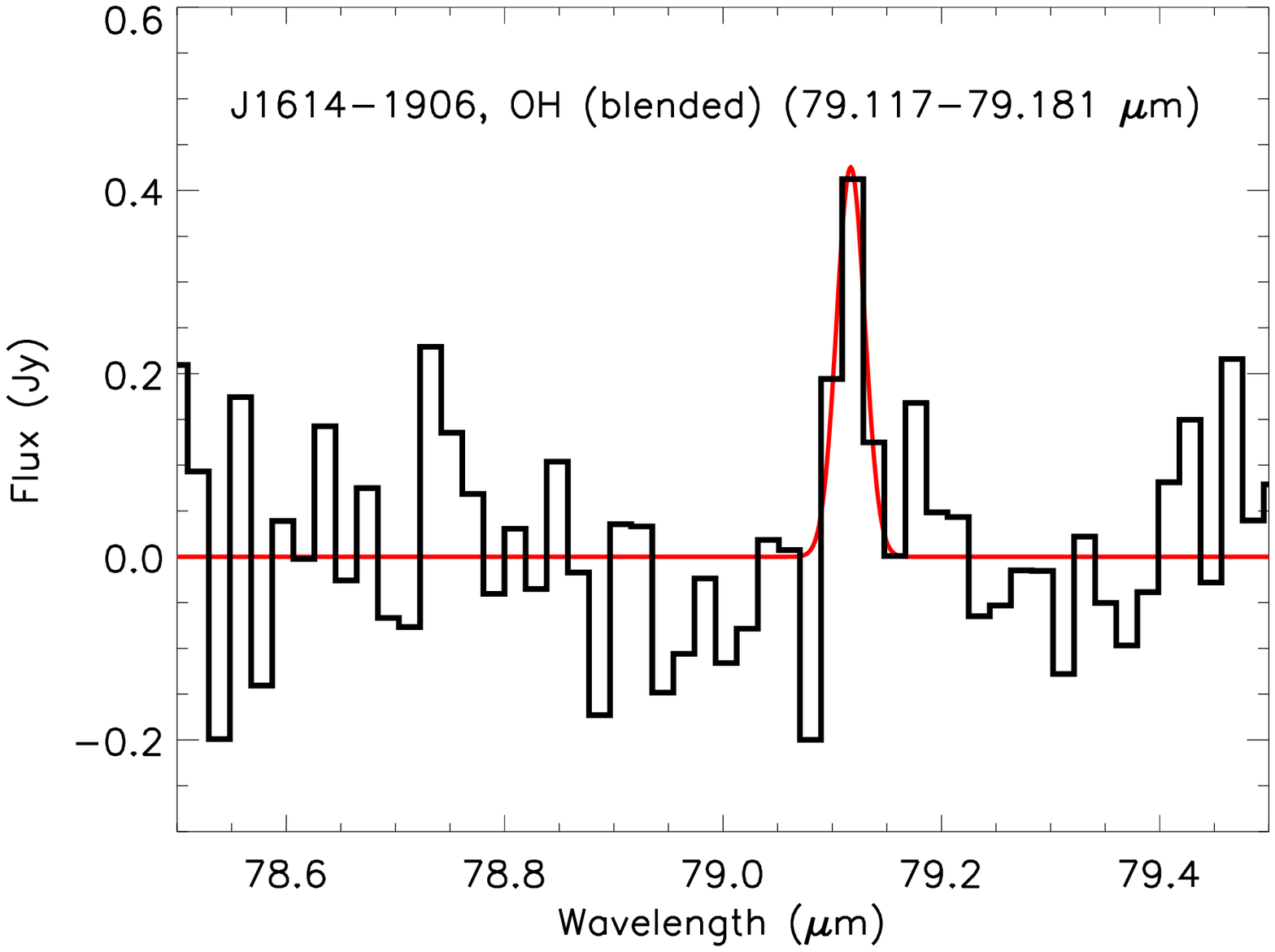} \\
  \caption{Continuum subtracted spectra of detected emission lines are shown in black.  Gaussian fits are overlaid in red. }
  \label{fig:spectra-detects}
\end{figure*}

\subsection{Photometry}

In the PACS blue passband at 70 $\mu$m, we detect 4/8 B\&A stars, 2/6 G\&K stars, and 8/21 M stars.  In our later analysis, we include 70 $\mu$m photometry of 3, 5, and 2 stars in each of these categories that were previously detected in \textit{Spitzer} MIPS band 2 observations, and were therefore not observed in the PACS blue passband.  At 100 $\mu$m, we detect 2/5 B\&A stars, 7/11 G\&K stars, and 6/17 M stars, and at 160$\mu$m, we detect 2/11 B\&A stars, 6/11 G\&K stars, and 6/23 M stars.  As none of our observations are sensitive enough to detect the stellar photosphere, all detections represent circumstellar disks, either primordial or debris.

\begin{landscape}
\begin{table*}[th]
\begin{center}
\caption{  \label{tab:photresults} Continuum measurements}
\hspace{-2in}\begin{tabular}{r ccc ccc ccc }

\hline\hline
  Object 						&  63    &  70    				    &  72    &  79    &   100    &  145    &  158    &  160      &  190     \\
  	   	 					&  mJy    &  mJy   			&  mJy &  mJy &   mJy   &  mJy    &  mJy    &  mJy      &  mJy	\\
\hline

1	&  472$\pm$52	&  373.1$\pm$11.3\tablefootmark{a} & ...	&  ...			&  354.9$\pm$6.8	& ...			& ...			& 159.7$\pm$27.3	& 197$\pm$72	\\  
2	& ...		& $<$ 15.3					& ... 		&  ...			& ... 			& ...			& ...			& $<$    63.6		& ...  	\\  
3	& $<$   105	& $<$41.4						& ...		&  ...			& $<$22.8			& ...			& ...			& $<$72.6		 	& $<$   158  	\\  
4	& ...		& $<$15.9						&  ...		&  ...			&  ...			&  ...			&  ...			& $<$59.4		  	&  ...	\\   
5	& ...		& 14.0$\pm$3.6				&  ...		&  ...			&  ...			&  ...			&  ...			& $<$   58.2  		&  ...	\\  
6	& ...		& $<$12.9						&  ...		&  ...			&  ...			&  ...			&  ...			& $<$    46.8  		&  ...	\\  
7	& ...		& 14.3$\pm$5.6				&  ...		&  ...			&  ...			&  ...			&  ...			& $<$    79.2  		&  ...	\\   
8	& $<$105		& $<$   10.5					&  ...		&  ...			&  ...			&  ...			&  ...			& $<$    48.6  		&  $<$   228		  	\\  
9	& $<$75		& $<$47.7\tablefootmark{a}		&  ...		&  ...			& $<$ 18.9		&  ...			&  ...			& $<$    77.4  		&  $<$117  			\\  
10	& 118$\pm$26	& 79.2$\pm$13.8\tablefootmark{a}	&  ...		&  ...			& 75.8 $\pm$    5.9	&  ...			&  ...			& 72.2 $\pm$   24.6  	& $<$89 	 	\\  
11	& ...		& $<$   16.8					&  ...		&  ...			& $<$     17.1		&  ...			&  ...			& $<$    65.4  		&  ...	\\  
12	& $<$162		& $<$   9.3	   				&  ...		&  ...			&  ...			&  ...			&  ...			& $<$    47.1  		&  $<$  317		 	\\   
13	& 134$\pm$41	& 296.7$\pm$11.5\tablefootmark{a}	&  149$\pm$22	& 127$\pm$26		& 252.3 $\pm$    7.1	&   117 $\pm$  8	&  87 $\pm$ 13		& 171.8 $\pm$ 24.1 	& $<$111  	\\   
14	& ...		& $<$14.1						&  ...		&  ...			& $<$    18.3		&  ... 	 	&  ...			& $<$  62.7  		&  ...	\\   
15	& ...		& $<$15.6   					&  ...		&  ...			& $<$     16.5		&  ...			&  ...			& $<$    64.8  		&  ...	\\  
16	& $<$66		& $<$16.2						&  ...		&  ...			& $<$18.9			&  ...			&  ...			& $<$45.9		  	&  $<$   104		  	\\   
17	& $<$141		& 16.9$\pm$6.3   				&  ...		&  ...			& 37.4 $\pm$     7.5	&  ...			&  ...			& 46.4 $\pm$  19.1 	& ... 		 	\\  
18	& ...		& $<$19.8   					&  ...		&  ...			&  ...			&  ...			&  ...			& $<$    72.9  		&  ...	\\  
19	& ...		& $<$16.2   					&  ...		&  ...			&   ...	  	&  ...			&  ...			& $<$    83.7  		&  ...	\\  
20	& $<$96		& 69.3$\pm$5.6   				&  ...		&  ...			&  ...			&  ...			&  ...			& 96.6 $\pm$   24.5 	& $<$   152  	\\  
21	& ...		& $<$   16.2   					&  ...		&  ...			& $<$     16.8		&  ...			&  ...			& $<$    51.3  		&  ...	\\  
22	& $<$183		& $<$  20.4   					&  ...		&  ...			& $<$    23.1		&  ...			&  ...			& $<$    58.2  		&  $<$   329		  	\\  
23	& $<$87		& 56.7$\pm$5.8   				& $<$54	  	& $<$114 			& 78.7 $\pm$     6.2	& 122 $\pm$    14	&  101 $\pm$  12	& 60.3 $\pm$   19.8 	&  $<$   145  			\\  
24	& $<$99		& $<$16.5   					&  ...		&  ...			& $<$    20.7		&  ...			&  ...			& $<$    49.2  		&  $<$   214  	\\   
25	& ...		& 44.0$\pm$5.1   				&  ...		&  ...			& 30.4 $\pm$     6.9	&  ...			&  ...			& 50.5 $\pm$   23.5  	&  ...	\\   
26	& ...		& 18.3$\pm$5.1   				&  ...		&  ...			& 15.4 $\pm$     5.9	&  ...			&  ...			& $<$    63.3  		&  ...	\\  
27	& ...		& $<$18.9   					&  ...		&  ...			& $<$ 23.1		&  ...			&  ...			& $<$     59.4  		&  ...	\\  \
28	& ...		& $<$17.7   					&  ...		&  ...			& $<$    21.3		&  ...			&  ...			& $<$ 54		 	&  ...	\\  
29	& 249$\pm$38	& 99.9$\pm$10.7\tablefootmark{a}	& 82$\pm$17	& 267$\pm$32		& 150.4 $\pm$    5.7	& 182 $\pm$ 9		&  271 $\pm$  12	& 191.7 $\pm$ 29.3  	& 213$\pm$53  	\\   
30	& ...		& 18.2$\pm$5.7  				&  ...		&  ...			& $<$20.1			&  ...			&  ...			& $<$     58.2  		&  ...	\\   
31	& ...		& 50.2$\pm$7.8   				&  ...		&  ...			& 41.1 $\pm$     8.9	&  ...			&  ...			& $<$ 86.1	  	&  ...	\\  
32	& ...		& 306.1$\pm$6.7	  			&  ...		&  ...			& 355.6 $\pm$   6.3	&  ...			&  ...			& 397.4 $\pm$ 24.9  	&  ...	\\  
33	& ...		& 18.6$\pm$5.2   				&  ...		&  ...			& $<$ 18.9		&  ...			&  ...			& $<$    61.5  		&  ...	\\   
34	& 146$\pm$34	& 127.6$\pm$15.2\tablefootmark{a}	&  239$\pm$19	& $<$93			& 110.1 $\pm$  6.4	& 69 $\pm$ 8	 	&  103 $\pm$    12  	& 106.2 $\pm$ 23.2 	& $<$152     	\\  
35	& ...		& 13.3$\pm$3.5 				&  ...		&  ...			& $<$     13.2		&  ...			&  ...			& $<$    47.1  		&  ...	\\  
36	&1134$\pm$39	& 668.3$\pm$25.2\tablefootmark{a}	& 1110$\pm$18 & 964$\pm$31	& 749.7 $\pm$   8.4	& 625 $\pm$    12  	&  689 $\pm$    15  	& 691.6 $\pm$   7.9  	& 521 $\pm$    70	\\  
37	& ...		& $<$13.2   					&  ...		&  ...			& $<$     21.3		&  ...			&  ...			& $<$    51.3  		&  ...	\\  
38	& ...		& $<$14.1   					&  ...		&  ...			& $<$     19.8		&  ...			&  ...			& $<$    60.0  		&  ...	\\  
39	& ...		& $<$17.7   					&  ...		&  ...			&  ...			&  ...			&  ...			& $<$    83.4  		&  ...	\\   
40	& 209 $\pm$39	& 187.8$\pm$12.7\tablefootmark{a}	& $<$63		& 127$\pm$27		& 172.6 $\pm$   7.0	& 43.0 $\pm$9.0	&  60$\pm$11		& 95.8 $\pm$  25.1  	& 137$\pm$21  	\\  
41	&1775$\pm$22	& 1917.4$\pm$24.8\tablefootmark{a}&2061$\pm$28 & 2218$\pm$42	& 3221.9 $\pm$  7.4	& 3126$\pm$12	&  3639$\pm$18	& 3796.7 $\pm$25.4	& 3439 $\pm$ 49 	\\   
42	& ...		& $<$18.9   					&  ...		&  ...			& $<$  18.9  		&  ...			&  ...			& $<$  61.8  		&  ...	\\  
43	& ...		& $<$16.5   					&  ...		&  ...			& $<$  19.5  		&  ...			&  ...			& $<$   64.5  		&  ...	\\  
44	& ...		& $<$19.2   					&  ...		&  ...			& $<$   17.7		&  ...			&  ...			& $<$  57.0  		&  ...	\\  
45	& $<$93		& 91.1$\pm$11.7\tablefootmark{a}  	&  ...		&  ...			& 87.7 $\pm$  6.9	&  ...			&  ...			& 63.2 $\pm$   27.9  	&  $<$125 	\\  
\hline

\end{tabular}

\tablefoot{
  \tablefoottext{a}{70 $\mu$m flux reported in \cite{Carpenter:2009b}.}
}

\end{center}
\end{table*}
\end{landscape}

We list our photometry in Table \ref{tab:photresults}, including 1$\sigma$ errors or 3$\sigma$ upper limits.  Our reported errors and upper limits do not include the calibration uncertainties.  In general, the center of emission of PACS emission lies within the 2$''$ positional uncertainty of the observations.  However, the 70 and 100$\mu$m emission for HIP 77911 is centered at a position $\sim$8$''$ to the northwest of the stellar position.  This star is known to have a distant companion with a separation of $8''$ \citep{Kouwenhoven:2007a}, suggesting the previously identified excess seen here is not a debris disk around HIP 77911, but a disk around the companion star.

Several stars (objects 13, 29, 34, 36, and 40) show factor of $\sim$2 discrepancies in continuum values between neighboring spectral regions.  While these differences are consistent at the 2$\sigma$ level when considering the calibration uncertainty of 30\%, they motivate us to make no further use here of the continuum values derived from the spectroscopic observations.

\subsection{Spectral energy distributions}
\label{sed}

We constructed spectral energy distributions (SEDs) using optical, 2MASS, and \textit{Spitzer} photometry from the literature and adopt uncertainties as described in \cite{Mathews:2011}.  We incorporated 3.4, 4.6, 11.6, and 22 $\mu$m photometry from the WISE Preliminary Data Release \citep{Cutri:2011}, using the zero magnitude fluxes of \cite{Wright:2010} to convert from magnitudes to Jansky.   We include the \textit{Spitzer} IRS spectroscopy of \cite{2009AJ....137.4024D}, covering a wavelength range from 5 to 35 $\mu$m, and bin the spectra as in \cite{Furlan:2006}.  We also include reported millimeter fluxes at 1.2 and 1.3 mm from \citep[][and references therein]{Mathews:2011}, and the 880 $\mu$m flux for J1604-2130 reported in \cite{Mathews:2012a}.

We show nondetections as 3$\sigma$ upper limits.  We assume large calibration uncertainty (40\%) for the optical photometry, which is largely drawn from scans of photographic plates.  For 2MASS photometry, we assume 15\% calibration uncertainty due to unknown NIR variability of the sources.  For \textit{Spitzer} IRAC and IRS peak-up photometry, we assume 10 and 20\% errors, respectively.  For the binned IRS spectra, we assume 10\% uncertainty, as in \cite{Furlan:2006}.  For WISE bands 1--4 at wavelengths of 3.4, 4.6, 11.6, and 22.1 $\mu$m, respectively, \cite{Jarrett:2011} report calibration uncertainties of 2.4\%, 2.8\%, 4.5\%, and 5.7\%.  However, in order to prevent these points from dominating our model fitting (discussed below), we adopt a minimum calibration uncertainty of 10\%.

Stellar radius estimates based on temperature and luminosity estimates from the literature \citep[PBB2002]{Hernandez:2005, Preibisch:1998} led to poor scaling of the stellar photosphere models.  As the stellar luminosity is particularly crucial to our later disk modeling, we have carried out new estimates of the stellar radius and extinction of the disk-bearing sources modeled below.  

We simultaneously fit the stellar radius and extinction using broadband photometry from B to H band.  We assume a stellar distance of $145\pm20$ pc, following parallax measurements of the distribution of high mass members \citep{de-Zeeuw:1999,de-Bruijne:1999}.  The stellar radius and distance are degenerate in the photospheric fitting, requiring that one parameter be fixed using external information.  While the measured effective temperatures could be used with stellar isochrones to select a stellar radius, the uncertain ages and large variations between isochrones make that approach far more uncertain. Though many sources appear to the eye to have photospheric K band emission, we omit this band from the fitting procedure as it could be potentially contaminated at small levels by dust at the sublimation radius.  Using effective temperatures ($T_{eff}$) from the literature, with typical uncertainties of 100 K, we select a closest match photospheric model \citep{Kurucz:1993}.   We also use the R=3.1 reddening law of \cite{Fitzpatrick:1999} to deredden photometry.  

Then, we carry out a $\chi^2$ minimization of the stellar radius and extinction using the Levenburg-Marquardt algorithm, as implemented in the IDL code MPFIT \citep{Markwardt:2009}.  Results of our fits for the extinction and stellar radius, are listed in Table \ref{tab:model-GA}, along with our later disk modeling results.  We present the total $\chi^2$ value for each object.

We show the SEDs of all members of our sample in Figures \ref{fig:seds-first} and \ref{fig:seds-too}, along with best fit models to the disk bearing K and M stars (discussed below).  All of our B\&A stars exhibit photometry or line fluxes consistent with being debris disk hosts or stars with no disk.  Therefore, while we present observational results for these objects, we do not include them in our modeling and discussion (Sect.\ref{sec:models} \& Sect.\ref{sec:discussion}), which focuses on potentially planet forming K\&M star disks.

\begin{table*}
  \caption{  \label{tab:model-GA} Stellar and disk properties}
  
\begin{tabular}{c c  c  c c c c c c c c}

\hline\hline
  Object 							& $A_{\rm{V}}$ & $R_{\rm{star}}$    & $R_{\rm{in}}$  & $\alpha$ & $M_{\rm{dust}}$  &      $H_{100}$   & $\beta$ &	$\chi^2$	&	degrees of       \\  
             	 						& mags      & $R_{\rm{sun}}$    &   AU          &                  & $10^{-6}$ $M_{\rm{sun}}$ &AU      &                &			&	freedom	\\
\hline

    13						&   0.2	&	1.47		&	1.586  &   -0.734 &      2.0  	&   12.7  	&   1.12  &      109.5  	&     24  \\     
14		 				&   0.9	&	0.76		&	0.032 &   -1.763 &      0.011	&    14.8  	&   1.14  &        18.1 	&     11  \\     
15						&   1.1	&	0.58		&	0.032  &   -1.791 &     7.8  	&    5.0  	&   1.08  &         8.7  	&     11  \\     
16						&   1.1	&	0.67		&	0.030  &   -1.250 &      1.0  	&    4.7  	&   1.13  &        17.5  	&     9  \\     
17						&   0.0	&	0.53		&	0.158  &   -0.835 &     2.6  	&   6.5  	&   1.12  &      100.5  	&     20  \\     
20		 				&   0.4	&	0.79		&	0.030  &   -0.597 &      1.1  	&   8.8  	&   1.13  &        54.2  	&     23  \\     
21		 				&   0.8	&	0.71		&	0.048  &   -1.733 &     0.051	&    3.3  	&   1.21  &        14.1  	&     9  \\     
22		 				&   0.4	&	0.54		&	0.033  &   -1.855 &    9.6  	&    3.4  	&   1.02  &        67.7  	&     23  \\     
23		 				&   0.1	&	0.73		&	0.067  &   -1.199 &     4.0  	&    9.3  	&   1.22  &        15.4  	&     9  \\     
24		 				&   0.4	&	0.87		&	0.038  &   -1.961 &     9.4  	&    2.0  	&   1.11  &        52.6  	&     21  \\     
25						&   0.1	&	0.69		&	0.400  &   -0.913 &      4.3  	&    5.6  	&   1.22  &        48.2  	&     22  \\     
26						&   0.7	&	1.17		&	0.032  &   -0.337 &  0.072  	&    3.3  	&   1.19  &        31.3  	&     22  \\     
27						&   0.4	&	0.56		&	0.030  &   -1.528 &      6.6  	&    7.8  	&   1.15  &        33.3  	&     23  \\     
29		 				&   0.4	&	1.01		&	0.031  &   -1.602 &     74 	 &   7.6  	&   1.15  &        36.2  	&     23  \\     
30		 				&   0.8	&	0.88		&	0.059  &   -1.754 &      0.46  	&    2.5  	&   1.11  &        22.0  	&     23  \\     
31						&   0.7	&	0.75		&	0.040  &   -0.567 &      2.0  	&   18.8 	&   1.25  &        51.0  	&     23  \\     
32		 				&   0.8	&	1.25		&	1.333  &   -0.914 &     39  	&    5.8  	&   1.13  &      232.8  	&     25  \\     
33		 				&   2.0	&	0.58		&	0.031  &   -1.177 &      1.1  	&    9.3  	&   1.18  &        15.9  	&     23  \\     
34						&   1.3	&	1.01		&	0.689  &   -0.647 &      3.5  	&   16.1  	&   1.15  &      91.4  	&     22  \\     
35						&   0.0	&	1.20		&	0.094  &   -1.275 &   0.041  	&    10.9  	&   1.11  &      45.2  	&     24  \\     
40	 					&   1.0	&	1.43		&	0.033  &   -1.681 &      2.8  	&   9.6  	&   1.13  &      147.0  	&     25  \\     
45	 					&   1.3	&	2.24		&	0.201  &   -0.847 &      1.0  	&    3.0  	&   1.05  &        45.9  	&     23  \\     

\hline
\hline
Upper Scorpius median\tablefootmark{a}			&  ---	&	0.79		&   0.041  &   -1.433 &      2.9  &    7.26  &   1.16  &        23.7  &     10  \\     
Taurus median\tablefootmark{a} 					& ---	&	0.79		&   0.044  &   -1.41 &   540  &    9.2  &   1.07  &        26.7  &      6  \\     
\hline

\end{tabular}

\tablefoot{
  \tablefoottext{a}{Adopted stellar properties are those of J1603-1942, an Upper Scorpius M2 star at the mid-range of the stellar mass and luminosity of disk bearing sources.}
}

\end{table*}


\begin{figure*}
  \includegraphics[angle=0,width=0.90\textwidth]{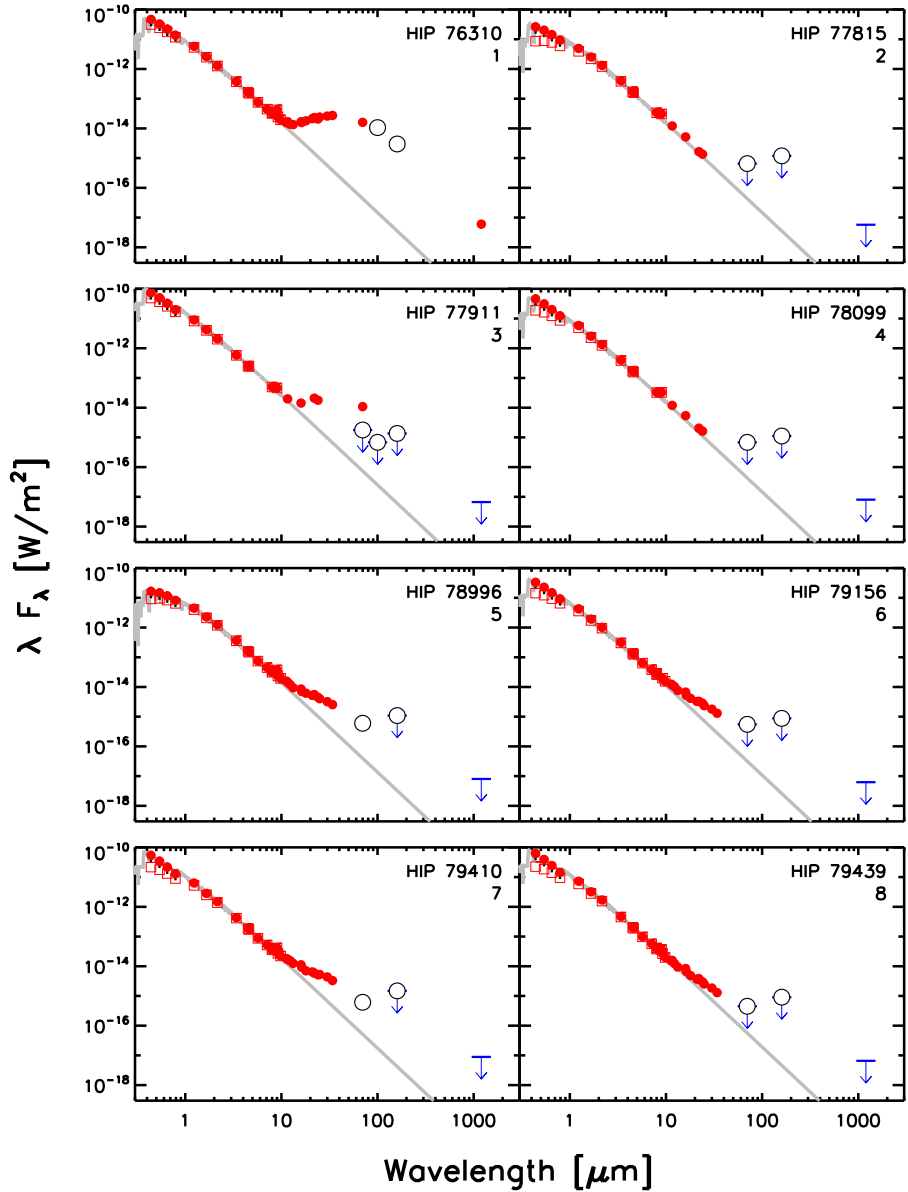}
  \caption{Spectral energy distributions for Upper Sco sources.  Dereddened photometry is shown as red circles, while original data is shown as open red squares.  PACS photometry is marked with open black circles, and 3$\sigma$ upper limits are shown as flat lines with downward facing arrows.  Stellar photospheres are shown as gray lines.  We have omitted upper limits when they have been superseded by a new observation at the same wavelength.  HIP 77911 is not detected in PACS photometry.  70 and 100 $\mu$m emission is detected at the location of a nearby companion (8''), suggesting past photometry may originate at that source.  }
  \label{fig:seds-first}
\end{figure*}

\begin{figure*}
  \includegraphics[angle=0,width=0.90\textwidth]{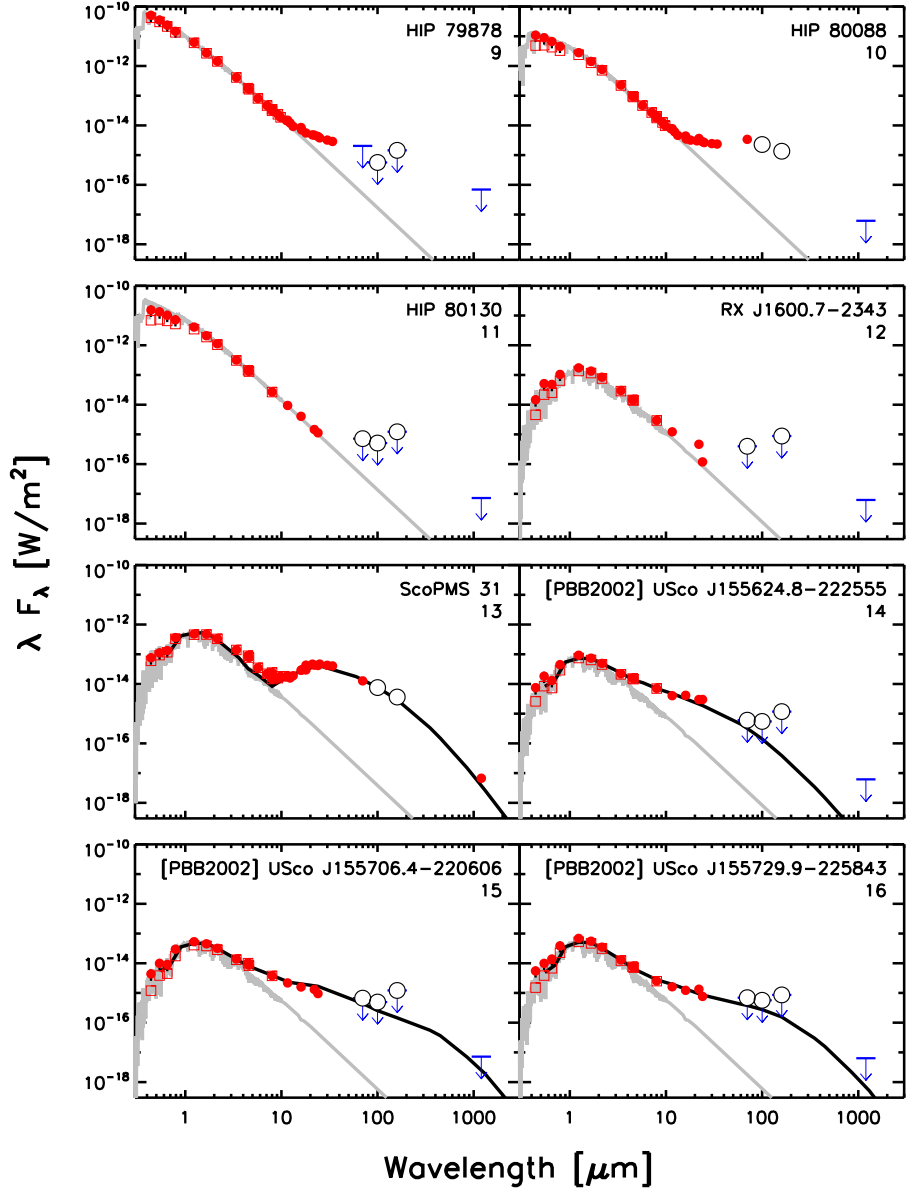}
  \caption{As in Figure \ref{fig:seds-first}, with the addition of black lines to indicate SEDs of best fit models from the genetic algorithm parameter search with MCFOST.}
  \label{fig:seds-too}
\end{figure*}

\setcounter{figure}{2}
\begin{figure*}
  \includegraphics[angle=0,width=0.90\textwidth]{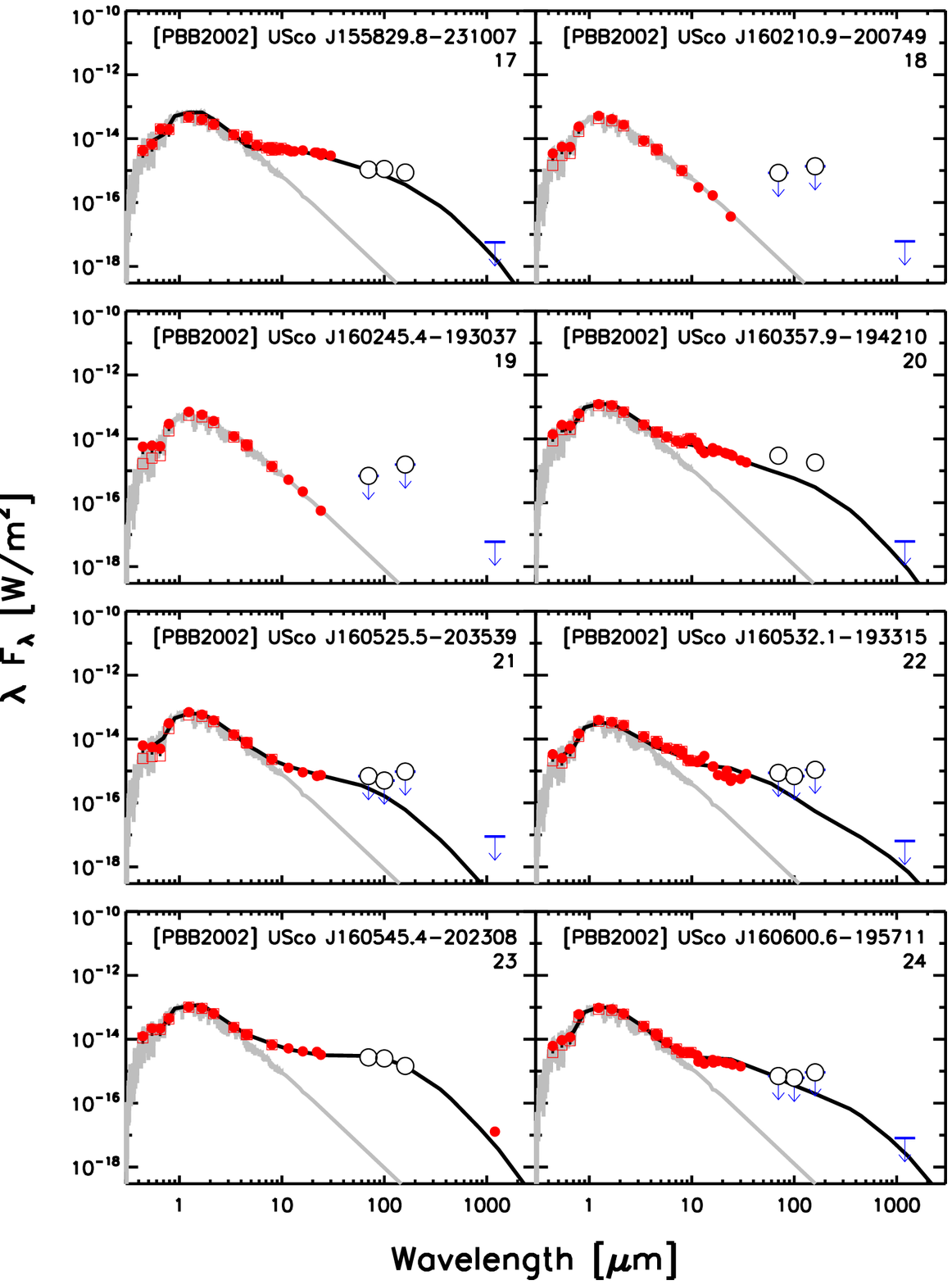}
  \caption{Continued}
\end{figure*}

\setcounter{figure}{2}
\begin{figure*}
  \includegraphics[angle=0,width=0.90\textwidth]{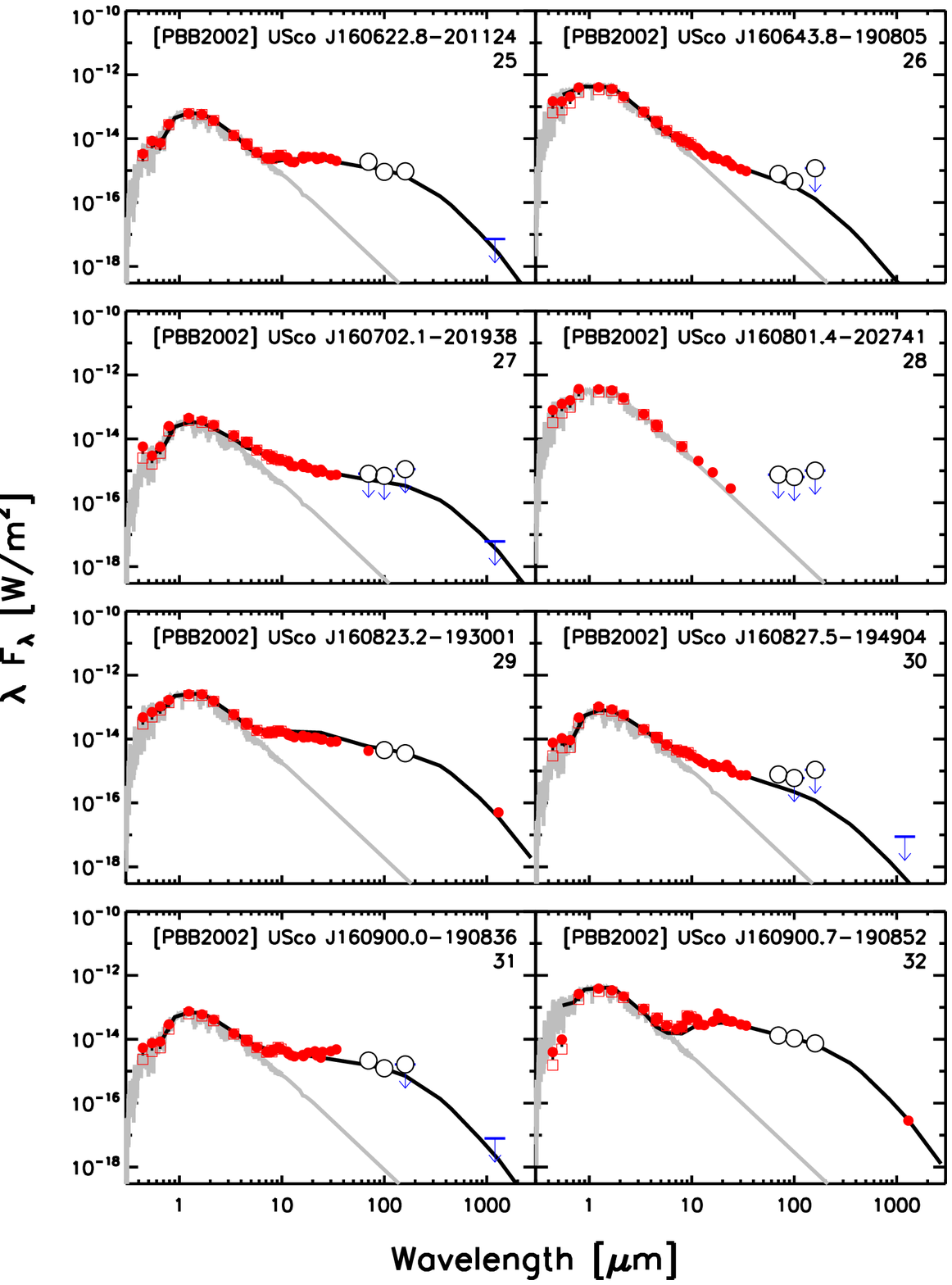}
  \caption{Continued}
\end{figure*}

\setcounter{figure}{2}
\begin{figure*}
  \includegraphics[angle=0,width=0.90\textwidth]{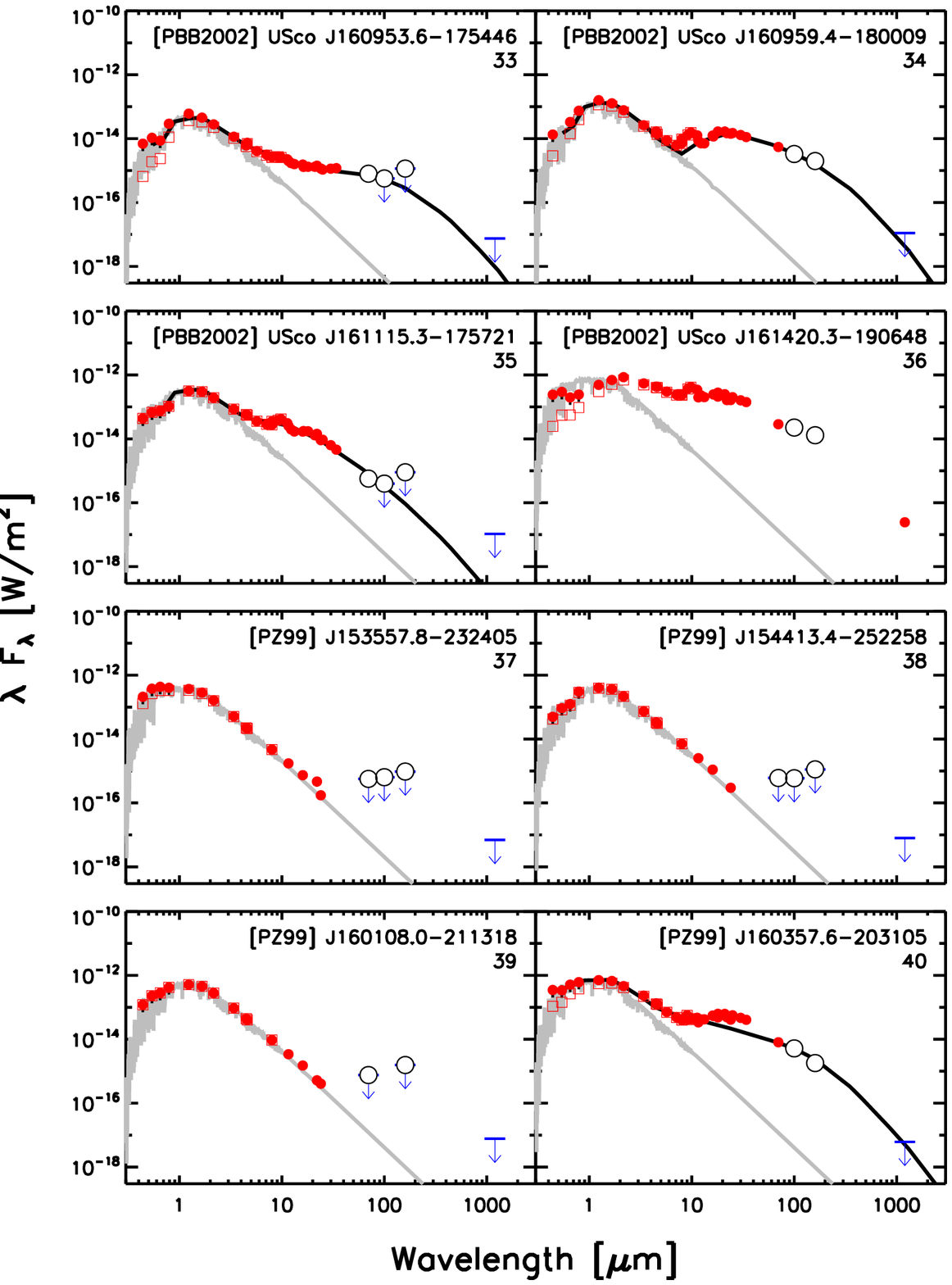}
  \caption{Continued}
\end{figure*}

\setcounter{figure}{2}
\begin{figure*}
  \includegraphics[angle=0,width=0.90\textwidth]{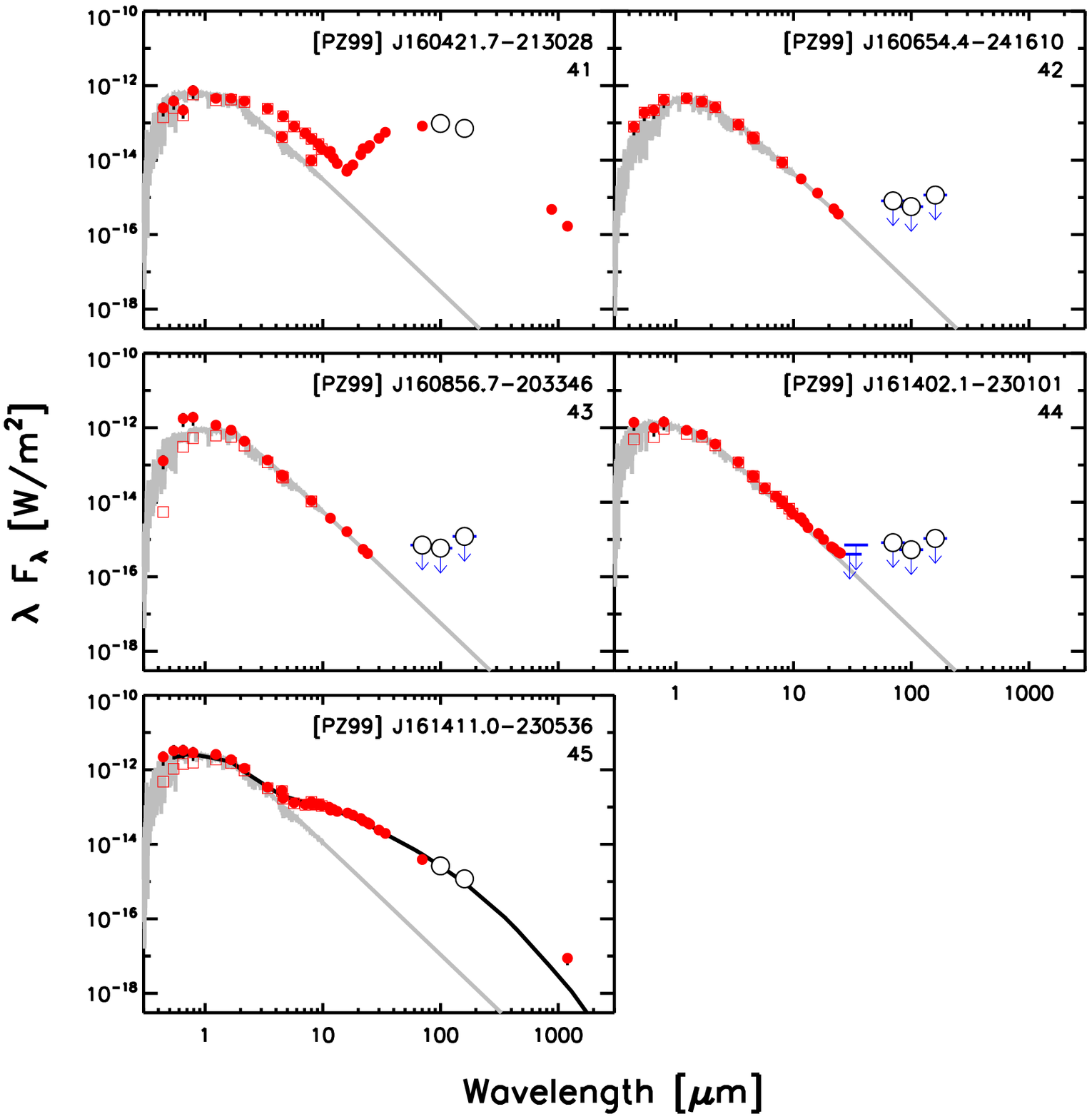}
  \caption{Continued}
  \label{fig:seds-last}
\end{figure*}


\section{Dust models}
\label{sec:models}

\label{sec:modphot}

To better constrain the dust content of the Upper Scorpius disks around K and M stars, we fit dust models to our SEDs using the radiative transfer code MCFOST \citep{Pinte:2006, Pinte:2009}.  We parameterize the radial surface density profile $\Sigma(R)$ of each object using the following simple geometry \citep[e.g.][]{Hartmann:1998}:
\begin{equation}
  \Sigma(R) = \Sigma_{100} \left ( \frac{R}{100 \rm{AU}} \right ) ^{\alpha}, 
\end{equation}
where $\Sigma_{100}$ is the surface density at 100 AU, and $\alpha$ describes the surface density variation of the disk.  $\Sigma_{100}$ is calculated by solving the relation: 
\begin{equation}
  M_{\rm{dust}} = \int_{R_{\rm{in}}}^{R_{out}} 2\pi R \Sigma(R) \,\mathrm{d}R .
\end{equation}
The dust mass ($M_{\rm{dust}}$), disk inner radius ($R_{\rm{in}}$), and surface density profile ($\alpha$) are free parameters, and the outer radius ($R_{\rm{out}}$) is set to 100 AU, 
as the far-infrared SED is insensitive to the outer disk radius.  
Five of the modeled objects are detected in millimeter emission (objects 13, 23, 29, 32, and 45), and thus may be sensitive to the choice of R$_{\rm{out}}$.  However, we do not allow this parameter to vary, choosing instead to emphasize fits to infrared emission.

The vertical dust distribution is modeled as a Gaussian with scale height:
\begin{equation}
H(R)=H_{100} (R/100 AU)^{\beta}, 
\end{equation}
where $H_{100}$ is the scale height at 100 AU and $\beta$ describes the power-law disk flaring.  For the dust properties, we adopt the silicate grain mineralogy of \cite{Draine:2006} with a minimum grain size of 0.01 $\mu$m.  Following \cite{Draine:2006}, we fix the exponent of the power law grain size distribution at $-3.5$ and the maximum grain size at 3.6 mm, three times the longest wavelength at which Upper Scorpius disks have been detected (with the exception of J1604-2130, which has been detected at 3 mm).  We must note, however, that most sources are not detected at this wavelength, and thus the modeling is most strongly affected by smaller grains that dominate emission at shorter wavelengths.  For the case of a smaller $a_{\rm{max}}$, the modeled dust mass is scaled down by a factor of $\sqrt{a_{\rm{max}} / 3600 ~\rm{\mu m}}$, where $a_{\rm{max}}$ may be reasonably set to three times the longest detected wavelength.  Therefore, for our objects lacking a detection at mm-wavelengths, the dust mass listed in Table \ref{tab:model-GA} should be treated as an upper limit.

We use a genetic algorithm (henceforth GA) to explore these 5 free disk parameters ($M_{\rm{dust}}$, $R_{\rm{in}}$, $\alpha$, $H_{100}$, and $\beta$).  In Table \ref{tab:model-ranges}, we show the ranges explored for each parameter, with variations noted below.  

\begin{table}
\caption{  \label{tab:model-ranges}  Model parameters}
\begin{tabular}{l c c }
\hline\hline
  Parameter &       Minimum            &      Maximum \\
  \hline
  $M_{\rm{dust}}$	& 	$10^{-6}$	$M_{\rm{sun}}$	&	$10^{-3}$	$M_{\rm{sun}}$	\\
  $R_{\rm{in}}$ 	&	0.03 AU				&	10 AU				\\
  $\alpha$		&	-2					&	0					\\
  $H_{100}$	&	2 AU			&	20 AU				\\
  $\beta$		&	1.05					&	1.25					\\
\hline

\end{tabular}
\end{table}

The GA carries out an iterative search for the lowest $\chi^2$ region of the parameter space.  For each object, we generate 20 generations of models.  In the first generation, 100 disk models are generated using disk parameters uniformly selected from within the ranges shown in Table \ref{tab:model-ranges}.  The $\chi^2$ values for each model are then calculated, and a model is selected to be the parent model of the next generation, with the selection probability being proportional to model $\chi^2$ ranking.  In addition, the lowest $\chi^2$ model is always carried over to the next generation.

The parameters of the child models in the new generation are randomly distributed around the values of the parent model.  A fraction of the models are then further mutated by varying all parameters by one-tenth the explored parameter ranges.  The mutation rate is adjusted during the course of the calculation: if the ratio of the median $\chi^2$ to lowest $\chi^2$ is greater than 2, the parameter variation is increased by a factor of 1.5 to accelerate the identification of the solution region.  If the ratio is below 1.1, then the mutation rate is reduced by a factor of 1.5, providing for denser sampling in the vicinity of the solution.   
For each object, we carry out this procedure for 20 generations (for a total of 2000 models per disk).  We adopt the parameters of the model with the lowest $\chi^2$ value for each object, which we report in Table \ref{tab:model-GA}.

\section{Discussion}
\label{sec:discussion}

\label{sec:DiskProperties}

\subsection{Median SED}
\label{sec:MedianSED}

To study the evolution of the typical T-Tauri star disk, we calculate the median SED for the 24 K\&M stars in Upper Scorpius with infrared excesses; these are the 22 sources modeled above, as well as J1604-2130 and J1614-1906.  20 of these are classified as Class II sources by the slope of their SEDs from K band to 24 $\mu$m \citep[for $\lambda F_{\lambda} \propto \lambda^n$, $-1.6 < n < -0.3$,][]{Greene:1994}.  4 are classified as class III with weaker but still clear infrared excess emission.  They are included here as our goal is to understand the median disk among K\&M stars in a single star forming region.  While we have collected optical, 2MASS, WISE, and \textit{Spitzer} photometry, we exclude \textit{Spitzer} IRS spectroscopy from our median as the sample was not uniformly observed with IRS; 5 low mass sources were omitted.  

Such medians have been determined before using \textit{Spitzer} IRAC, IRS, and MIPS observations in numerous young regions \citep[e.g.][]{Furlan:2006}.  These have typically adjusted for the varying stellar luminosity of the stars by normalizing individual SEDs to the same H band flux and then determining the medians of the resulting photometric points.  This approach may lead to an underestimation of variation within the population (i.e. the upper and lower quartiles) by suppressing excesses at the shortest wavelengths.  This is unlikely to be an important issue for the stars of Upper Scorpius, but it may be a significant issues for the youngest, most massive disks which can exhibit excesses at H and K bands.  Therefore, we adopt a new approach which more directly compensates for the variation in stellar luminosity.

For each source, we carry out the following procedure:  
\begin{enumerate}
\item Deredden the SED using the values found in fits to the stellar photospheres. 
\item Subtract the stellar photosphere, as determined in our previous fitting (Sect. \ref{sed}).
\item Scale each set of excesses to the luminosity of a median spectral type Upper Scorpius M2 star, $[$PBB2002$]$ J160357.9-194210 (object 20, 0.09 \Lsun).
\item Add the excesses to the stellar SED of this M2 star.  
\end{enumerate}

We choose object 20 as the representative star as it lies near both the mid-range of the spectral types represented here and the mean of the stellar luminosities, as well as having a low extinction.  This will minimize error contributions in the scaling process.  

At wavelengths up to 160 $\mu$m, we determine the median flux, as well as upper and lower quartiles.  We omit (sub)millimeter photometry from this analysis as it is determined almost entirely by the disk dust mass, and we seek to determine what constraints the far-infrared photometry can provide in its place. 

Since all of these sources are detected at wavelengths up to 24 $\mu$m, the determination of median and quartile fluxes are straightforward at short wavelengths.  However, at the PACS wavelengths of 70, 100, and 160 $\mu$m, from 6 to 10 sources have fluxes falling below our three sigma detection limits.  Therefore, we adopt a new procedure for the determination of median and upper and lower quartile fluxes.  We use the Kaplan-Meier product limit estimator -- designed for work with censored data sets --  to construct the scaled cumulative flux distribution at each wavelength \citep[e.g.][]{Feigelson:1985}.  As we did not carry out PACS 70 $\mu$m observations for objects detected in \textit{Spitzer} MIPS band 2 observations, we merge those detections with our 70 $\mu$m observations.  We show the distributions for 70, 100, and 160 $\mu$m emission in Figure \ref{fig:PACS-KM-distr}.  From these distributions, we then estimate the median and quartile fluxes for inclusion in the median SED, which we show as the vertical red lines.  In all three cases, the lower-quartile is an upper limit, while in the case of the 160 $\mu$m flux, the median is also an upper limit.

\begin{figure*}
\vspace{2cm}

  \subfloat[]{ \includegraphics[angle=90,width=0.33\textwidth]{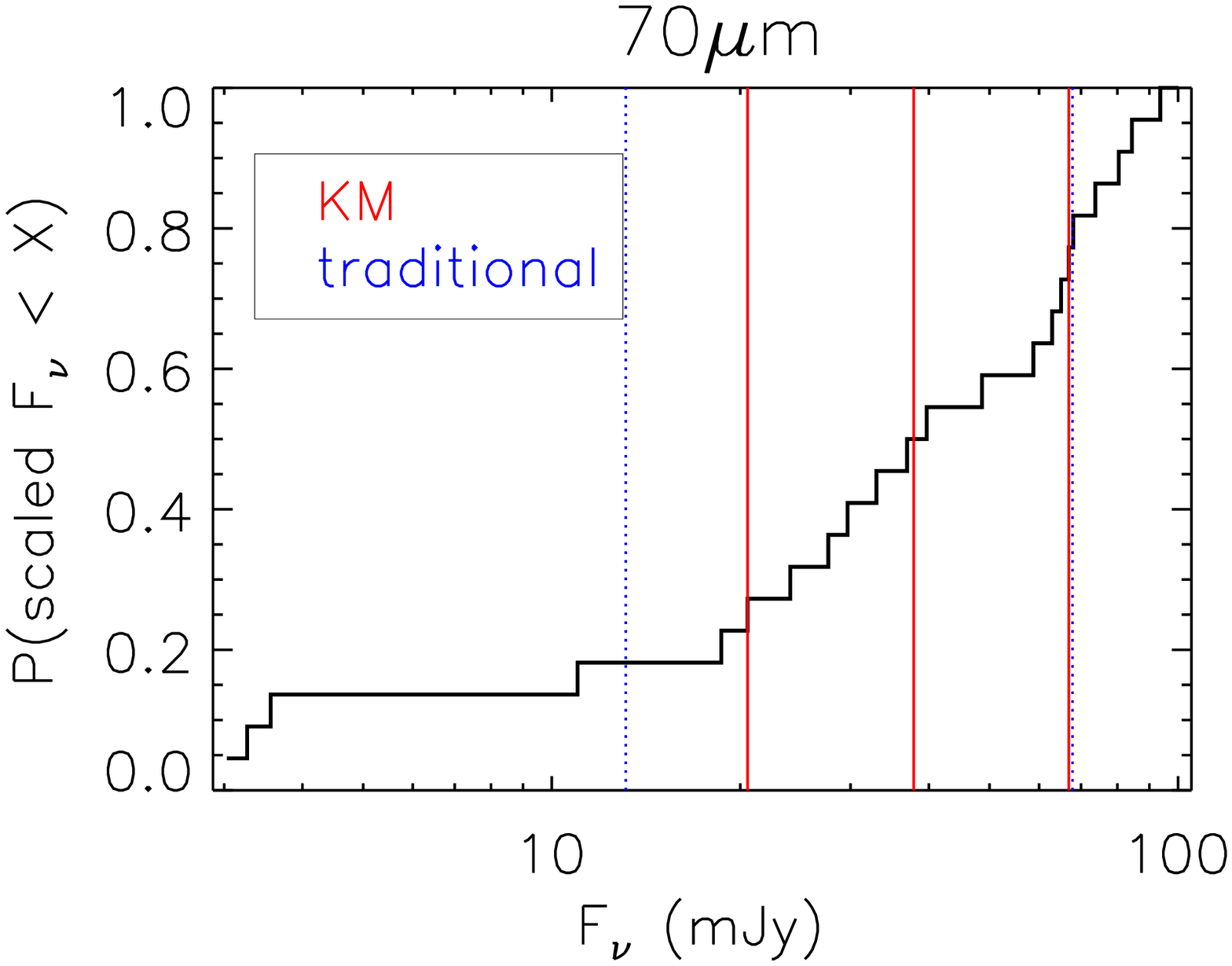}}
  \subfloat[]{ \includegraphics[angle=90,width=0.33\textwidth]{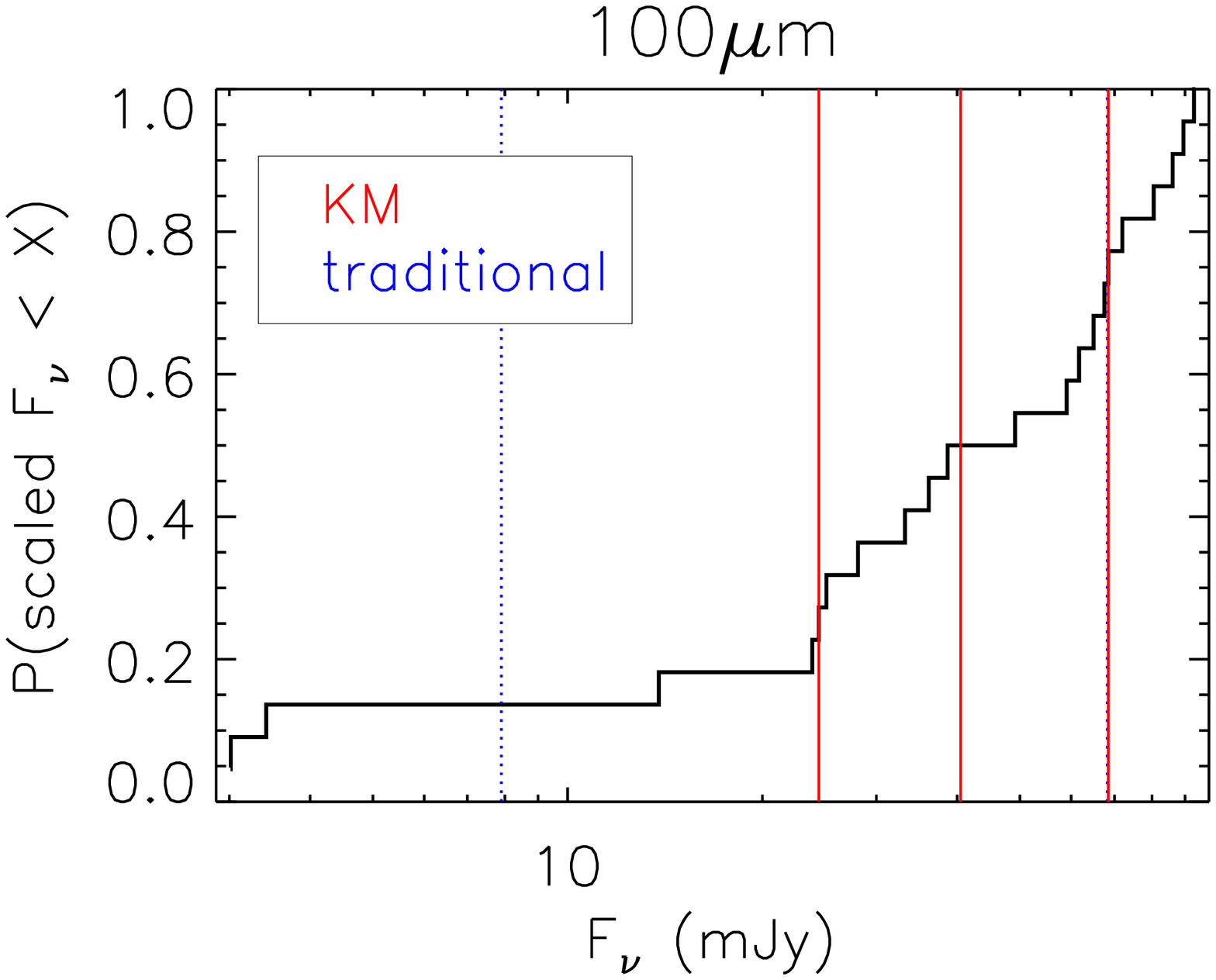}}
  \subfloat[]{ \includegraphics[angle=90,width=0.33\textwidth]{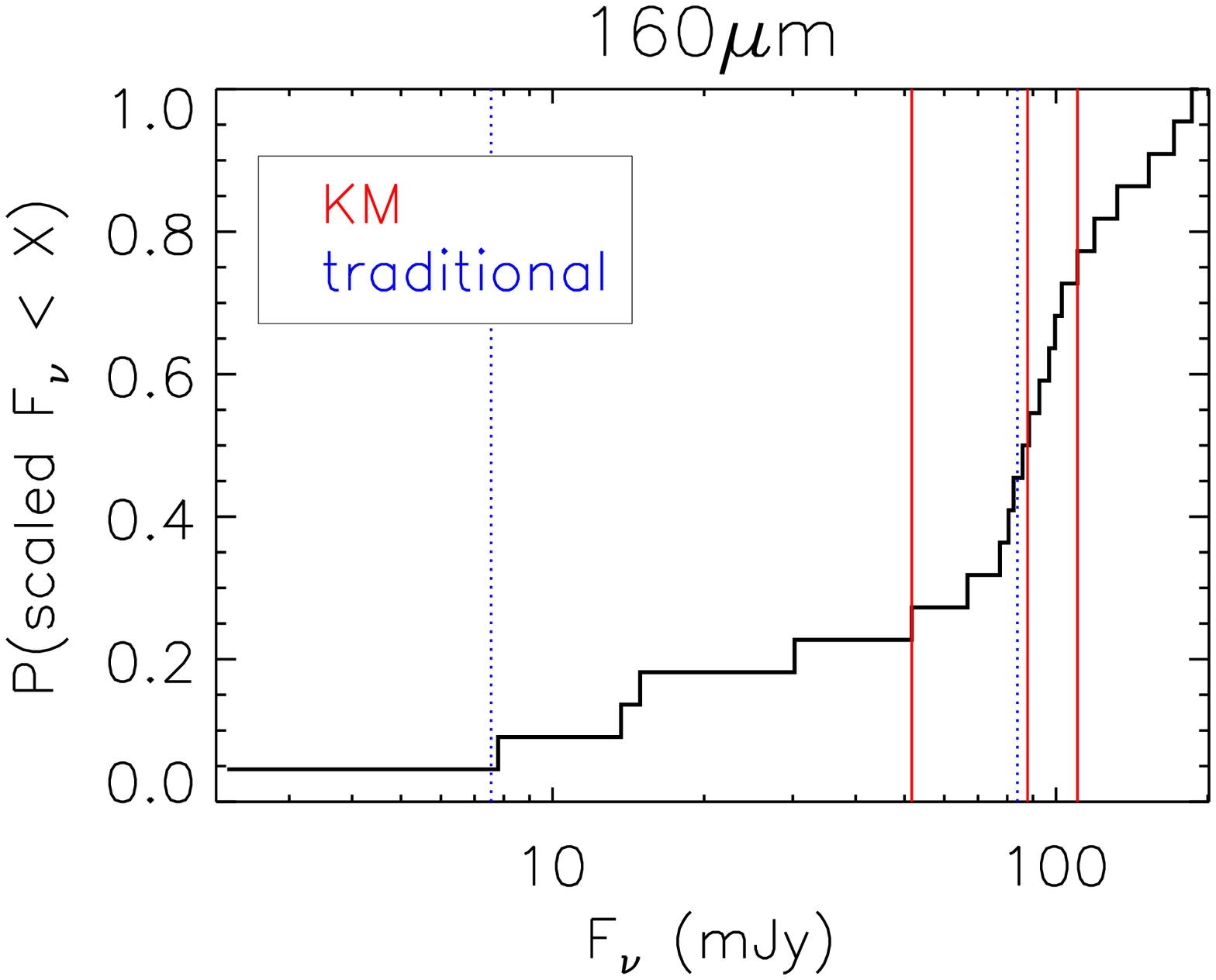}}
  \caption{Cumulative distributions of the luminosity scaled flux for PACS photometry of K and M Class II sources modeled in this paper.  Distributions are built using the Kaplan-Meier estimator.  The median and quartile values are shown with the vertical red lines, and the traditional median and quartiles are shown with the blue lines.  The traditional method for determining lower quartiles fails in the situation where more than $\sim$25\% of observations result in nondetection, whereas this approach allows the determination of an upper limit to the quartile.  The 160 $\mu$m median presented here represents an upper limit, as more than half the sources are undetected at this wavelength.}
  \label{fig:PACS-KM-distr}
\end{figure*}


For comparison, we determine the median SED for a sample of Taurus stars in the same fashion, using the 26 Class II sources of spectral types K2 to M5 observed with PACS photometry in C.D. Howard, et al., (2013, ApJ).  For these sources, we have collected B and I band photometry from SDSS, J, H, and K band photometry from 2MASS, and 3.4, 4.6, 11.6, and 22.1 $\mu$m photometry from WISE.  The spectral type distribution is slightly skewed toward warmer sources, including 10 K stars and 16 M stars.  The wavelength distribution of disk emission is a function of the temperature distribution in the disk, which is in turn more strongly dependent on the distribution of disk material rather than stellar luminosity.  Therefore, we consider this a useful comparison sample.

We present the median SEDs for Upper Scorpius and Taurus in Tables \ref{tab:MedianUpSco} and \ref{tab:MedianTau}, respectively, and compare the median SEDs in Figure \ref{fig:median-SED}.  We note that our determination of the median SED exhibits a much larger range between upper and lower quartiles at short wavelengths than was found in previous determinations.  This highlights the need to directly scale excesses by the stellar luminosity, rather than by a proxy.  The common practice of scaling all fluxes to a common H band value artificially reduces the contribution of sources with short wavelength excesses.  We also point out that Taurus sources show both a significant optical excess and a large dispersion in optical excess values, which we attribute to the much higher accretion rates found among Taurus sources.

\begin{table}
\caption{   \label{tab:MedianUpSco} Upper Scorpius median SED}
\begin{tabular}{l c c c }
\hline\hline
  Wavelength &       Lower quartile            &      Median         &      Upper quartile                \\
  $\mu$m           &       $W/m^2$                 &      $W/m^2$      &      $W/m^2$        \\
\hline
   0.44 &      1.17E-14 &      1.65E-14 &      1.82E-14  \\
   0.54 &      1.65E-14 &      2.13E-14 &      2.56E-14  \\
   0.65 &      2.02E-14 &      2.31E-14 &      2.82E-14  \\
   0.79 &      5.27E-14 &      6.16E-14 &      7.57E-14  \\
   1.24 &      1.13E-13 &      1.21E-13 &      1.37E-13  \\
   1.66 &      1.04E-13 &      1.13E-13 &      1.26E-13  \\
   2.16 &      6.65E-14 &      7.04E-14 &      7.45E-14  \\
   3.40 &      2.36E-14 &      2.72E-14 &      3.17E-14  \\
   4.50 &      1.40E-14 &      1.55E-14 &      2.23E-14  \\
   8.00 &      4.40E-15 &      6.68E-15 &      7.71E-15  \\
  11.6 &      3.34E-15 &      5.29E-15 &      6.15E-15  \\
  16.0 &      2.51E-15 &      4.90E-15 &      5.81E-15  \\
  24.0 &      1.55E-15 &      3.21E-15 &      4.62E-15  \\
  70.0 &      9.26E-16 &      1.67E-15 &      2.91E-15  \\
 100. &      7.65E-16 &      1.25E-15 &      2.09E-15  \\
 160. &      9.74E-16 &      1.65E-15 &      2.07E-15  \\
\hline
\end{tabular}
\end{table}

\begin{table}
\caption{  \label{tab:MedianTau}  Taurus median SED}
\begin{tabular}{l c c c}
\hline\hline
  Wavelength &       Lower quartile            &      Median         &      Upper quartile               \\
  $\mu$m           &       $W/m^2$                 &      $W/m^2$      &      $W/m^2$        \\
\hline  
  0.48  &  9.49E-15  &  2.15E-14  &  1.38E-13    \\   
   0.63 &      3.31E-14 &      5.73E-14 &      9.36E-14  \\	
  0.77  &  1.60E-14  &  6.04E-14  &  9.97E-14    \\   
   0.91 &      3.90E-14 &      6.04E-14 &      1.56E-13  \\	
  1.24  &  1.03E-13  &  1.15E-13  &  1.28E-13    \\   
  1.66  &  9.36E-14  &  1.06E-13  &  1.19E-13    \\   
  2.16  &  6.21E-14  &  8.60E-14  &  1.11E-13    \\   
  3.35  &  2.47E-14  &  5.02E-14  &  7.95E-14    \\   
 4.60  &  1.76E-14  &  3.43E-14  &  5.72E-14    \\   
 11.6  &  6.43E-15  &  1.07E-14  &  2.04E-14    \\   
 22.1  &  4.84E-15  &  1.19E-14  &  2.12E-14    \\   
 70.0  &  2.37E-15  &  4.23E-15  &  7.87E-15    \\   
100.  &  1.74E-15  &  2.48E-15  &  5.13E-15    \\   
160.  &  7.31E-16  &  1.36E-15  &  2.84E-15    \\   
\hline
\end{tabular}
\end{table}

\begin{figure}
\vspace{1cm}\hspace{0.5cm}  \includegraphics[angle=90,width=0.9\columnwidth]{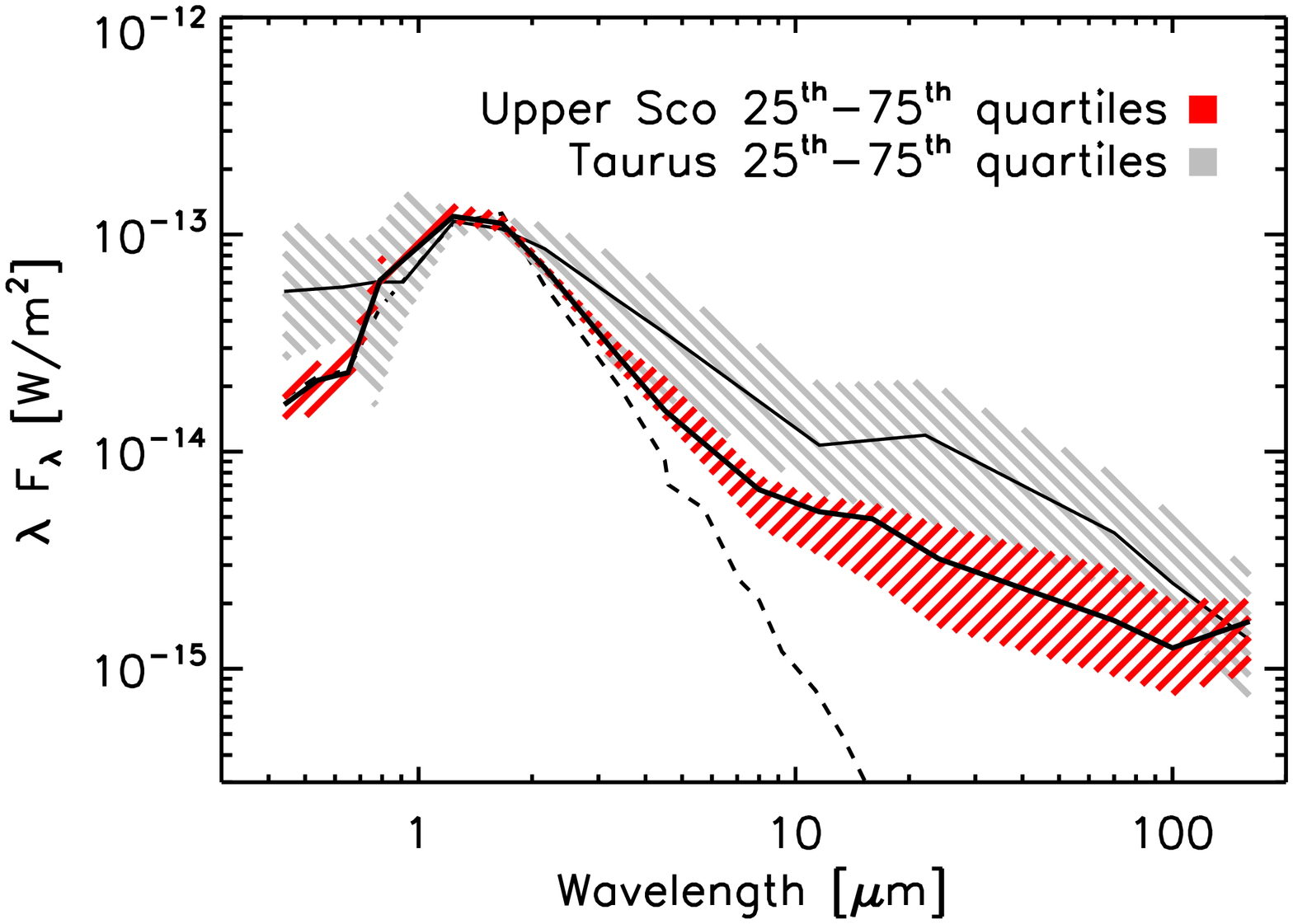}
  \caption{Comparison of the Upper Scorpius and Taurus median SEDs.  While the lower quartile of Taurus fluxes and upper quartile of Upper Scorpius fluxes overlap, the medians are separated by a factor of $\sim3$, suggesting the typical surviving Upper Scorpius disk is more evolved (e.g. dust settling, grain growth).  In addition, Upper Scorpius sources show little to no optical excess, attributable to their low or nondetectable accretion.}
  \label{fig:median-SED}
\end{figure}

For these median SEDs, we make initial parameter estimates by carrying out model fits to the SEDs using the genetic algorithm as described above.  We show these best fit models overlaid on the SEDs in Figure \ref{fig:medianfits}.  We list the model parameters for the two median SEDs at the end of Table \ref{tab:model-GA}.  For all five properties, the Upper Scorpius median disk lies within the central 68\% of the distribution of properties, suggesting it serves as a reasonable proxy for discussion of the whole population.  

\begin{figure}
  \includegraphics[angle=0,width=0.95\columnwidth]{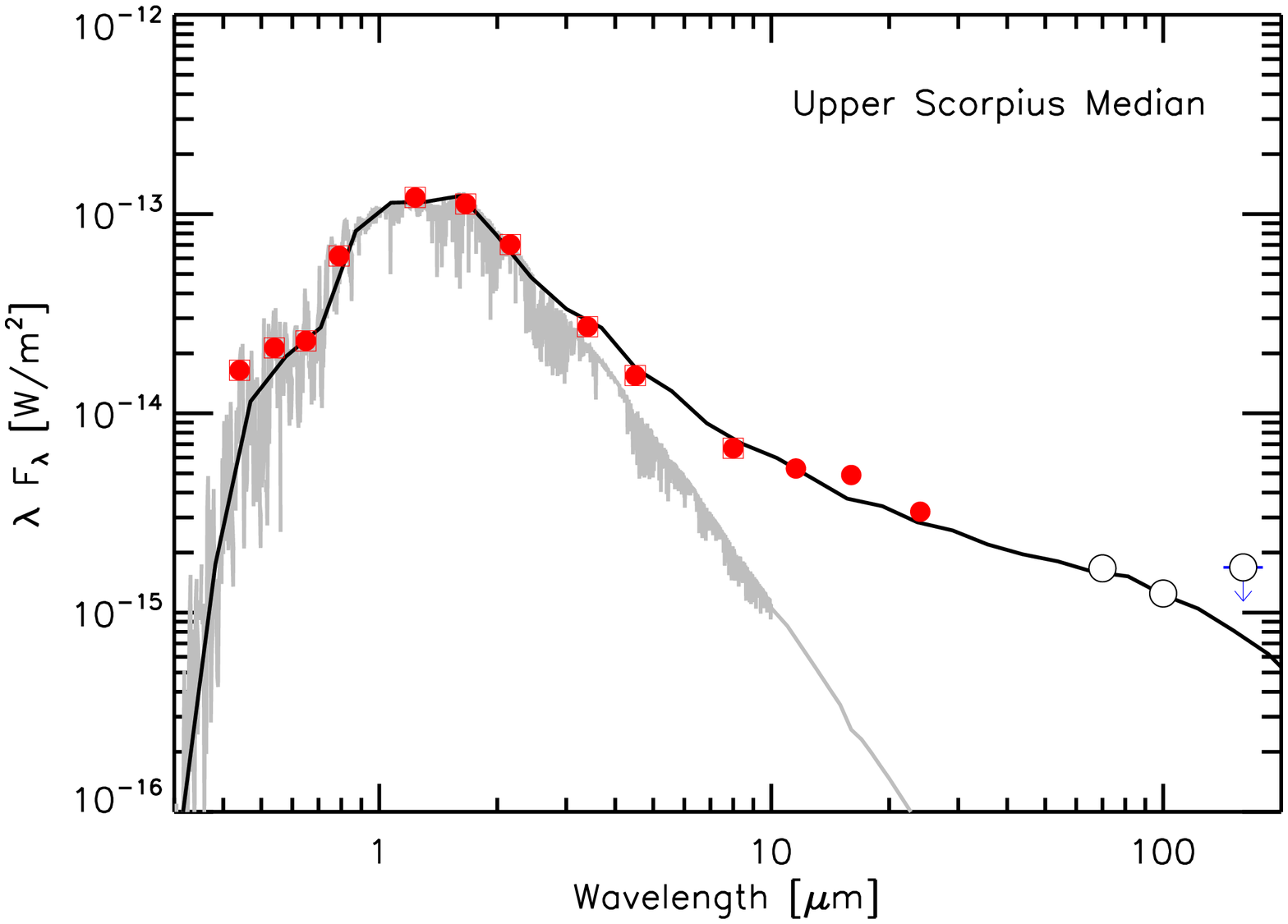}    \\   \includegraphics[angle=0,width=0.95\columnwidth]{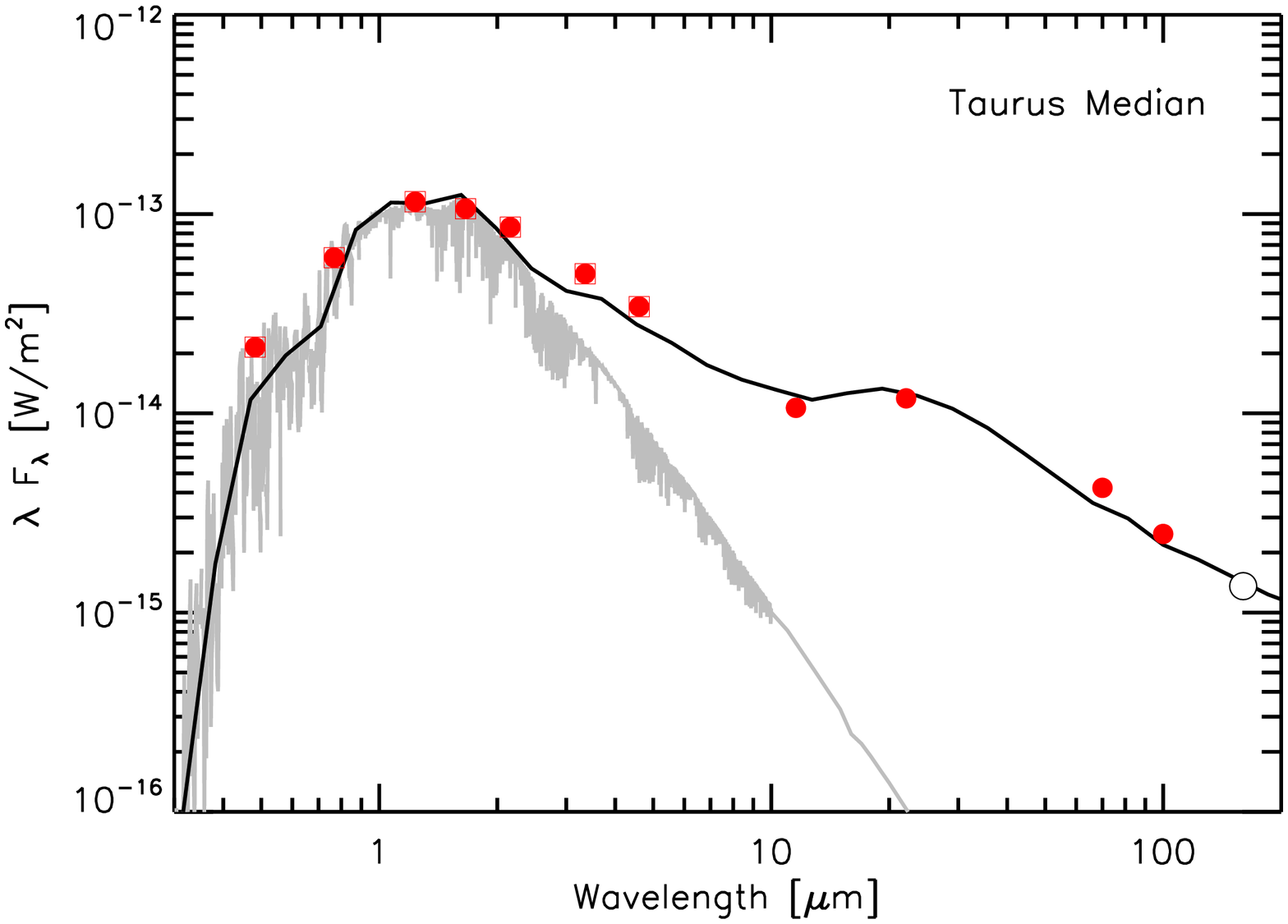}
  \caption{Best fit models to the median Upper Scorpius (top) and Taurus (bottom) SEDs. }
  \label{fig:medianfits}
\end{figure}

To estimate the parameter uncertainties, we have generated an evenly sampled grid around the best fit GA model for each median disk.  From the 5-dimensional grid of models, we generated the one-dimensional probability distributions by assigning each model $i$ a probability $P_i \propto \exp [-\chi^2 / 2]$, normalizing such that the model probabilities sum to 1, and then binning along each dimension.  We show the property probability distributions in Fig. \ref{fig:median-params}.  The distribution of properties for the Upper Scorpius median SED is much flatter than that for Taurus, due to the 160~$\mu$m point being an upper limit.  We have carried out further examination of the parameter sensitivities and their covariance in Appendix \ref{sec:chisq-maps}.

\begin{figure*}[]
\vspace{2cm}
  { \includegraphics[angle=0,width=0.45\textwidth]{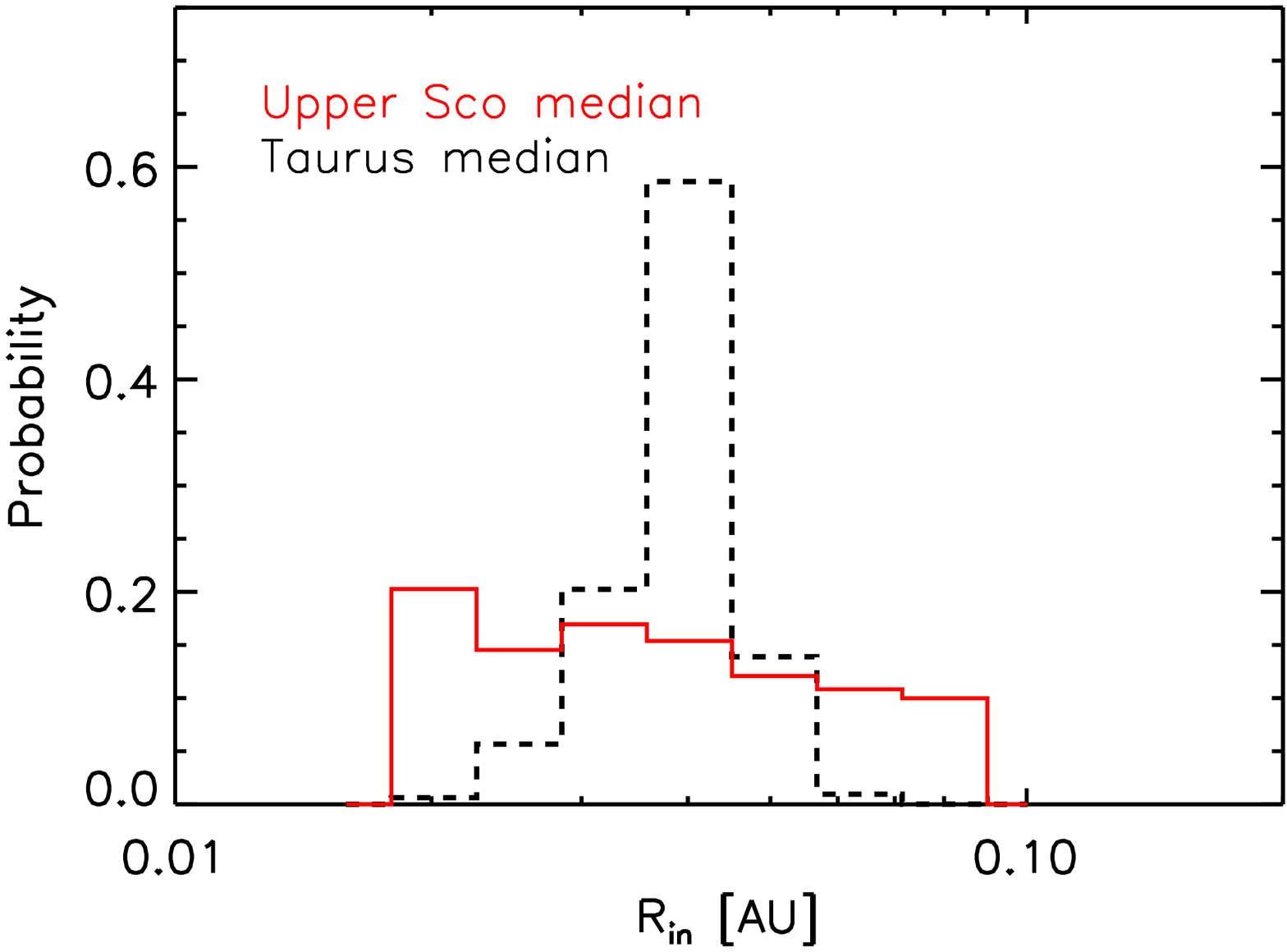}}	{ \includegraphics[angle=0,width=0.45\textwidth]{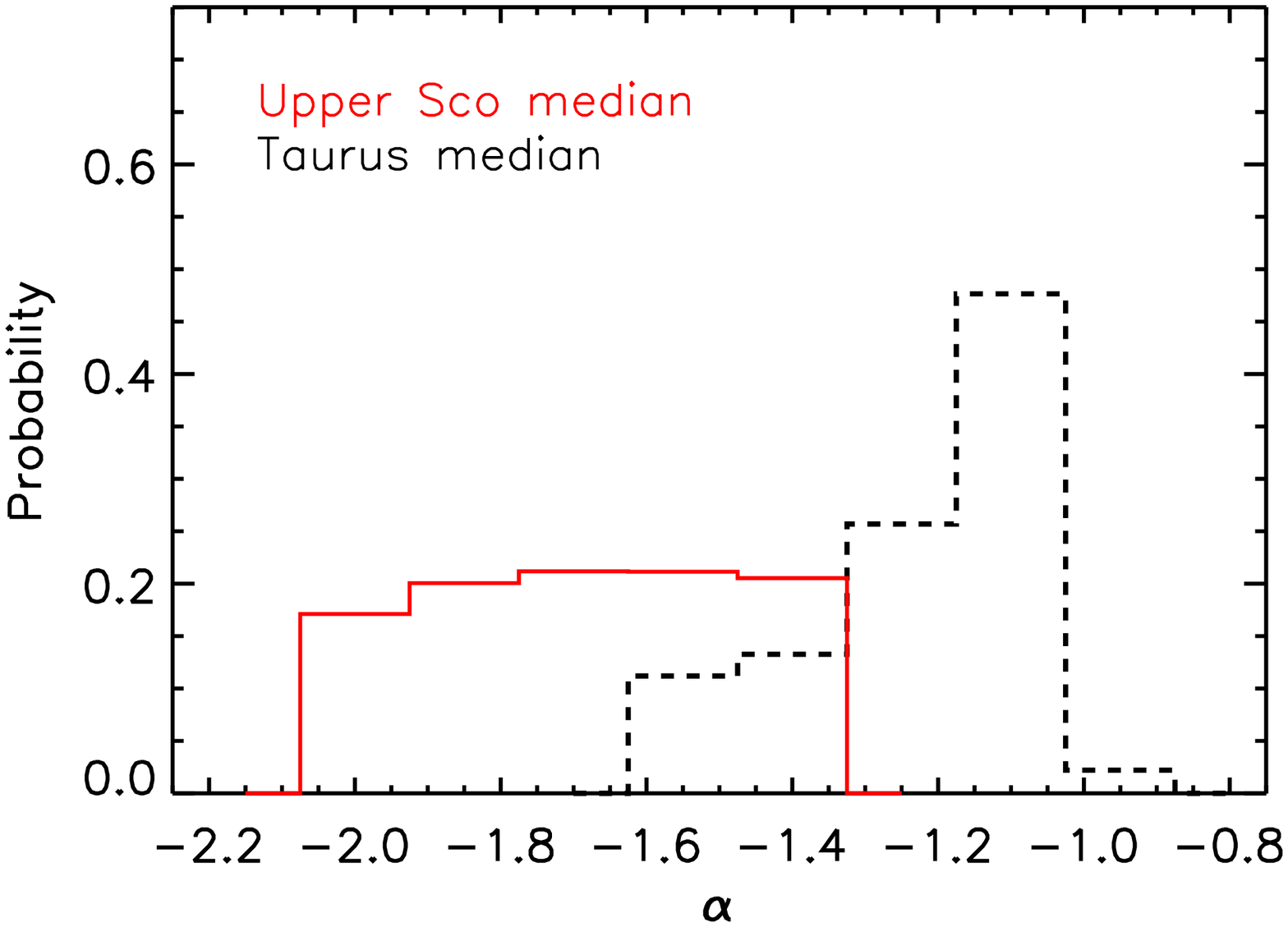}} \\
  { \includegraphics[angle=0,width=0.45\textwidth]{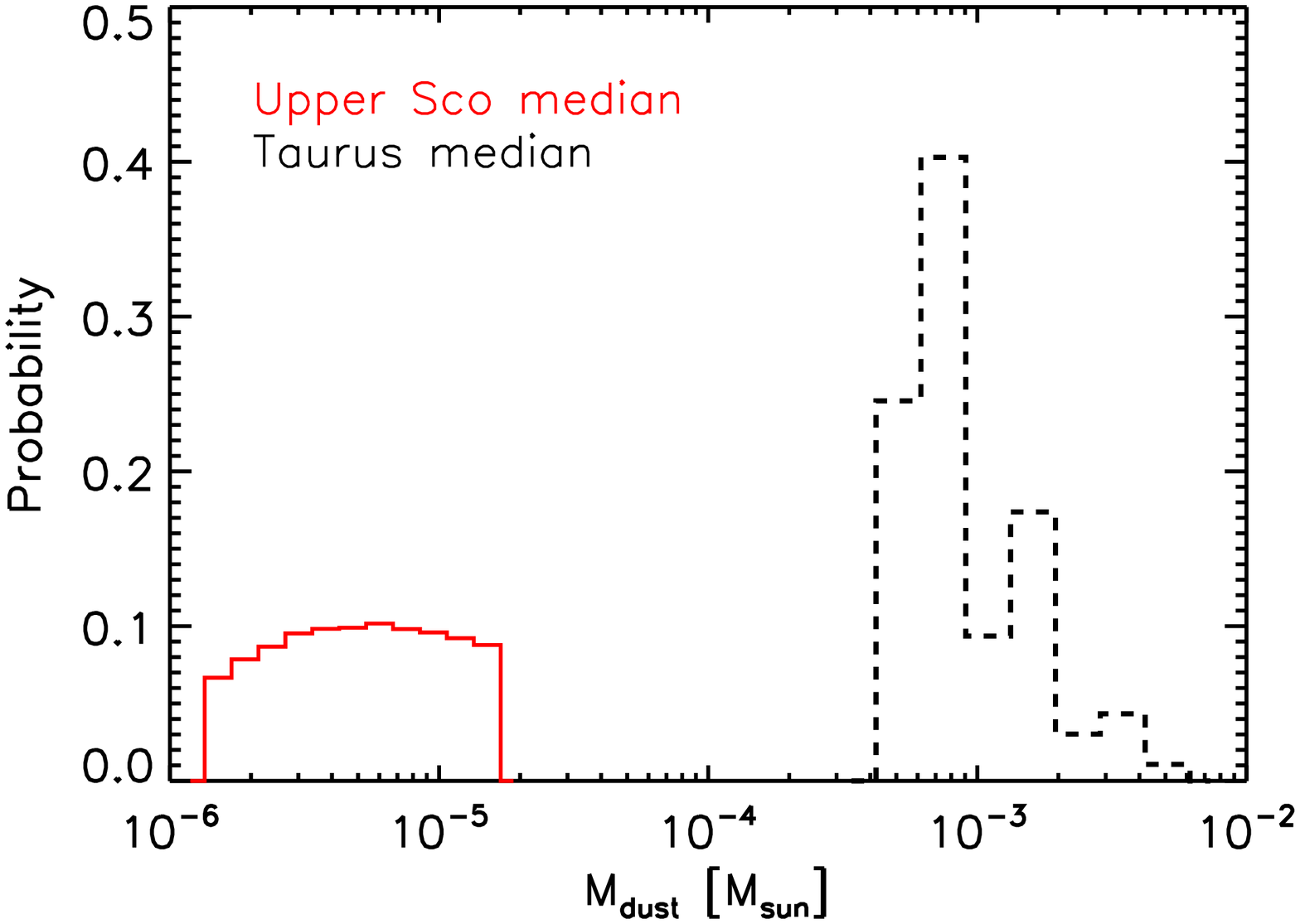}}  { \includegraphics[angle=0,width=0.45\textwidth]{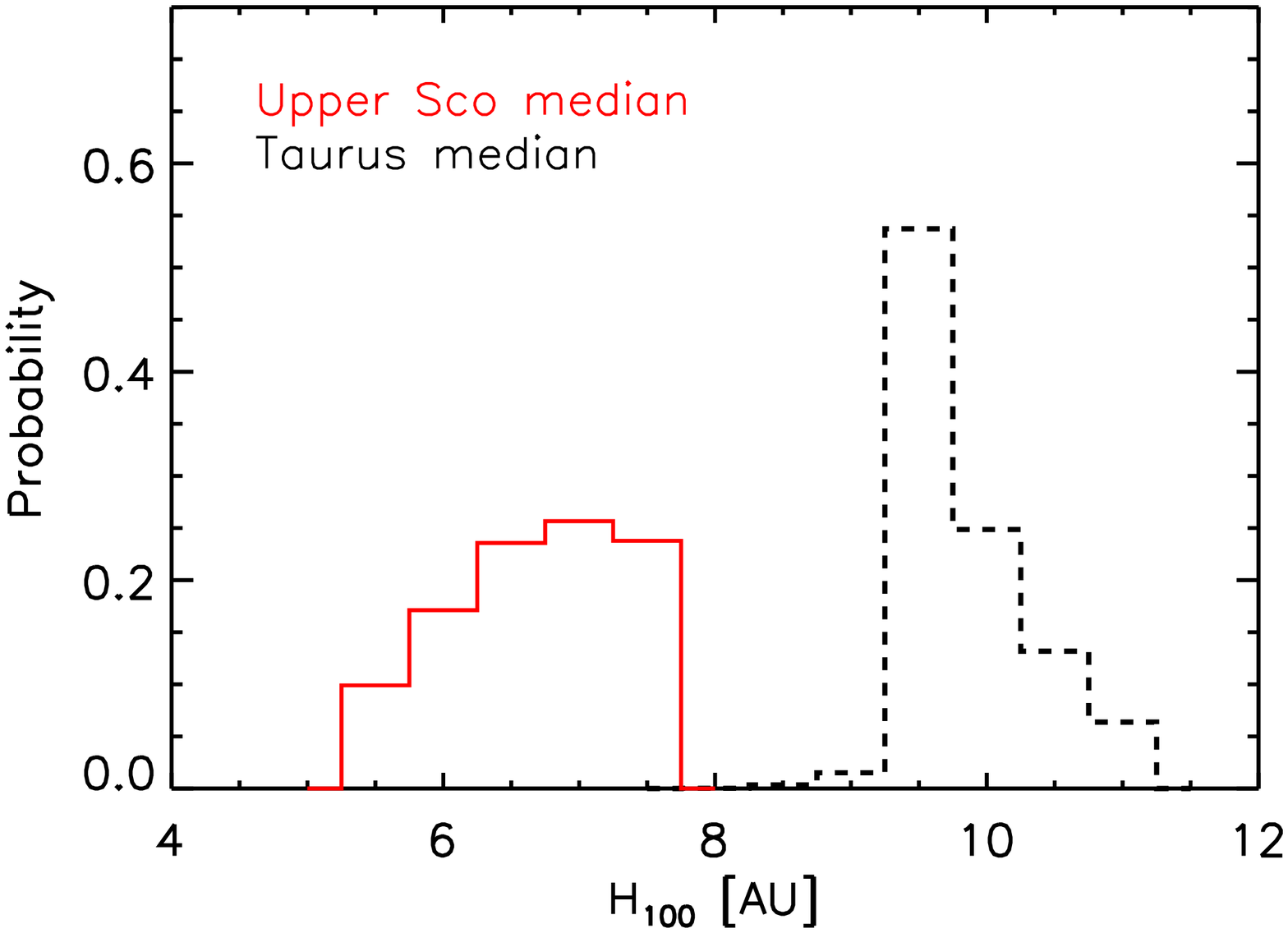}}  \\
  { \includegraphics[angle=0,width=0.45\textwidth]{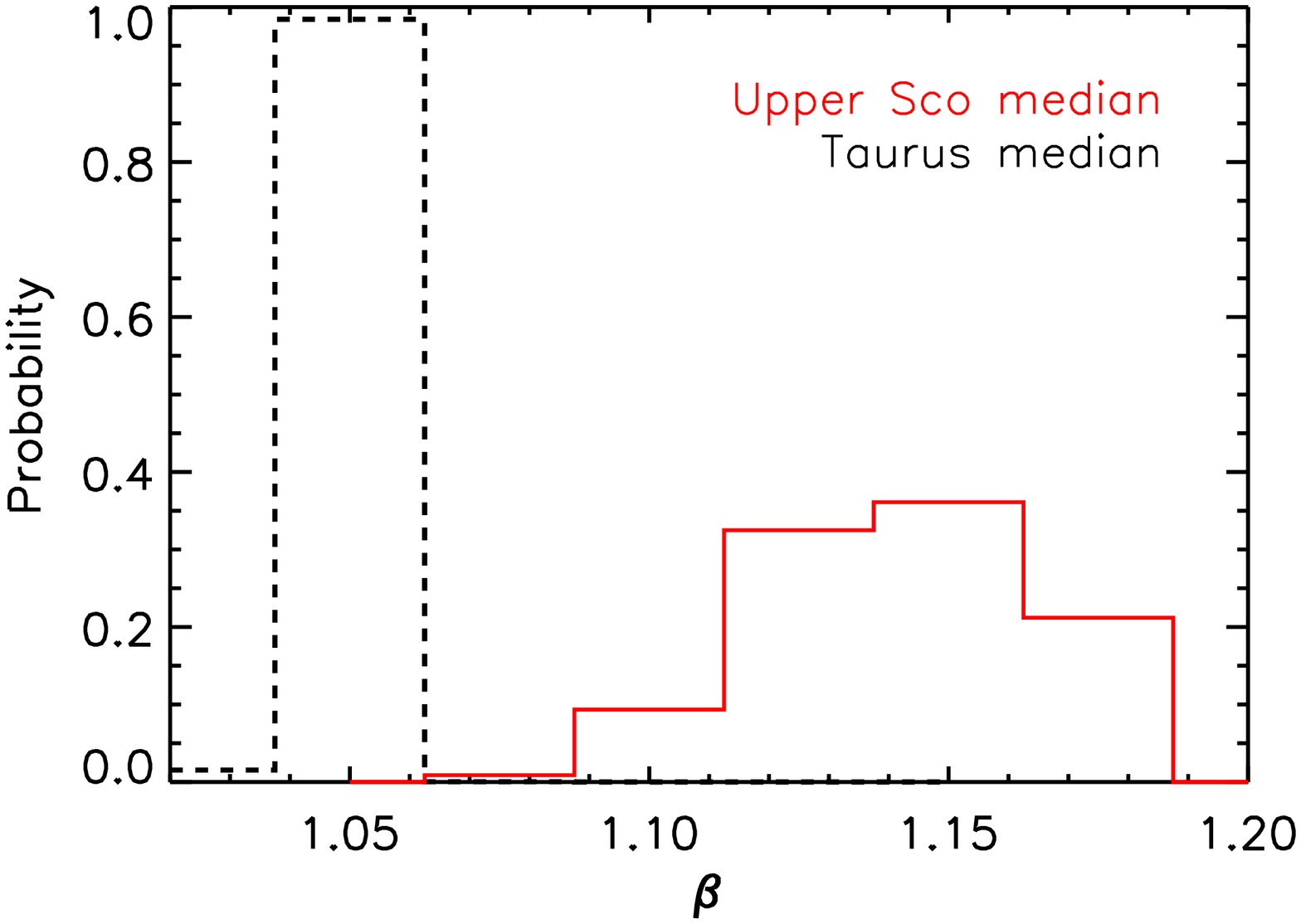}} \\  
  \caption{Comparison of the probability distributions for the Upper Scorpius and Taurus median disks.  The striking differences in the disk scale height and flaring index show that the median Upper Scorpius SED represents a disk that has experienced more dust settling than in Taurus.  The low dust mass hints at the potential importance of grain growth in reducing opacity and hence detectable dust mass.  The sharp edges to the Upper Scorpius distributions are due to the sampling of the parameter space. }
  \label{fig:median-params}
\end{figure*}

The typical Upper Scorpius disk has both a smaller scale height and greater flaring than the typical Taurus disk, with a scale height at 100 AU, $H_{100}$, of 7.3 vs. 9.2 AU, and flaring of 1.16 vs. 1.07.  These factors combine to give a ratio of disk height to radius at 1 AU of 0.035 for Upper Scorpius, and 0.067 for Taurus.  This suggests dust that is more settled than Taurus disks at all radii, and the degree of settling is greater at smaller radii.  

The median Upper Scorpius SED corresponds to a disk with an inner disk edge at 0.041 AU, compared to that of Taurus with an inner disk edge at 0.044 AU.  These inner disk radii are close to the sublimation radius, suggesting that the typical observed disk has not yet begun inside out clearing.  This is consistent with previous findings that transitional disks showing evidence of inner disk clearing account for $\lesssim10$\% of circumstellar disks \citep{Williams:2011}, a low enough frequency as to not affect the median of the ensemble.  This low frequency is evidence that disks dissipate rapidly once clearing begins.  However, the Upper Scorpius median SED has a tentatively steeper surface density density profile.  Imaging of dust emission is necessary to better constrain the surface density power-law index.

The total amount of material seen in these disks, however, has decreased, with the inferred dust mass of the Upper Scorpius median disk being a factor of $\sim$100 lower than that of the median Taurus disk.  We must note that the infrared SED which we use here is at least a factor of 5 less sensitive to the dust mass than is millimeter photometry.  Our previous comparison of millimeter emission in Upper Scorpius and Taurus \citep{Mathews:2011} showed that the high end threshold of millimeter-dust masses had decreased by a factor of 20.  Our modeling here suggests a similar trend across all disk masses, but sensitive (sub)millimeter photometry is necessary to confirm this scenario.

Our results suggest that the few disks that remain at the age of Upper Scorpius are highly evolved in their other dust properties, as well.  The combination of parameters suggesting a high degree of dust settling, low observable dust mass, yet an inner radius consistent with the sublimation radius may reinforce the idea that most disks experience homologous evolution (i.e. processes occurring at all radii, rather than inside-out or outside-in depletion mechanisms) through most of their lifetimes.  Below, we discuss observations that inform our understanding of the gas evolution of these disks.

\subsection{The nature of [OI] 63 $\mu$m emission}

In cases where line emission is detected, it is important to clearly establish the nature of the emission source.  Upper Scorpius is old enough that background CO emission is largely dissipated, and the IFU observations allowed us to identify background contamination in [CII] emission.  But in addition to disks, jets are a prominent source of [OI] line emission from young stars \citep{Liseau:2006}.  In an earlier study \citep{Mathews:2010}, we compared the emission line ratios of [OI]63, [OI]145, and [CII]157 of four GASPS sources to the locus of jet sources observed by \citeauthor{Liseau:2006}, and determined that these sources were unlikely to overlap.  However, comparisons to this locus are useful only in cases where one or several lines are detected.  Among our sample here only 2 sources are detected in line emission.

To classify sources as potential jet or disk sources, we make use of a recently developed diagnostic using the [OI]63$\mu$m line emission and continuum.  Howard, et al. (in review) have identified the ratio of the [OI] 63.184 $\mu$m line flux and the 63 $\mu$m continuum flux as a powerful diagnostic of the nature of the [OI] emission, finding that
disk sources lacking a jet lie along a narrow locus, and sources with jets range up to an order of magnitude higher.  In Fig. \ref{fig:OIvsCont}, we show our Upper Scorpius sample in comparison to Taurus objects and the trend found by Howard et al.  Two of our sources are detected in [OI] 63 $\mu$m line emission.  One of these, J1614-1906, lies above the nonjet locus, and may potentially host a jet.  The other detected source, J1604-2130 lies in the nonjet regime, as may the rest of the sample which is undetected in [OI] line emission.

 \begin{figure}
\vspace{2cm}
  \includegraphics[angle=0,width=1.0\columnwidth]{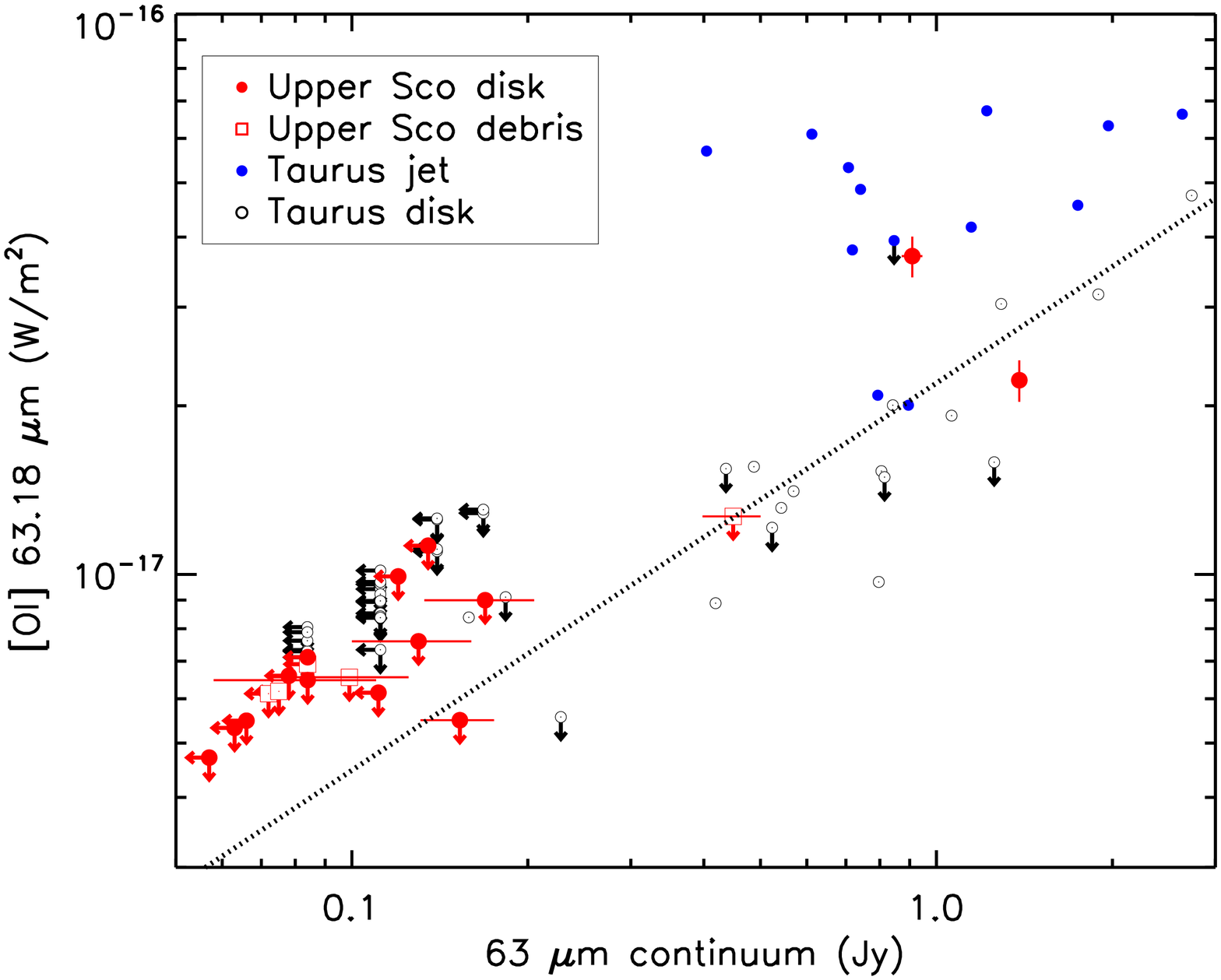}
  \caption{Comparison of [OI]63$\mu$m and 63 $\mu$m continuum fluxes to those of jet and nonjet sources in Taurus.  The empirical relationship discovered by Howard et al. (in review) is shown with a dotted line and provides a useful diagnostic.  From its placement among Taurus jet sources, we tentatively identify [OI] emission from J1614-1906 as originating in a jet.  Similarly, we identify J1604-2130 as showing emission from a disk.}
  \label{fig:OIvsCont}
\end{figure}

\subsection{Gas mass}

We may use recently developed gas mass diagnostics to make initial estimates of disk gas mass in the Upper Scorpius sample.  \cite{Kamp:2011a} outline the use of the [OI] 63 $\mu$m emission line and millimeter CO line emission to constrain the disk gas mass based on identification of empirical relations seen in the DENT grid of models \citep{2010MNRAS.405L..26W}.  In that work, they discuss the use of the [OI] 63 / CO 2-1 ratio to constrain the mean gas temperature, which then allows the direct conversion of the observed [OI] emission to the gas mass.  They point out that this works for disks with gas masses less than about 1 \Mjup ~in which the [OI] 63 $\mu$m line is optically thin.  Furthermore, it may improve at even lower gas masses at which the CO $J$=2-1 line becomes optically thin ($\sim$0.1 \Mjup).  With the detection of both lines, the gas mass can be estimated to within an order of magnitude, based on comparison of estimated gas masses to input model gas masses from the DENT grid.  They note that this large uncertainty is likely due to many other variables affecting the emission of the two lines, including the disk outer radius, disk flaring, and the amount of UV radiation.

In the case of Upper Scorpius, however, where few stars are detected in either CO or [OI] emission, the DENT grid still allows for limits to be placed on the disk gas mass.  We calculate the contours of mean gas mass as a function of [OI] 63 $\mu$m and CO J=2-1 flux for stars with masses of 0.5 and 1.0 solar mass.  In selecting models from the DENT grid, we applied the additional constraints of including stars older than 3 Myr, with dust masses $\leq0.1 M_{Jup}$, and observed at inclinations $\leq$ 80\degr, in order to avoid edge on disk models where even low mass disks can reach large column densities.  All fluxes were scaled to a distance of 145 pc.  For each of our stars, we then compare their flux or 3$\sigma$ upper limits to the grid of gas masses (Figure \ref{fig:OIvsCO}), using bilinear interpolation to determine the values.  The grid of mean gas masses has a typical uncertainty of at least 0.5 dex, suggesting a factor of at least $\sim3$ uncertainty in the estimated gas masses and mass limits.  Of the 8 K\&M stars observed for both [OI] 63 $\mu$m and CO J=2-1 emission, only J1604-2130 is detected in both lines.  It has an estimated gas mass of 2\Mjup, consistent with the mass of $\sim1$\Mjup ~found by comparison to figure 24 in \cite{Kamp:2011a}.  The other 7 sources have upper limits to the gas mass of 0.4 to 1.8 \Mjup, with a median upper limit of 1.1 \Mjup.  We report gas mass estimates and upper limits in Table \ref{tab:gasmass}.

\begin{figure}
\vspace{2cm}
  \includegraphics[angle=0,width=0.95\columnwidth]{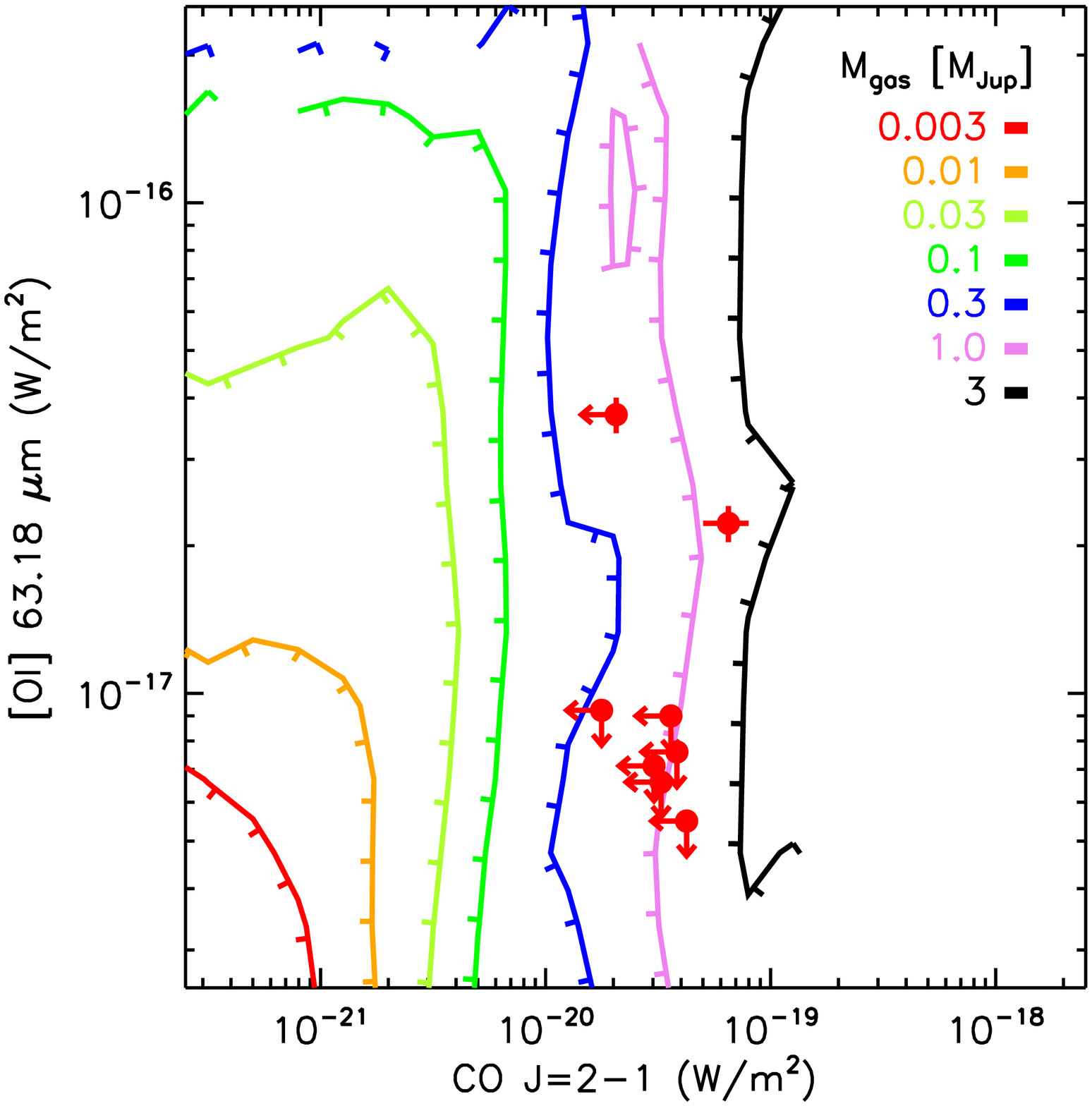}

  \caption{Comparison of [OI] 63$\mu$m and CO 2-1 fluxes to mean gas masses from the DENT grid.  Observed values are shown with red circles, with 3$\sigma$ upper limits indicated with arrows.  Another order of magnitude sensitivity to CO will be necessary to distinguish most gas masses.}
  \label{fig:OIvsCO}
\end{figure}

\begin{table}
\caption{  \label{tab:gasmass}  Gas mass estimates}
\begin{tabular}{l c }

\hline\hline
  Name &      $M_{gas}$        \\
              &      M$_{Jup}$        \\ 		           
\hline

ScoPMS 31                                &            $<1.3$	\\
     $[$PBB2002$]$ J155829.8-231007     &      $<0.9$        	\\
     $[$PBB2002$]$ J160545.4-202308     &       $<1.1$  	\\
     $[$PBB2002$]$ J160823.2-193001     &      $<0.4$         \\
     $[$PBB2002$]$ J160959.4-180009     &        $<1.8$     	\\
         $[$PBB2002$]$ J161420.3-190648 &       $<0.5$      	\\
     $[$PZ99$]$ J160357.6-203105             &           $<1.1$  	\\
     $[$PZ99$]$ J160421.7-213028             &          2.1     	\\
\hline
\end{tabular}
\end{table}

Examination of the mean gas mass contours in Figure \ref{fig:OIvsCO} indicates that the gas mass estimates are limited in this regime by our CO line sensitivity.  The large collection area of new telescopes, such as ALMA, will make it possible to probe deeper at millimeter wavelengths.  With another order of magnitude of sensitivity to CO emission, the upper limits on [OI] emission from these sources will still allow for a factor of ten improvement in gas mass estimates.  

\subsection{Gas-to-dust ratio}

To attempt to place a limit on the gas to dust ratio of the typical Upper Scorpius disk, we use the thermochemical modeling code ProDiMo \citep{Woitke:2009,2010A&A...510A..18K} to predict [OI], [CII], and CO line emission for the median SED disk for gas-to-dust ratios of 100, 10, and 1.  These values represent ratios ranging from that of ISM material and the presumed ratio in the youngest disks, down to the regime where disks may be considered to be transitioning to debris disks.  We assume the disks have well mixed gas and dust, and adopt the disk geometry found by SED fitting (Sect. \ref{sec:MedianSED}).

High energy UV photons play an important role in heating gas in the disk atmosphere, but the ultraviolet environment around these stars remains unknown.  To compensate, we explored two UV emission scenarios, with a fractional luminosity of 0.05 and 0.1, and adjusting the spectral slope so as the smoothly intersect the photospheric emission.  In neither case, however, was line emission strong enough to be detected at the sensitivity of our observations.  Hence, the gas-to-dust ratio of the typical Upper Scorpius disk is not directly constrained by our observations.  

For comparison, we carried out similar modeling for the Taurus median SED, finding that only in the case of a 100:1 gas to dust ratio and a high UV luminosity did the typical disk have [OI] 63 $\mu$m emission bright enough to be detected.  All other lines remain below our detection limits.  We present the results of the gas line models in Tables \ref{tab:gasLineUpSco} and \ref{tab:gasLineTaurus}.  A high fraction of Class II sources in Taurus exhibit [OI] 63 $\mu$m line emission (C. Howard et al., in review), suggesting that many young disks do have a gas-to-dust ratio near 100 or host jets.

\begin{table*}
\caption{  \label{tab:gasLineUpSco}  Upper Sco median predicted gas line strength by gas-to-dust ratio}
\begin{tabular}{l l  c c c  c c c}

\hline\hline
  		&							&       \multicolumn{3}{ c }{f$_{UV}$ = 0.05, UV slope = 2.2\tablefootmark{a}}             			&       \multicolumn{3}{ c }{f$_{UV}$ = 0.1, UV slope = 1.6\tablefootmark{a}}                \\
		&		Gas-to-dust ratio:		&		1:1              &       10:1               &       100:1                &      1:1               &      10:1               &      100:1  \\
  Line &       Wavelength ($\mu$m)           &	 $W/m^2$	  &	 $W/m^2$	  &	 $W/m^2$	  &	 $W/m^2$	  &	 $W/m^2$	  &	 $W/m^2$	 \\
  
\hline
  $[$OI$]$		& 	63.184				&	$2.55\times10^{-19}$ &   $6.96\times10^{-19}$&   $1.83\times10^{-18}$		&   $4.84\times10^{-19}$&   $1.52\times10^{-18}$&   $3.76\times10^{-18}$   \\
  $[$OI$]$		&	145.525				&	$7.37\times10^{-21}$&   $1.31\times10^{-20}$&   $4.04\times10^{-20}$		&   $1.42\times10^{-20}$&   $3.17\times10^{-20}$&   $9.93\times10^{-20}$   \\
  $[$CII$]$			&	157.741			&	$4.50\times10^{-20}$&   $9.77\times10^{-20}$&   $2.28\times10^{-19}$		&   $6.81\times10^{-20}$&   $1.63\times10^{-19}$&   $3.29\times10^{-19}$   \\
  CO 3-2		&	866.963				&	$5.53\times10^{-21}$&   $1.02\times10^{-20}$&   $1.44\times10^{-20}$		&   $5.58\times10^{-21}$&   $1.11\times10^{-20}$&   $1.34\times10^{-20}$   \\  
  CO 2-1		&	1300.404				&	$1.75\times10^{-21}$&   $3.31\times10^{-21}$&   $4.50\times10^{-21}$		&   $1.68\times10^{-21}$&   $3.56\times10^{-21}$&   $4.25\times10^{-21}$   \\
\hline

\end{tabular}
\tablefoottext{a}{The fractional UV luminosity is indicated by f$_{UV}$, and the UV slope indicates the spectral slope of the UV excess.  The UV slope is set to achieve a smooth intersection with the SED of the stellar photosphere.}
\end{table*}

\begin{table*}
\caption{  \label{tab:gasLineTaurus} Taurus median predicted gas line strength by gas-to-dust ratio}
\begin{tabular}{l l  c c c  c c c}

\hline\hline
  		&							&       \multicolumn{3}{ c }{f$_{UV}$ = 0.05, UV slope = 2.2\tablefootmark{a}}             			&       \multicolumn{3}{ c }{f$_{UV}$ = 0.1, UV slope = 1.6\tablefootmark{a}}                \\
		&		Gas-to-dust ratio:		&		1:1               &       10:1               &       100:1               &      1:1               &      10:1               &      100:1  \\
  Line &       Wavelength ($\mu$m)           &	 $W/m^2$	  &	 $W/m^2$	  &	 $W/m^2$	  &	 $W/m^2$	  &	 $W/m^2$	  &	 $W/m^2$	\\

  $[$OI$]$		& 	63.184				&	  $6.45\times10^{-19}$&   $1.49\times10^{-18}$&   $4.30\times10^{-18}$ &	  $1.63\times10^{-18}$&   $3.26\times10^{-18}$&   $7.67\times10^{-18}$   \\
  $[$OI$]$		&	145.525				&	$1.17\times10^{-20}$&   $2.01\times10^{-20}$&   $7.23\times10^{-20}$   	&	  $3.65\times10^{-20}$&   $5.13\times10^{-20}$&   $1.33\times10^{-19}$      \\
  $[$CII$]$			&	157.741			&       $5.14\times10^{-20}$&   $1.26\times10^{-19}$&   $3.10\times10^{-19}$   	&	  $8.17\times10^{-20}$&   $2.14\times10^{-19}$&   $5.35\times10^{-19}$      \\
  CO 3-2		&	866.963				&	$6.67\times10^{-21}$&   $1.30\times10^{-20}$&   $1.06\times10^{-20}$   	&	  $6.82\times10^{-21}$&   $1.13\times10^{-20}$&   $1.57\times10^{-20}$      \\  
  CO 2-1		&	1300.404				&	$2.23\times10^{-21}$&   $3.86\times10^{-21}$&   $3.68\times10^{-21}$   	&	  $2.29\times10^{-21}$&   $3.40\times10^{-21}$&   $5.55\times10^{-21}$      \\
  
  \hline

\end{tabular}
\tablefoottext{a}{The fractional UV luminosity is indicated by f$_{UV}$, and the UV slope indicates the spectral slope of the UV excess.  The UV slope is set to achieve a smooth intersection with the SED of the stellar photosphere.}
\end{table*}

Two sources in Upper Scorpius have gas mass upper limit estimates and millimeter determinations of the dust mass (objects 34 and J1614-1906).  In both cases, the low gas mass estimates of $\lesssim 0.5$ \Mjup ~indicate gas-to-dust ratios much less than the primordial value of 100.  With mm-dust masses of 0.041 and 0.014 \Mjup \citep{Mathews:2011}, respectively, these disks have gas-to-dust ratios of $\lesssim10$ and $\lesssim30$.  

The one source detected in both [OI] 63 $\mu$m and CO emission, J1604-2130, has an estimated gas mass of 2 $M_{Jup}$.  In previous work we found this disk has a dust mass of 0.1 $M_{Jup}$ \citep{Mathews:2011,Mathews:2012a}, suggesting a gas to dust ratio of $\sim20$.  However, this disk is an unusual one in the Upper Scorpius sample, a transition disk with a $\sim$70 AU hole in millimeter emission and smaller dust grains within the hole \citep{Mathews:2012a, 2012ApJ...760L..26M}.  It likely represents the tail end of the disk survival function, and the unique dust configuration will likely lead to different excitation conditions than assumed in the DENT grid, raising the uncertainty in the gas mass estimation.  However, as this is the sole source for which we can determine a gas-to-dust ratio, it may be reasonable to treat it as an indicator of the upper limit to the gas-to-dust ratio in Upper Scorpius.  

Combined with our previous finding that the highest dust masses have decreased by a factor $\sim$20 from the age of Taurus to that of Upper Scorpius and the extension of that conclusion to the general disk population via our modeling here, this suggests an additional factor of $\sim$5 decrease in gas masses.  Several mechanisms could play a role in preferentially preserving dust compared to gas.  As dust grows beyond micron sizes or settles to the midplane, it  becomes decoupled from the gas.  There is a net transfer of material from the gas phase to solids by freeze-out in the disk interior, though some of this material may return to the gas phase through later heating.  In addition, detectable dust is likely replenished by collisions among larger objects which are undetectable due to low opacity.  Future observations with instruments offering high sensitivity and resolution at cm wavelengths, such as the eVLA and Square-Kilometer Array, will provide a crucial test of replenishment models, by directly probing the population of pebbles.  

Our finding of a low gas to dust ratio, in combination with the typically geometrically thin structure, suggests the importance of dust settling for allowing gas depletion to proceed more rapidly than depletion of dust.  

\subsection{Planet formation}

Recent decades have seen the discovery of hundreds of gas giant extrasolar planets, and the \textit{Kepler} mission is leading to an explosion in the number of candidates.  \cite{Borucki:2011} estimate approximately 20\% of stars host giant planets, and recent observations have nearly doubled the number of candidates, allowing for further refinement of this estimate \citep{Batalha:2013qf}.  However, far less than 20\% of stars in Upper Scorpius have sufficient disk gas mass to form a giant planet if that process were to begin today.  In \cite{Mathews:2011}, we argued that if disks retained the 100 to 1 gas to dust ratio, the 4 stars with dust masses of 0.01 \Mjup ~or greater could have at least 1 \Mjup ~of gas.  Assuming a 10--20\% planet formation efficiency for a Jupiter composition planet, based on the disk mass required to form Jupiter in the minimum mass solar nebula \citep{Weidenschilling:1977a,Desch:2007}, such a disk may be sufficient to form a gas giant planet with a mass ranging from 2 Neptune masses to approximately a Saturn mass.  

Furthermore, as our sample was built using all disk bearing sources from the statistically complete parent sample of 218 stars from \cite{2006ApJ...651L..49C}, the potential gas giant planet forming disk fraction ($f_{Nep}$) is approximately 1.8$\pm$0.9\% (4/218).  If $f_{Nep}$ decays as an exponential in time, and assuming that all stars form with a potentially giant planet forming disk of 1 \Mjup ~or greater, then Upper Scorpius has experienced $5.8\pm0.7$ half-lives.  The 5 Myr age for Upper Scorpius \citep[PZ99]{Blaauw:1978} would imply that giant planet formation must be a rapid process, with a half-life for potentially planet forming disks of $0.9\pm0.1$ Myr.  That in turn would imply that giant planet formation must generally begin within $\sim2.0$ Myr of formation, as that is when approximately 80\% of disks would drop below 1 \Mjup.  However,  if Upper Scorpius is 11 Myr old \citep{Pecaut:2012}, the half-life of the $f_{Nep}$ would be 1.9 Myr, suggesting that giant planet formation could be correspondingly slower.  

This implied 2--4 Myr limit to the time available for giant planet formation lies at the low to mid-range of ages needed to form giant planets via core accretion models \citep{Pollack:1996,Alibert:2005}.  However, if the gas-to-dust ratio of 20 of J1604-2130 represents the upper limit among Upper Scorpius disks, then this is the only star which could support further giant planet formation.  In either case, the fraction of stars hosting potentially giant planet forming disks is lower than the gas giant fraction in the field by at least a factor of 10, suggesting that gas giant planet formation is essentially finished in Upper Scorpius.  

Regarding the formation of small, rocky planets, the median dust mass in Upper Sco, $3.5\times10^{-6}$ \Msun, corresponds to $\sim$1 \Mearth.  The rocky planets of the solar system only require about a factor of 3 enrichment from primordial solids \citep{Weidenschilling:1977a}, suggesting that a dust mass of $\sim10^{-5}$\Msun ~would be the minimum dust mass necessary to form a 1 \Mearth ~planet.  Including J1604-2130 and J1614-1906, which we did not model here but which have dust masses determined by millimeter photometry, 5 / 24 disk bearing K\&M stars in Upper Scorpius have this minimum dust mass for the formation of earth mass planets.  These represent $2.3\pm1.0$\% of the stars in Upper Scorpius.  In the field, $\sim$15\% of stars host 0.5 -- 2 $R_{\oplus}$ planets \citep{Borucki:2011}.  Assuming these planets have similar densities to Earth, the fraction of Upper Sco stars which could form an Earth-mass planet if the formation process was to begin now is $\sim$0.33 that of the fraction of Earth-mass planet hosts in the field.   This suggests that, much like giant planet formation, the processes that lead to Earth-mass planet formation are well underway in Upper Sco, and much of the dust has already grown to large sizes that are unobservable without longer wavelength observations.

\subsection{Notes on selected sources}
\label{sec:notes}

\label{sec:J1604} 
\textbf{J1604-2130:}  The outer disk structure of the K2 star J1604-2130 was described in \cite{Mathews:2012a}, based on sub-millimeter interferometry.  The disk has a dust mass $\sim0.1$ \Mjup ~and exhibits a 70 AU hole in millimeter continuum emission, but rising 16 and 24 $\mu$m excesses imply the presence of small amounts of dust as close as 20 AU.  Furthermore, the system has exhibited a factor of $\sim$4 variation in emission at shorter wavelengths \citep{2009AJ....137.4024D}, with NIR spectroscopy indicating the presence of 900 K dust (corresponding to $\sim 0.1$ AU).  After merging WISE photometry and \textit{Spitzer} spectroscopy with the previously studied SED, we find the presence of an excess at all wavelengths from 3.4 $\mu$m to 16 $\mu$m (the wavelength at which an excess was first apparent in previous work).  The sharply rising excess from 16 $\mu$m is still apparent and is consistent with the presence of dust at 20 AU, and does not affect the gap in millimeter emission required by SMA observations.  This disk structure has since been broadly verified by high resolution H-band imaging of the disk by \cite{2012ApJ...760L..26M}, who found small grain dust features extending inside the mm-disk inner edge.  

This complex dust disk structure -- a lack of millimeter-sized grains at radii less than 70 AU, yet the presence of dust near the sublimation radius and locally within the cavity -- can only be produced by dynamical interaction.  J1604-2130 will provide a test bed for examining interactions between giant planets and disks, as well as for the complex modeling techniques needed to simultaneously fit these multiple disk regions.

\label{sec:J1614} 
\textbf{J1614-1906:}  The second source detected in our line survey also has the fourth largest dust mass in Upper Sco as determined by millimeter photometry \citep{Mathews:2011}, $\sim 0.015 M_{Jup}$.  However, the large extinction suggests the star is being viewed edge on through a disk, and large NIR excess suggests we may be directly viewing the disk inner edge.  In addition, the high ratio of the [OI] 63 $\mu$m line to 63 $\mu$m continuum indicates this system may be driving a jet. 
This star has the second highest H$\alpha$ equivalent width in Upper Sco, -52\AA, which suggests an accretion rate $\sim10^{-9}$ \Msun/yr using the conversion presented in \cite{Dahm:2008}.  In addition, strong emission has been detected in the [OI] 6300 \AA ~forbidden line (Montesinos, in prep), a common indicator of jets \citep[e.g.][]{Cabrit:1990}.  We defer modeling of this system to a later paper.

\label{sec:binaries}

\textbf{Binaries:}  ScoPMS 31 and $[$PZ99$]$ J161411.0-230536 (objects 13 and 45) have binary companions at projected separations of 84 and 32 AU, respectively \citep{Kohler:2000,Metchev:2009}.  These sources are are also detected in millimeter continuum and exhibit Class II infrared excesses.  Mid-infrared excesses indicate, in both cases, the presence of dust at radii of $\sim1$ AU or less.  The disks will necessarily be truncated at outer radii of a few tens of AU by dynamical interactions with the companions.  While the millimeter emission could originate from these disks or from circumbinary material, there are no abrupt changes in the SED shape indicating gaps in emission, as may be seen in a combination circumstellar / circumbinary emission system.  High sensitivity, high spatial resolution millimeter imaging, such as may be accomplished with ALMA, will allow for the measurement of the disk outer radii and determination of the presence of circumbinary disks.

\section{Summary}
\label{sec:summary}

We have carried out a survey for far-infrared emission lines and photometry, as well as millimeter line emission, among the stars of Upper Scorpius.  From these observations, we have found that:

\begin{enumerate}
\item  17 of 23 K and M stars with 16 or 24 $\mu$m excess exhibit continuum emission at 70, 100, and / or 160 $\mu$m.
\item  The median of Upper Scorpius emission represents an early M star bearing a geometrically thin (H(1 AU) = 0.035 AU), depleted (M$_{dust} \sim 3\times10^{-6}$ \Msun ) disk with an inner edge near the sublimation radius.  The typical Upper Scorpius disk has only half the scale height and much lower dust mass compared to the typical disk in Taurus.
\item  No B or A stars show evidence for hosting circumstellar gas, and their SEDs are all consistent with either having no disk or a debris disk.  
\item  One K star, $[$PBB2002$]$ J161420.3-190648, has far-infrared line emission consistent with origin from a jet.
\item  The K star $[$PZ99$]$ J160421.7-213028 has a disk gas mass of $\sim$2 $M_{Jup}$.  All 6 other stars with both [OI] and CO observations have line upper limits consistent with gas masses less than 1 \Mjup.  Combined with dust mass measurements, these imply an upper limit gas-to-dust ratio of 0.2 among the remaining Upper Scorpius disks. 
\end{enumerate}

In general, the stars of Upper Scorpius lack sufficient material for forming giant or earth-size planets if this formation process were to begin now.  By the age of Upper Scorpius, giant planets must have completed formation, as must the precursor bodies of Earth-size planets.  

\begin{acknowledgements} 
This work is supported by the EU A-ERC grant 291141 CHEMPLAN.
C. Pinte acknowledges funding from the European Commission's 7$^\mathrm{th}$ Framework Program (contract PERG06-GA-2009-256513) and from Agence Nationale pour la Recherche (ANR) of France under contract ANR-2010-JCJC-0504-01. 
We acknowledge the Service Commun de Calcul Intensif de l'Observatoire de Grenoble (SCCI) for computations on the super-computer funded by ANR (contracts ANR-07-BLAN-0221, ANR- 2010-JCJC-0504-01 and ANR-2010-JCJC-0501-01). Financial support was also provided by the Milenium Nucleus P10-022-F from the Chilean Ministry of Economy and from the European Community 7th Framework programme (Grant Agreement 284405).
\end{acknowledgements}

\bibliographystyle{aa}

\clearpage
\newpage

\Online
\begin{appendix}

\section{Observation list}


\begin{longtable}{l c l r}

\caption{\label{tab:obs}Observations}  \\

\hline\hline
Name &  Date / ObsID  &  Setting / Instrument  &  Obs. Time (s) \\
\hline
\endfirsthead

\caption{Continued.} \\
\hline
Name &  Date / ObsID  &  Setting / Instrument  &  Obs. Time (s) \\
\hline
\endhead

\hline
\endfoot

\hline
\endlastfoot

HIP 76310				&      	1342215621	      &     	PacsPhoto, green	     &     	276	      \\
	    				    &      	1342215622	      &     	PacsPhoto, green	     &     	276	      \\
		    			    &      	1342191303	      &     	PacsLineSpec	     &     	1252	      \\
					    &		2009-10-26		&	JCMT RxA		&	1200		\\
						&	2011-02-18		&	JCMT RxA		&	1200	 	\\
						&	2011-06-16		&	JCMT RxA		&	1200	 	\\
HIP77815	 			       &      	1342189658	      &     	PacsPhoto	     &     	220	      \\   	
	    				    &      	1342215474	      &     	PacsPhoto, blue	     &     	276	      \\   	
	       				 &      	1342215475	      &     	PacsPhoto, blue	     &     	276	      \\   	
					 &		2008-02-22		&	JCMT HARP	&		600		\\
						&	2011-04-15		&	JCMT RxA		&	1200	 	\\
HIP77911	   			     &      	1342189656	      &     	PacsPhoto	     &     	220	      \\   	
	       				 &      	1342214223	      &     	PacsLineSpec	     &     	3316	      \\   	
	       				 &      	1342215480	      &     	PacsPhoto, green	     &     	276	      \\   
	        				&    	  	1342215481	      &     	PacsPhoto, green	     &     	276	      \\   
					&		2008-02-22		&	JCMT HARP		&	600		\\
					&		2008-07-09		&	JCMT HARP		&	600		\\
HIP 78099			        &      	1342215486	      &     	PacsPhoto, blue	     &     	276	      \\   
	    				    &      	1342215487	      &     	PacsPhoto, blue	     &     	276	      \\   
HIP 78996	        			&      	1342215502	      &     	PacsPhoto, blue	     &     	558	      \\   
						&      	1342215503	      &     	PacsPhoto, blue	     &     	558	      \\   
HIP 79156	       			&      	1342215414	      &     	PacsPhoto, blue	     &     	558	      \\   
						&      	1342215415	      &     	PacsPhoto, blue	     &     	558	      \\   
HIP 79410			        &      	1342215404	      &     	PacsPhoto, blue	     &     	276	      \\   
					        &      	1342215405	      &     	PacsPhoto, blue	     &     	276	      \\   
HIP 79439			        &      	1342215402	      &     	PacsPhoto, blue	     &     	558	      \\   
					        &      	1342215403	      &     	PacsPhoto, blue	     &     	558	      \\   
					        &      	1342216190	      &     	PacsLineSpec	     &     	3316	      \\   
HIP 79878			        &      	1342215615	      &     	PacsPhoto, green	     &     	276	      \\   
					        &      	1342215616	      &     	PacsPhoto, green	     &     	276	      \\   
					        &      	1342216170	      &     	PacsLineSpec	     &     	3316	      \\   
HIP 80088			        &      	1342215514	      &     	PacsPhoto, green	     &     	276	      \\   
					        &      	1342215515	      &     	PacsPhoto, green	     &     	276	      \\   
					        &      	1342216168	      &     	PacsLineSpec	     &     	3316	      \\   
HIP 80130	   		     &      	1342215510	      &     	PacsPhoto, blue	     &     	276	      \\   
					        &      	1342215511	      &     	PacsPhoto, blue	     &     	276	      \\   
					        &      	1342215512	      &     	PacsPhoto, green	     &     	276	      \\   
					        &      	1342215513	      &     	PacsPhoto, green	     &     	276	      \\   
						&	2011-04-15		&	JCMT RxA		&	1200	 	\\
RX J1600.7-2343	        &      	1342215496	      &     	PacsPhoto, blue	     &     	840	      \\   
				        &      	1342215497	      &     	PacsPhoto, blue	     &     	840	      \\   
				        &      	1342213760	      &     	PacsLineSpec	     &     	1252	      \\   
ScoPMS 31	       			&      	1342215638	      &     	PacsRangeSpec	     &     	8316	      \\   
					        &      	1342216196	      &     	PacsLineSpec	     &     	3316	      \\   
					        &      	1342215422	      &     	PacsPhoto, green	     &     	276	      \\   
					        &      	1342215423	      &     	PacsPhoto, green	     &     	276	      \\   
							&	2011-02-13		&	JCMT RxA		&	1200	 	\\
$[$PBB2002$]$ J155624.8-222555	        &      	1342189657	      &     	PacsPhoto	     &     	220	      \\   
						        &      	1342215476	      &     	PacsPhoto, blue	     &     	276	      \\   
						        &      	1342215477	      &     	PacsPhoto, blue	     &     	276	      \\   
						        &      	1342215478	      &     	PacsPhoto, green	     &     	276	      \\   
						        &      	1342215479	      &     	PacsPhoto, green	     &     	276	      \\   
 $[$PBB2002$]$ J155706.4-220606	        &      	1342215466	      &     	PacsPhoto, blue	     &     	276	      \\   
	  				      &      	1342215467	      &     	PacsPhoto, blue	     &     	276	      \\   
					        &      	1342215468	      &     	PacsPhoto, green	     &     	276	      \\   
					        &      	1342215469	      &     	PacsPhoto, green	     &     	276	      \\   
						&	2008-02-22		&	JCMT HARP	&		600		\\
 $[$PBB2002$]$ J155729.9-225843	        &      	1342214222	      &     	PacsLineSpec	     &     	3316	      \\   
					        &      	1342215488	      &     	PacsPhoto, blue	     &     	276	      \\   
					        &      	1342215489	      &     	PacsPhoto, blue	     &     	276	      \\   
					        &      	1342215490	      &     	PacsPhoto, green	     &     	276	      \\   
					        &      	1342215491	      &     	PacsPhoto, green	     &     	276	      \\   
$[$PBB2002$]$ J155829.8-231007	        &      	1342203460	      &     	PacsLineSpec	     &     	1660	      \\   
					        &      	1342215492	      &     	PacsPhoto, blue	     &     	276	      \\   
					        &      	1342215493	      &     	PacsPhoto, blue	     &     	276	      \\   
					        &      	1342215494	      &     	PacsPhoto, green	     &     	276	      \\   
	        					&      	1342215495	      &     	PacsPhoto, green	     &     	276	      \\   
						&	2011-02-18		&	JCMT RxA		&	1200	 	\\
 $[$PBB2002$]$ J160210.9-200749	        &      	1342215434	      &     	PacsPhoto, blue	     &     	276	      \\   
									        &      	1342215435	      &     	PacsPhoto, blue	     &     	276	      \\   
$[$PBB2002$]$ J160245.4-193037	        &      	1342214580	      &     	PacsPhoto, blue	     &     	276	      \\   
									        &      	1342214581	      &     	PacsPhoto, blue	     &     	276	      \\   
										&	2008-02-22		&	JCMT HARP		&	600		\\
$[$PBB2002$]$ J160357.9-194210	        &      	1342214226	      &     	PacsLineSpec	     &     	3316	      \\   
									        &      	1342215432	      &     	PacsPhoto, blue	     &     	276	      \\   
									        &      	1342215433	      &     	PacsPhoto, blue	     &     	276	      \\   
										&	2008-07-09		&	JCMT HARP		&	600		\\
 $[$PBB2002$]$ J160525.5-203539 	        &      	1342215458	      &     	PacsPhoto, blue	     &     	276	      \\   
									        &      	1342215459	      &     	PacsPhoto, blue	     &     	276	      \\   
									        &      	1342215460	      &     	PacsPhoto, green	     &     	276	      \\   
									        &      	1342215461	      &     	PacsPhoto, green	     &     	276	      \\   
										&	2008-07-09		&	JCMT HARP		&	600		\\
 $[$PBB2002$]$ J160532.1-193315	        &      	1342215637	      &     	PacsLineSpec	     &     	1252	      \\   
									        &      	1342215424	      &     	PacsPhoto, blue	     &     	276	      \\   
									        &      	1342215425	      &     	PacsPhoto, blue	     &     	276	      \\   
									        &      	1342215426	      &     	PacsPhoto, green	     &     	276	      \\   
									        &      	1342215427	      &     	PacsPhoto, green	     &     	276	      \\   
$[$PBB2002$]$ J160545.4-202308 	        &      	1342216198	      &     	PacsRangeSpec	     &     	8316	      \\   
									        &      	1342216197	      &     	PacsLineSpec	     &     	3316	      \\   
									        &      	1342215438	      &     	PacsPhoto, blue	     &     	276	      \\   
									        &      	1342215439	      &     	PacsPhoto, blue	     &     	276	      \\   
									        &      	1342215440	      &     	PacsPhoto, green	     &     	276	      \\   
									        &      	1342215441	      &     	PacsPhoto, green	     &     	276	      \\   
										&	2011-02-13		&	JCMT RxA		&	1200	 	\\
 $[$PBB2002$]$ J160600.6-195711	        &      	1342215736	      &     	PacsLineSpec	     &     	3316	      \\   
									        &      	1342215428	      &     	PacsPhoto, blue	     &     	276	      \\   
									        &      	1342215429	      &     	PacsPhoto, blue	     &     	276	      \\   
									        &      	1342215430	      &     	PacsPhoto, green	     &     	276	      \\   
									        &      	1342215431	      &     	PacsPhoto, green	     &     	276	      \\   
 $[$PBB2002$]$ J160622.8-201124	        &      	1342215442	      &     	PacsPhoto, blue	     &     	276	      \\   
										&      	1342215443	      &     	PacsPhoto, blue	     &     	276	      \\   
										&      	1342215444	      &     	PacsPhoto, green	     &     	276	      \\   
										&      	1342215445	      &     	PacsPhoto, green	     &     	276	      \\   
$[$PBB2002$]$ J160643.8-190805 	        &      	1342215410	      &     	PacsPhoto, blue	     &     	276	      \\   
										&      	1342215411	      &     	PacsPhoto, blue	     &     	276	      \\   
										&      	1342215412	      &     	PacsPhoto, green	     &     	276	      \\   
										&      	1342215413	      &     	PacsPhoto, green	     &     	276	      \\   
 $[$PBB2002$]$ J160702.1-201938	        &      	1342215448	      &     	PacsPhoto, green	     &     	276	      \\   
										&      	1342215449	      &     	PacsPhoto, green	     &     	276	      \\   
										&      	1342215446	      &     	PacsPhoto, blue	     &     	276	      \\   
										&      	1342215447	      &     	PacsPhoto, blue	     &     	276	      \\   
 $[$PBB2002$]$ J160801.4-202741	        &      	1342215450	      &     	PacsPhoto, blue	     &     	276	      \\   
										&      	1342215451	      &     	PacsPhoto, blue	     &     	276	      \\   
										&      	1342215452	      &     	PacsPhoto, green	     &     	276	      \\   
										&      	1342215453	      &     	PacsPhoto, green	     &     	276	      \\   
$[$PBB2002$]$ J160823.2-193001	        &      	1342215416	      &     	PacsPhoto, green	     &     	276	      \\   
										&      	1342215417	      &     	PacsPhoto, green	     &     	276	      \\   
										&      	1342216195	      &     	PacsRangeSpec	     &     	8316	      \\   
										&      	1342216193	      &     	PacsLineSpec	     &     	3316	      \\   
										&      	1342216194	      &     	PacsLineSpec	     &     	3316	      \\   
										&	2011-02-02		&	JCMT RxA		&	1200	 	\\
$[$PBB2002$]$ J160827.5-194904	        &      	1342215420	      &     	PacsPhoto	     &     	276	      \\   
										&      	1342215418	      &     	PacsPhoto, blue	     &     	276	      \\   
										&      	1342215419	      &     	PacsPhoto, blue	     &     	276	      \\   
										&      	1342215421	      &     	PacsPhoto, green	     &     	276	      \\   
    $[$PBB2002$]$ J160900.0-190836	        &      	1342215406	      &     	PacsPhoto, blue	     &     	276	      \\   
 \& $[$PBB2002$]$ J160900.7-190852	        &      	1342215407	      &     	PacsPhoto, blue	     &     	276	      \\   
										&      	1342215408	      &     	PacsPhoto, green	     &     	276	      \\   
										&      	1342215409	      &     	PacsPhoto, green	     &     	276	      \\   
$[$PBB2002$]$ J160900.0-190836		&	2011-02-25		&	JCMT RxA		&	1200	 	\\
 $[$PBB2002$]$ J160953.6-175446	        &      	1342215394	      &     	PacsPhoto, blue	     &     	276	      \\   
										&      	1342215395	      &     	PacsPhoto, blue	     &     	276	      \\   
										&      	1342215396	      &     	PacsPhoto, green	     &     	276	      \\   
										&      	1342215397	      &     	PacsPhoto, green	     &     	276	      \\   
 $[$PBB2002$]$ J160959.4-180009 	        &      	1342215398	      &     	PacsPhoto, green	     &     	276	      \\   
										&      	1342215399	      &     	PacsPhoto, green	     &     	276	      \\   
										&      	1342216188	      &     	PacsLineSpec	     &     	3316	      \\   
										&      	1342216189	      &     	PacsRangeSpec	     &     	8316	      \\   
										&	2011-02-25		&	JCMT RxA		&	1200	 	\\
$[$PBB2002$]$ J161115.3-175721	        &      	1342215393	      &     	PacsPhoto, green	     &     	276	      \\   
										&      	1342215390	      &     	PacsPhoto, blue	     &     	276	      \\   
										&      	1342215391	      &     	PacsPhoto, blue	     &     	276	      \\   
										&      	1342215392	      &     	PacsPhoto, green	     &     	276	      \\   
$[$PBB2002$]$ J161420.3-190648	        &      	1342215400	      &     	PacsPhoto, green	     &     	276	      \\   
										&      	1342215401	      &     	PacsPhoto, green	     &     	276	      \\   
										&      	1342216191	      &     	PacsLineSpec	     &     	3316	      \\   
										&      	1342216192	      &     	PacsRangeSpec	     &     	8316	      \\   
										&	2011-02-02		&	JCMT RxA		&	1200	 	\\
    $[$PZ99$]$ J153557.8-232405      &      1342189592	      &     	PacsPhoto	     &     	220	      \\
						       &      	1342215623	      &     	PacsPhoto, blue	     &     	276	      \\
						       &      	1342215624	      &     	PacsPhoto, blue	     &     	276	      \\
						        &      	1342215625	      &     	PacsPhoto, green	     &     	276	      \\
						       &      	1342215626	      &     	PacsPhoto, green	     &     	276	      \\
						       &	2008-02-22		&	JCMT HARP			&	600		\\
							&	2011-02-13		&	JCMT RxA		&	1200	 	\\
 $[$PZ99$]$ J154413.4-252258       &      	1342189659	      &     	PacsPhoto	     &     	220	      \\
	 					       &      	1342215482	      &     	PacsPhoto, blue	     &     	276	      \\
	   					     &      	1342215483	      &     	PacsPhoto, blue	     &     	276	      \\
	     					     &      	1342215484	      &     	PacsPhoto, green	     &     	276	      \\
	   					     &      	1342215485	      &     	PacsPhoto, green	     &     	276	      \\
							&	2008-02-22		&	JCMT HARP			&	600		\\
 $[$PZ99$]$ J160108.0-211318 	        &      	1342215464	      &     	PacsPhoto, blue	     &     	276	      \\   
							        &      	1342215465	      &     	PacsPhoto, blue	     &     	276	      \\   
								&	2008-02-22		&	JCMT HARP		&	600		\\
$[$PZ99$]$ J160357.6-203105	        &      	1342214227	      &     	PacsLineSpec	     &     	3316	      \\   
							        &      	1342216199	      &     	PacsRangeSpec	     &     	8316	      \\   
							        &      	1342215436	      &     	PacsPhoto, green	     &     	276	      \\   
							        &      	1342215437	      &     	PacsPhoto, green	     &     	276	      \\   
								&	2011-02-26		&	JCMT RxA		&	1200	 	\\
$[$PZ99$]$ J160421.7-213028	        &      	1342215462	      &     	PacsPhoto, green	     &     	276	      \\   
							        &      	1342215463	      &     	PacsPhoto, green	     &     	276	      \\   
							        &      	1342215680	      &     	PacsRangeSpec	     &     	10279	      \\   
							        &      	1342215681	      &     	PacsLineSpec	     &     	3316	      \\   
							        &        2008-03-09             &          JCMT HARP	     &           600        \\                                   
							        &        2010-05-18             &          JCMT HARP	     &           1200        \\                                   
							        &        2010-05-19             &          JCMT HARP	     &           500        \\                                   
							        &        2010-05-20             &          JCMT HARP	     &           3000        \\                                   
							        &	2009-07-27	      &          JCMT RxA              &           1200      \\
 $[$PZ99$]$ J160654.4-241610   	     &      	1342215501	      &     	PacsPhoto, green	     &     	276	      \\   
								&      	1342215498	      &     	PacsPhoto, blue	     &     	276	      \\   
								&      	1342215499	      &     	PacsPhoto, blue	     &     	276	      \\   
								&      	1342215500	      &     	PacsPhoto, green	     &     	276	      \\   
 $[$PZ99$]$ J160856.7-203346	        &      	1342215454	      &     	PacsPhoto, blue	     &     	276	      \\   
								&      	1342215455	      &     	PacsPhoto, blue	     &     	276	      \\   
								&      	1342215456	      &     	PacsPhoto, green	     &     	276	      \\   
								&      	1342215457	      &     	PacsPhoto, green	     &     	276	      \\   
$[$PZ99$]$ J161402.1-230101	        &      	1342215506	      &     	PacsPhoto, blue	     &     	276	      \\   
								&      	1342215507	      &     	PacsPhoto, blue	     &     	276	      \\   
								&      	1342215508	      &     	PacsPhoto, green	     &     	276	      \\   
								&      	1342215509	      &     	PacsPhoto, green	     &     	276	      \\   
$[$PZ99$]$ J161411.0-230536	        &      	1342215504	      &     	PacsPhoto, green	     &     	276	      \\   
								&      	1342215505	      &     	PacsPhoto, green	     &     	276	      \\   
								&      	1342216169	      &     	PacsLineSpec	     &     	3316	      \\   
								&	2011-02-03		&	JCMT RxA		&	1200	 	\\
								&	2011-06-27		&	JCMT RxA		&	1200	 	\\
  
\end{longtable}

\clearpage
\newpage

\section{[CII] maps}
\label{sec:CII-maps}

\begin{figure*}[]
\includegraphics[angle=0,width=0.95\textwidth]{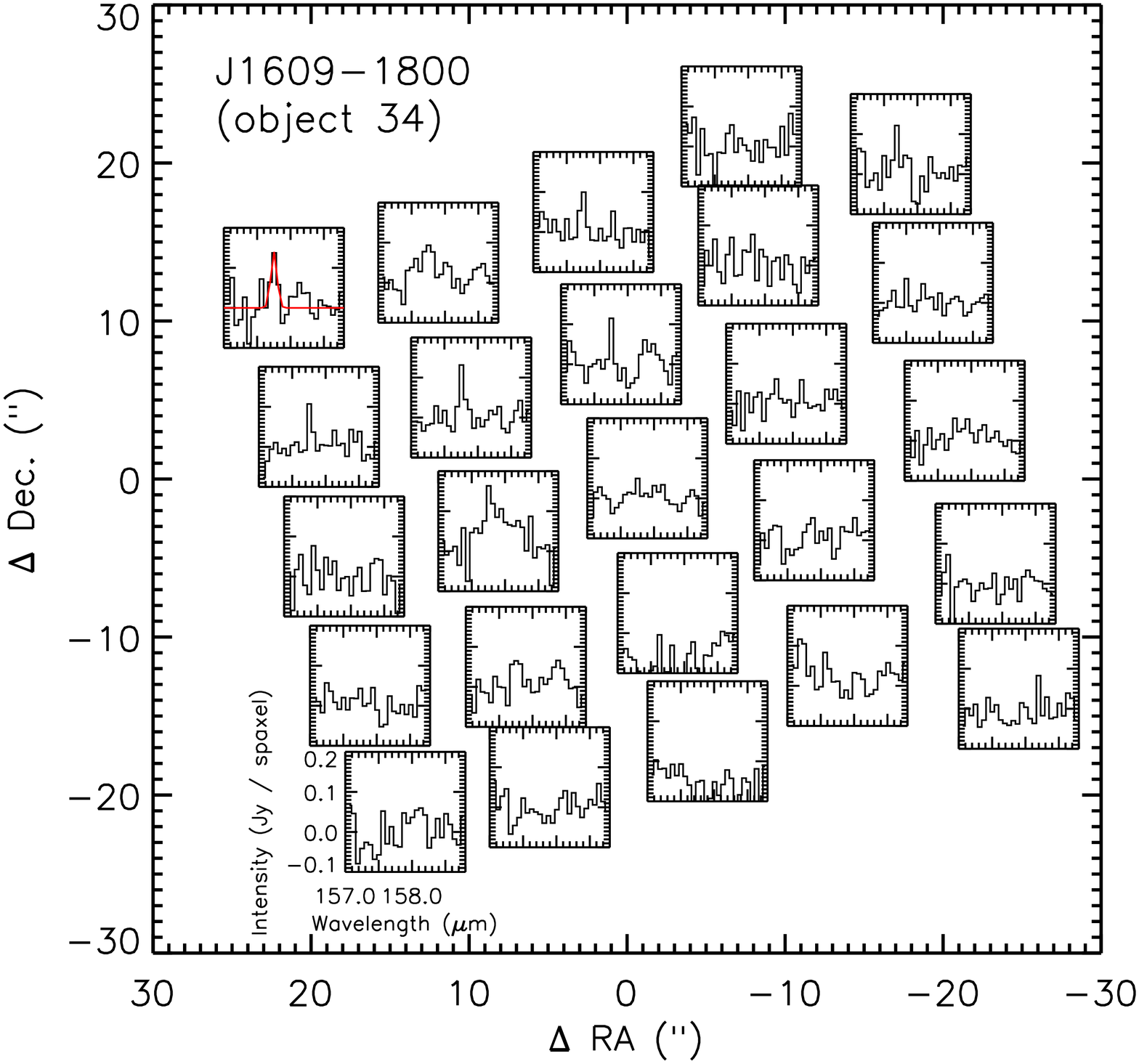}
  \caption{Footprint continuum-subtracted spectra of [CII] 157.7 $\mu$m emission toward object 34, without aperture correction.  Each box is centered at the reported position of the spaxel for the observation.  Red lines show Gaussian fits to spectra which are detected at the 3$\sigma$ level or higher (typical limit is $\sim2.5\times10^{-18}$ W m$^{-2}$ per spaxel).  Additional spaxels may have emission at $\sim2\sigma$ level.  }
  \label{fig:CII-J1609}
\end{figure*}

\begin{figure*}[]
\includegraphics[angle=0,width=0.95\textwidth]{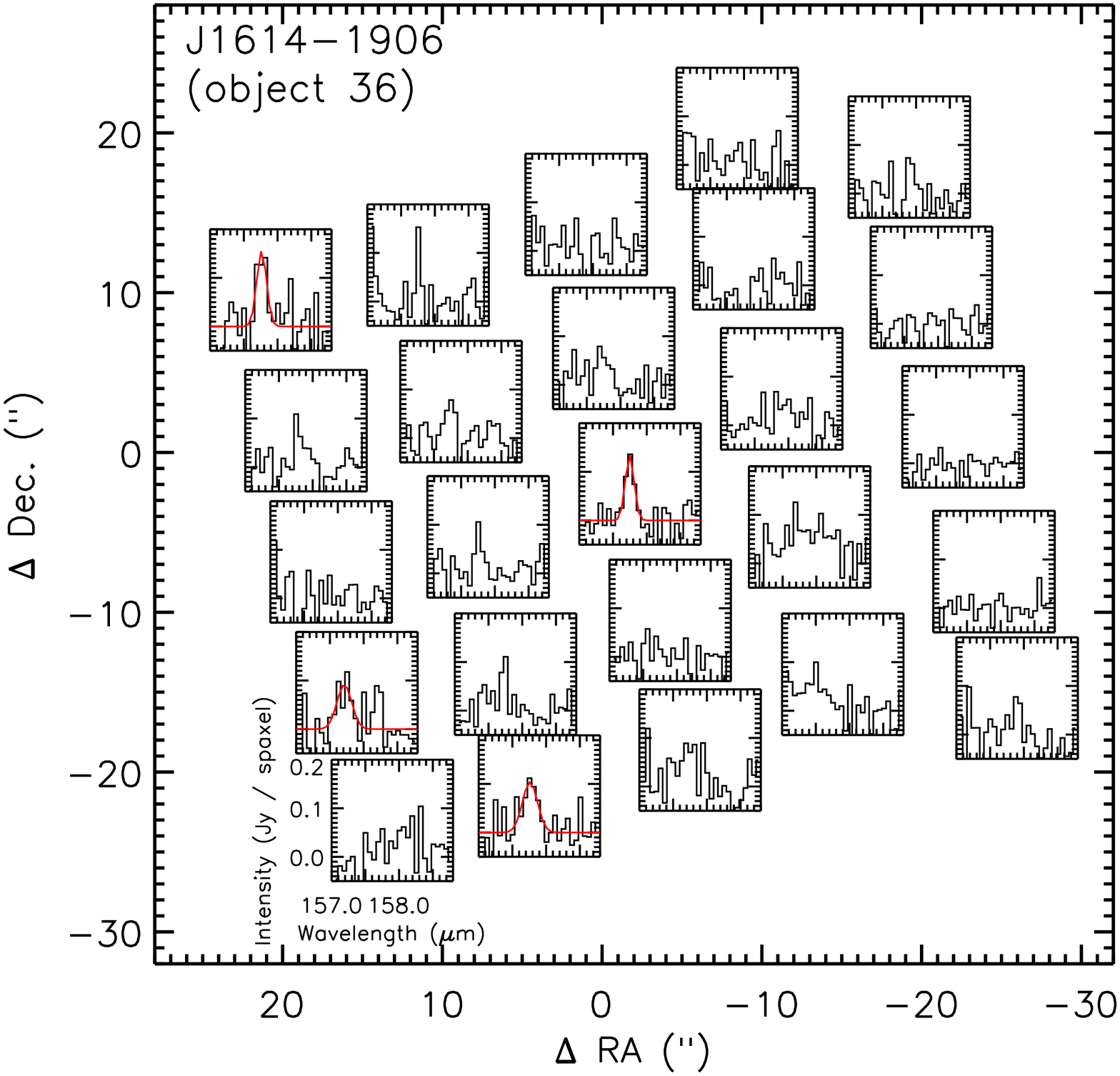}
  \caption{As Fig. \ref{fig:CII-J1609}, but for J1614-1906.}
  \label{fig:CII-J1614}
\end{figure*}

\begin{figure*}[]
\includegraphics[angle=0,width=0.95\textwidth]{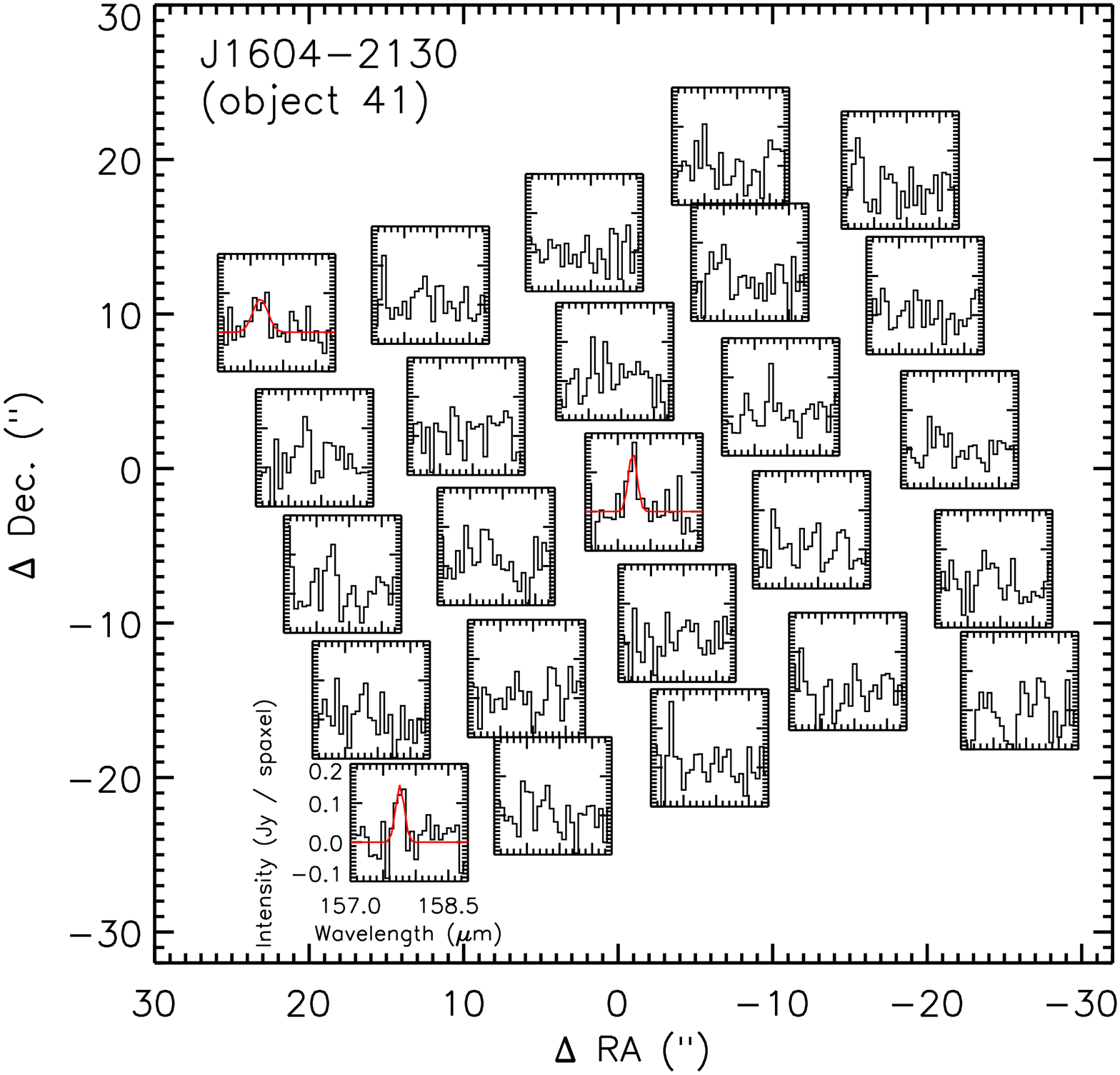}
  \caption{As Fig. \ref{fig:CII-J1609}, but for J1604-2130.}
  \label{fig:CII-J1604}
\end{figure*}

\clearpage
\newpage

\section{Parameter covariance}
\label{sec:chisq-maps}

To confirm the sensitivities to our parameters implied by the probability distributions shown in Fig. \ref{fig:median-params} and to test the assumptions of our fiducial model, we varied the parameters of the Upper Scorpius median model to examine changes in the SED.  We set the values of the free parameters to reflect the size of the explored ranges shown in Fig. \ref{fig:median-params}, generated SEDs for these modified values, and calculated the ratio of the modified SEDs to that of the adopted median model (Figs. \ref{fig:vary-params}).

While each free parameter has effects throughout the SED, we discuss the wavelengths at which the effects are largest.  The free parameter with the least overall effect is $\alpha$.  Varying its value by $\pm$0.4 changes fluxes at wavelengths from $\sim20 ~ \mu$m to 3 mm by 10--20\%, with smaller changes at shorter wavelengths.  Variations in dust mass have, as expected, a nearly linear effect on the mm-SED.  However, factor of 10 variations in dust mass still cause factor of 2 variations in the SED at 70--160 $\mu$m.  ~$\beta$ and $H_{100}$ have complementary effects at the 20--40\% level for variations of $\pm0.05$ and $\pm2$, respectively, and $R_{\rm{in}}$ causes factor of 2 changes throughout the infrared SED for a factor of 10 increase in its value.

\begin{figure*}[]
\includegraphics[angle=0,width=0.45\textwidth]{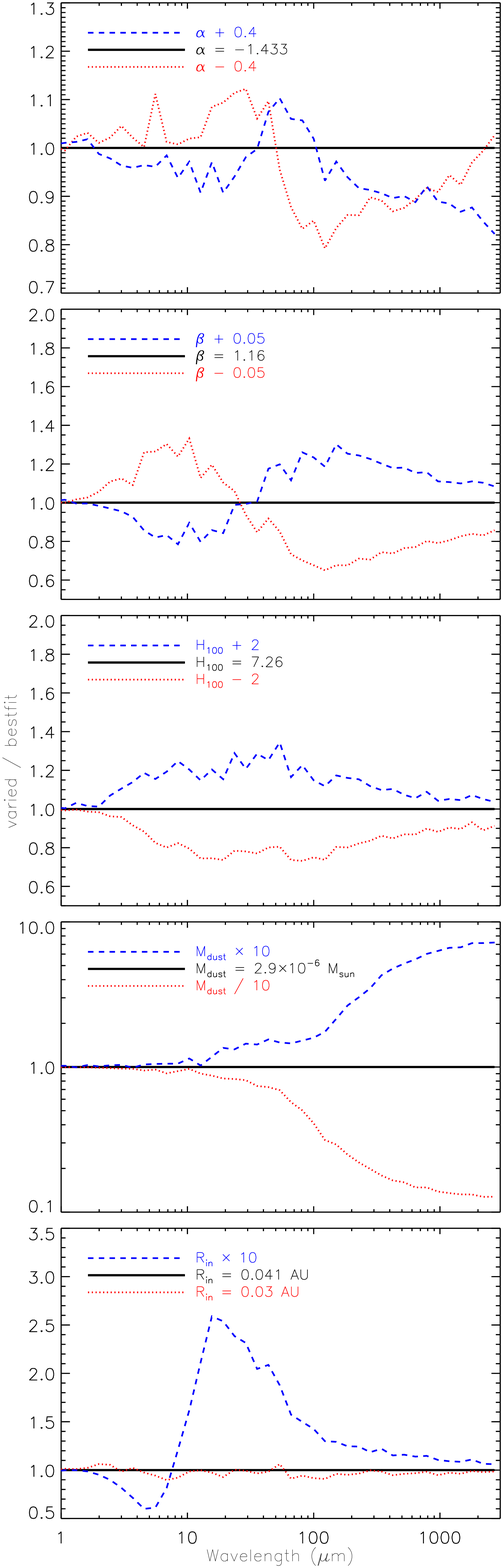}  \includegraphics[angle=0,width=0.45\textwidth]{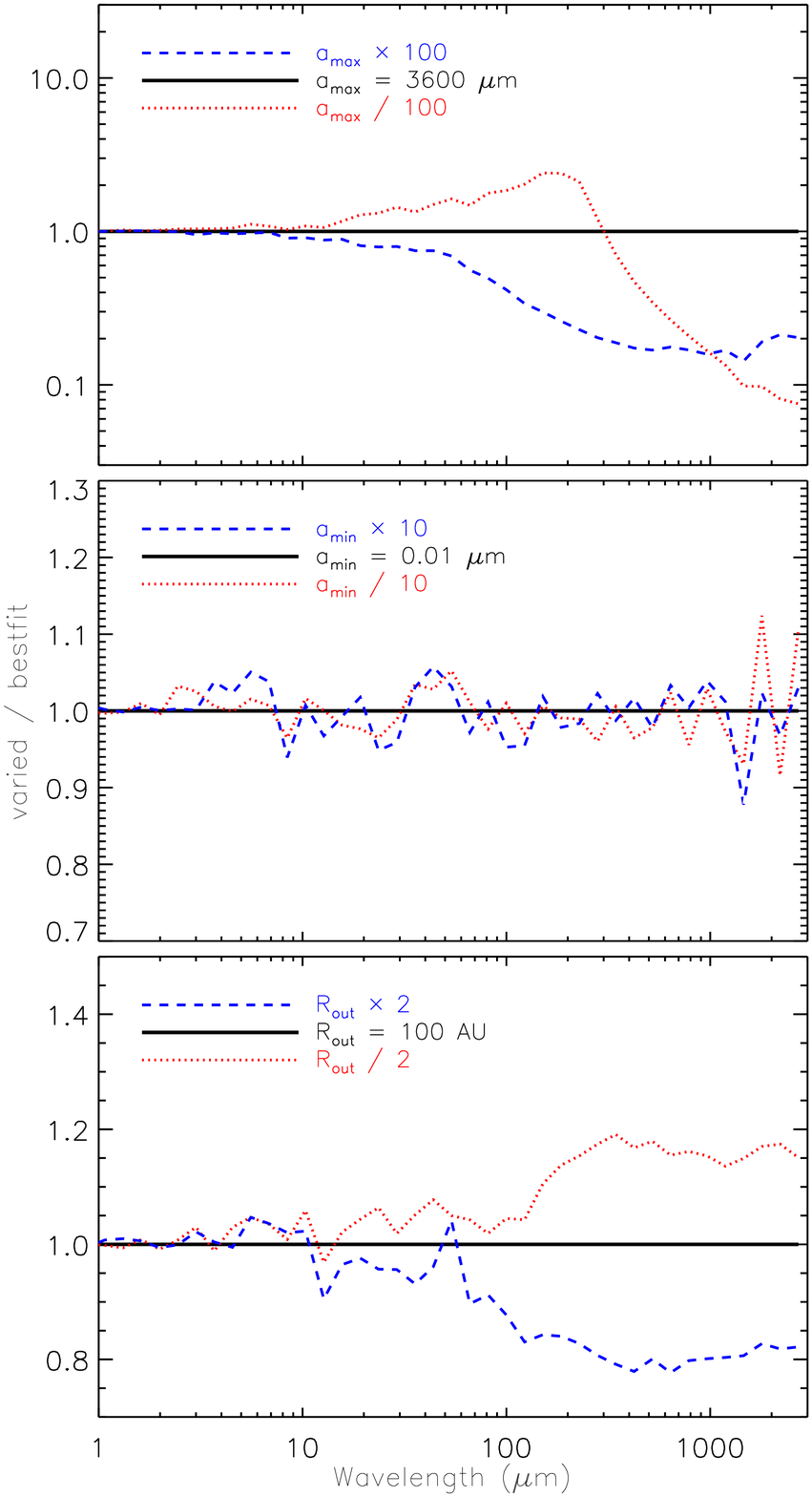} \\
  \caption{Ratio of SEDs for models with varied parameters compared to the adopted parameters for the Upper Scorpius median disk.  The variation of the SED with the free parameters is shown on the left, while the variation of three selected fixed parameters is shown on the right.}
  \label{fig:vary-params}
\end{figure*}

We also tested three key assumptions of the models, that of fixing the minimum and maximum dust grain sizes ($a_{\rm{min}}$ = 0.01 $\mu$m and $a_{\rm{max}}$ = 3600 $\mu$m, respectively), and the outer radius of the disk ($R_{\rm{out}}$ = 100 AU).  A factor of 100 increase in $a_{\rm{min}}$ leads to a few percent decrease in flux at J, H, and K bands, $\sim$10\% flux increase at $\sim5$--50$\mu$m, and $\sim$10\% decrease at longer wavelengths.  Decreasing $a_{\rm{min}}$ has negligible effects.  Increasing $a_{\rm{max}}$ by a factor of 100 for a constant dust mass leads to a factor of $\sim$5 drop in model fluxes from 400 to 1000 $\mu$m, an effect smoothly increasing from about 30\% drop in flux at 70 $\mu$m.  This change is due to shifting dust mass into large, undetectable $\sim10$ cm pebbles and the corresponding decrease in opacity.  Decreasing $a_{\rm{max}}$ by a factor of 100, on the other hand, leads to a similar factor of $\sim$10 drop in the SED at long wavelengths, but a comparatively slight factor of 2 increase in flux at 100 to 200 $\mu$m.  Increasing and decreasing $R_{\rm{out}}$ by a factor of 2 has an effect ranging from $\sim10\%$ at 100 $\mu$m to 20\% at 1000 $\mu$m.

We can then consider how variations in these parameters impact each other.  The simplest case to examine is that of impacts on the dust mass, which varies in essentially a one-to-one fashion with the millimeter flux due to low opacity.  Variation of the magnitude explored here for $\alpha$, $\beta$, $H_{100}$, and $a_{\rm{min}}$ ($\pm$0.4, $\pm$0.05, $\pm$2, and $\times$100, respectively) lead to $\sim$10--20\% variations across a wide range of the SED.  These would in turn lead to at most 10--20\% variations in dust mass, and could be less since variations in the other parameters can together compensate.  An increase in $M_{\rm{dust}}$ for a given observed SED could be driven by large increases or decreases in $\alpha$, or by decreasing values of $\beta$ or $H_{100}$.  Conversely, increasing values of $\beta$ or $H_{100}$, both of which would suggest more effective heating at large radii, would require a decrease in dust mass.  None of these parameters could singly vary with mass, however, due to the effects they have at short wavelengths.

Several parameters, however, primarily affect the long wavelength SED, and thus their impact on the dust mass can be more directly examined.  For a given observed millimeter flux, a factor of 100 increase in $a_{\rm{max}}$ or factor of 2 increase in $R_{\rm{out}}$ will require a factor of 5 or a 20\% increase in the disk dust mass, respectively.  

In order to carry out a broader exploration of the parameter space than is possible with either the single 5D-grid used in section 6.1 to generate the parameter probability distributions or with a single GA search, we have carried out 10 GA searches of 10 generations each for the Upper Scorpius median disk.  Each independent search begins in a random region of the parameter space, thus ensuring a broader sampling.

Using the models from all 10 searches, we have constructed 2D $\chi^2$ maps comparing the two parameters that dominate variation in the disk opacity, $M_{\rm{dust}}$ and $\alpha$, and the parameters which dominate variation in the interception of the stellar luminosity, $H_{100}$ and $\beta$ (Fig. \ref{fig:chisq-maps}).  The number of bins in each dimension (15) was chosen to ensure a mean of $\sim$50 points per bin.  Each model's contribution to its bin is weighted according to its likelihood based on 10 degrees of freedom.  Comparisons with $R_{\rm{in}}$ showed no significant covariances, with 99\% likelihood contours extending uniformly from values of 0.03 to 0.1 AU. 

The shapes of the $\chi^2$ spaces reinforce our discussion of parameter covariance.   $M_{\rm{dust}}$ and $\alpha$ are positively correlated to increasing values of $\alpha$, while there is no or a slight negative correlation to decreasing values of $\alpha$ (Fig. \ref{fig:chisq-maps}, upper left).  The surface density distribution parameter, $\alpha$, appears poorly constrained by our observations (upper right), as was previously shown by the 1D probability distribution.  The negative correlation between $H_{100}$ and $M_{\rm{dust}}$ is consistent with the idea that the observed long-wavelength flux is determined by both the temperature and mass of the outer disk, and these two parameters are the primary drivers of those properties.  The strong positive correlation between $H_{100}$ and $\beta$ suggests that low infrared fluxes require a relatively low scale height in the inner disk region.  Increases in these two parameters will primarily drive increased heating in the outer disk, which can be compensated for by a decreased disk mass.  

Our adopted parameters lie within the 99\% contours in all cases.  However, comparison of these 2D explorations of the parameter space with the results of single GA searches and with the 1D probability distribution produced from the uniformly sampled 5D grid (Section 6.1) suggest improvements for future work.  The use of several short GA searches achieves wider sampling of the parameter space, due largely to the start of the searches in random locations.  The convergence of several searches to a limited range of values lends confidence that a global $\chi^2$ minimum has been found.  The combination of these results can then lead to adoption of a better representative model, which in turn will serve as an improved central point for a uniform grid. 

\begin{figure*}[]
\includegraphics[angle=0,width=0.95\textwidth]{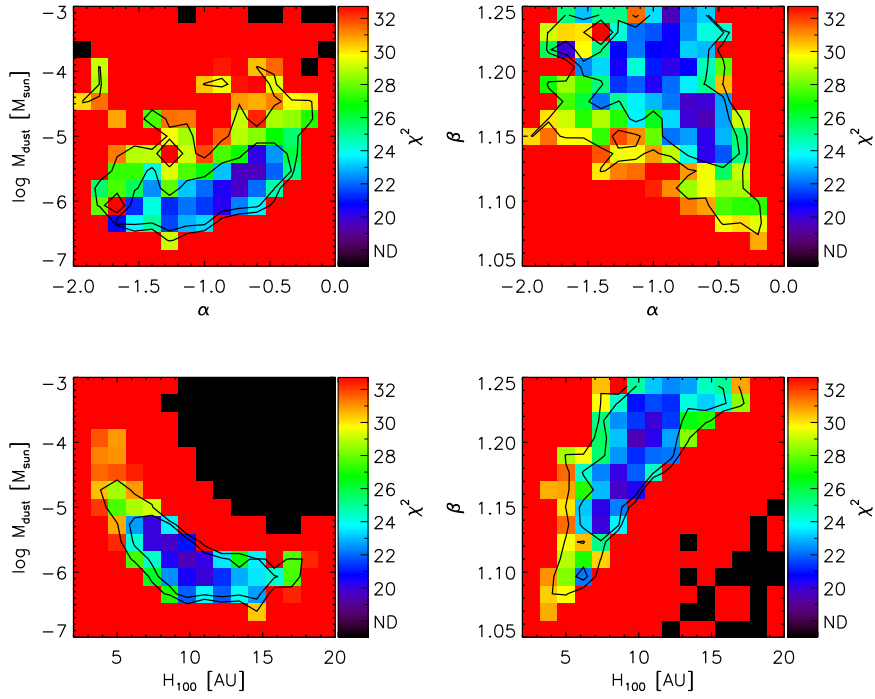}
  \caption{2-dimensional $\chi^2$ maps for $M_{\rm{dust}}$, $\alpha$, $H_{100}$ and $\beta$, generated from 10 genetic-algorithm parameter searches of the Upper Scorpius median disk.  Contours are drawn at levels of 4.6 and 9.21 higher than the minimum $\chi^2$, corresponding to the 90\% and 99\% probability boundaries.  Black regions have sample numbers too low (n $<$ 10) for calculation of $\chi^2$ value.}
  \label{fig:chisq-maps}
\end{figure*}

\end{appendix}

\end{document}